\documentclass[seceqn,preprintnumbers,superscriptaddress,nobalancelastpage]{elsart1p}

\usepackage[hypertex]{hyperref}
\usepackage{graphicx}
\usepackage{dcolumn} 
\usepackage{bm,bbm} 
\usepackage{curves}
\usepackage{epic}
\usepackage{epsfig}
\usepackage{amsmath}
\usepackage{amssymb}
\usepackage[mathcal]{euscript}
\usepackage{epsf,times}
\usepackage[dvips]{color}

\def\re#1{(\ref{#1})}
\def\Tr{\mathop{\mathrm{Tr}}}
\def\widebar{\overline}

\journal{Nuclear Physics B}

\begin{document}

\begin{frontmatter}

\title{
The superspin approach to a disordered quantum wire
in the chiral-unitary symmetry class
with an arbitrary number of channels
      }

\author[kitp]{Andreas P.\ Schnyder},
 \author[psi]{Christopher Mudry},
\author[uch]{Ilya A.\ Gruzberg}

\address[kitp]{
Kavli Institute for Theoretical Physics, University of California,
Santa Barbara, California 93106, USA
}

\address[psi]{
Condensed Matter Theory Group, Paul Scherrer Institute,
CH-5232 Villigen PSI, Switzerland
 }

\address[uch]{
The James Franck Institute, The University of Chicago,
5640 S.\ Ellis Avenue, Chicago, Illinois 60637, USA
}

\begin{abstract}

We use a superspin Hamiltonian defined on an
infinite-dimensional Fock space with positive definite scalar
product to study localization and delocalization of
noninteracting spinless quasiparticles in quasi-one-dimensional
quantum wires perturbed by weak quenched disorder. Past works
using this approach have considered a single chain. Here, we
extend the formalism to treat a quasi-one-dimensional system:
a quantum wire with an arbitrary number of channels
coupled by random hopping amplitudes. The
computations are carried out explicitly for the case of a
chiral quasi-one-dimensional wire with broken time-reversal
symmetry (chiral-unitary symmetry class). By treating the space
direction along the chains as imaginary time, the effects of
the disorder are encoded in the time evolution induced by a single site
superspin (non-Hermitian) Hamiltonian.
We obtain the density of states near the band
center of an infinitely long quantum wire. Our results agree
with those based on the Dorokhov-Mello-Pereyra-Kumar equation
for the chiral-unitary symmetry class.

\end{abstract}

\begin{keyword}
Disordered systems
\sep Localization
\sep Supersymmetry
\sep Path integral
\sep Mesoscopics
\PACS 03.65.Pm \sep 71.23.An \sep 72.15.Rn \sep 71.23.-k \sep 11.30.Rd
\end{keyword}

\end{frontmatter}

\section{Introduction}
\label{sec: Introduction}

It has been 50 years since Anderson wrote his seminal paper
that came to define the problem of Anderson localization, a
tight-binding model (in three-dimensional space) for
noninteracting quasiparticles subject to a static disorder
(random uncorrelated on-site energies) \cite{Anderson58}. The
insight of Anderson was to realize that disorder, however weak,
can have dramatic consequences on the nature of the
eigenfunctions. In three dimensions,
disorder turns extended states of the clean system into
exponentially localized states when the characteristic kinetic
energy of the extended states, measured with respect to edge of
the band, is of the same order or smaller than the
characteristic energy of the disorder.

For fermionic quasiparticles, this ratio can be tuned by
changing the number of occupied single-particle states, i.e.,
by changing the chemical potential. The value of this ratio at
which single-particle eigenstates are neither exponentially
localized nor extended, if it exists, is called a mobility
edge. The mobility edge realizes a quantum critical point,
i.e., there exists a diverging length scale as it is
approached~\cite{Lee85,Kramer93,Kramer05,Evers08}. On the
metallic side of this continuous phase transition, disorder can
be treated perturbatively thereby defining the diffusive
regime. On the insulating side of this continuous transition,
it is the kinetic energy that can be treated perturbatively
\cite{Anderson58, Frohlich83}. A solid analytical grasp of the
quantum criticality at a mobility edge remains an outstanding
problem in Anderson localization.

The existence and characterization of a quantum critical point
at a mobility edge for fermionic single-particle states was
argued on the basis of a scaling hypothesis combined with
perturbative calculations
(see Refs.\ \cite{Lee85,Kramer93,Evers08} for reviews).
The tool of choice
to implement the scaling hypothesis in a system of infinite
size has been to encode the effects of weak disorder on the
fermionic quasiparticles into an effective interacting field
theory called a nonlinear-sigma model (NLSM). The space spanned
by the fields of the NLSM, the target manifold, is determined
by the intrinsic symmetries obeyed by the random microscopic
Hamiltonian
\cite{Dyson62,Verbaarschot94,Zirnbauer96,Altland97,Caselle04,Heinzner05}.
These intrinsic symmetries are the presence or absence of the
symmetry under time-reversal, spin-rotation, or
charge-conjugation, respectively \cite{Heinzner05}. By
demanding that the time-evolution generated by the random
microscopic Hamiltonian is unitary and that it preserves the
fermionic anticommutation relations, the target manifold of the
corresponding NLSM belongs to one of the 10 classical symmetric
spaces \cite{Helgason78} in either their compact, noncompact,
or supersymmetric incarnations \cite{Zirnbauer96}. The notion
of 10 symmetry classes has thus emerged as a corollary to the
scaling hypothesis, here implemented by 10 families of NLSMs.
The local and global properties of a target manifold fixes the
number of independent coupling constants that enters in a NLSM.
From this point of view, a (nonexhaustive) classification of
the mobility edges for fermionic quasiparticles in
one, two, and three dimensions
\cite{footnote mobility edge} into universality classes can be
deduced from the curvature and the homotopy group of the 10
classical symmetric spaces
\cite{Wegner79,Hikami80,Schafer80,Efetov80,Houghton80,Hikami81,%
Efetov83,Pruisken84,%
Oppermann90,Gade93,Zirnbauer92,%
Bundschuh98,Senthil98,Senthil00,Zirnbauer00,Read00,Guruswamy00,%
Fendley01,Altland01,Lamacraft04,Ryu06,Ostrovsky07,Ryu07}. This
construction of a NLSM starting from a random microscopic
Hamiltonian relies on the existence of a diffusive regime
\cite{Efetov83}. However, mobility edges in $d=1,2,3$
dimensions are usually far away from the
diffusive regime. The predictive power of NLSM beyond the
existence of mobility edges is therefore severely limited by
ones ability to study NLSM at their strong coupling fixed
points.

The validity of the scaling hypothesis has been tested with the
help of nonperturbative numerical calculations on random
tight-binding or network models. The hypothesis of
one-parameter scaling in quantum wires as well as the existence
of mobility edges in $d=1,2,3$ dimensions
for different intrinsic symmetries of the microscopic models is
consistent with diverse numerical calculations in systems of
finite-sizes
(see Refs.\ \cite{Kramer93,Kramer05,Evers08} for reviews). In this
context, random network models play an important role
\cite{Chalker88,DKKLee94,Kim95,Chalker95,Klesse95,Ho96,Cho97,Kagalovsky97,%
Merkt98,Kagalovsky99,Chalker01,Bocquet03,Obuse07}. Consider for
example the plateau transition in the integer quantum Hall
effect (IQHE) (for a review see Ref.\ \cite{Huckestein95}).
Chalker and Coddington (CC) have proposed a random network
model to study the effects of disorder on the highly degenerate
single-particle states in the first Landau level of a
two-dimensional electronic gas in a strong magnetic field when
the cyclotron length is much smaller than the characteristic
length over which the random potential varies significantly
\cite{Chalker88}. The CC network model simplifies greatly the
study of the plateau transition in the first Landau level. For
example, the position of the mobility edge that separates two
integer Hall insulating states in the CC network model follows
from a duality argument. Properties of the mobility edge in the
IQHE can thus be simulated numerically for large system sizes
and large statistical ensembles for the disorder with the help
of the CC network model. In addition, the CC network model has
been used as a regularization of the corresponding NLSM.
Namely, the CC model has been mapped to a certain
one-dimensional antiferromagnetic (super)spin chain, with the
space of states being the tensor product of alternating highest
and lowest weight infinite-dimensional irreducible
representations (irreps) of a Lie (super)algebra, also loosely
called ``(super)spins'', and the Hamiltonian (an infinitesimal
version of the transfer matrix of the CC model) built out of
generators of the same algebra. The (super)symmetry of the spin
chain was the same as that of the target space of the NLSM
\cite{Read91,DHLee94,Zirnbauer94,Lee96,Kondev97,Zirnbauer97,Marston99}.
This was an important conceptual step in the sense that it
provided the hope that the flow to strong coupling of the
corresponding NLSM could be captured at the level of the
(super)spin chain in a way similar as the flow from the
$O(3)$-NLSM to the $SU(2)^{\ }_{1}$ Wess-Zumino-Witten (WZW)
conformal field theory is captured by the Bethe Ansatz solution
to the quantum spin$-1/2$ Heisenberg chain \cite{Affleck86}.
Although, this hope proved ephemeral in the context of the
plateau transitions (for a review see Ref.\
\cite{Zirnbauer97}), it lead to remarkable exact results when
the same approach was used in the case of the two-dimensional
Bogoliubov-de-Gennes (BdG) symmetry class with spin-rotation
invariance but with broken time-reversal symmetry, and the
corresponding network model
\cite{Kagalovsky99,Gruzberg99,Cardy00,Beamond02,Mirlin03,Cardy05}.

Unfortunately, this success story is the exception rather than
the rule at the present. For example, consider the cases of
spinless fermionic tight-binding models with either random
real-valued or random complex-valued hopping amplitudes,
whereby the hopping amplitude is restricted to connecting pair
of sites with each site belonging to a different sublattice of
a two-dimensional bipartite lattice. Such tight-binding
Hamiltonians have the property that their spectra of energy
eigenvalues are symmetric under charge conjugation, i.e., they
are left invariant under sign reversal whatever the realization
of the random hopping amplitudes. This is most easily seen by
changing the sign of the single-particle eigenfunction for all
sites belonging to one sublattice. As this property is also
shared by the Dirac Hamiltonian in quantum chromodynamics for a
given configurations of the gluonic gauge fields, it is called
a chiral symmetry (chiral-orthogonal symmetry when the
Hamiltonian is real-valued and chiral-unitary symmetry
otherwise) \cite{Verbaarschot94}. In one and two dimensions,
the band center in the chiral-unitary or
chiral-orthogonal symmetry classes realizes a mobility edge
separating two insulating phases very much in the same way as
is the case with the plateau transitions in the IQHE.

The study of localization in the chiral-unitary symmetry class
in one-dimensions was pioneered by Dyson \cite{Dyson53}. Dyson
found a diverging density of states on approaching the band
center for a random microscopic Hamiltonian describing a
single-particle hopping with a random amplitude between the
nearest-neighbor sites of an infinitely long open chain. The
band center realizes a quantum critical point that separates
two insulating phases. This quantum critical point has been
investigated using various methods such as a mapping onto a
random $XY$ model \cite{Smith70}, solving a recurrence relation
\cite{Theodorou76,Fleishman77}, solving a stationary
Fokker-Planck equation \cite{Ovchinnikov77}, the Berezinskii
diagram technique \cite{Berezinskii73,Gogolin77}, solving a
one-dimensional diffusion process with reflecting and absorbing
boundary conditions at the left and right ends
\cite{Eggarter78}, respectively, solving the evolution equation
of a transfer matrix \cite{Stone81}, a mapping onto Witten's
supersymmetric quantum mechanics \cite{Witten81,Comtet95}, a
mapping to Liouville quantum mechanics \cite{Shelton98}, and,
at last, a mapping onto a spin-Hubbard-like model
\cite{Balents97,Bocquet99,Gruzberg01}. (We note here that in
this case the spin-like model involves two sites with a highest
and lowest weight representations of a Lie
superalgebra paired in a supersymmetric fashion. In this
setting one has to solve the problem of the two couples
superspins, analogous to the ``spin addition'' in quantum
mechanics, albeit more complicated due to the infinite
dimensionality of the superspins.) At the band center, the
typical wavefunction is not exponentially localized
\cite{Fleishman77}, while the logarithm of the conductance is
not a selfaveraging random variable \cite{Stone81}, as would
otherwise occur on both sides of the mobility edge . A weaker
divergence of the density of states on approaching the band
center survives in the two-dimensional chiral-orthogonal and
chiral-unitary symmetry classes
\cite{Gade93,Guruswamy00,Motrunich02,Mudry03}. This divergence
is captured by effective field theories for weak disorder that
include two-dimensional NLSMs with compact, noncompact, or
supersymmetric target manifolds, respectively. These
two-dimensional NLSMs can be regularized by two-dimensional
random network models \cite{Bocquet03}. In turn, the latter can
be mapped onto spin-Hubbard-like chains, which, however, have
been intractable to this date. One of the reasons is that a
straightforward generalization of the methods of Refs.
\cite{Read91,DHLee94,Zirnbauer94,Lee96,Kondev97,Zirnbauer97,Marston99,
Gruzberg99,Balents97,Bocquet99,Gruzberg01} leads to spaces of
states on the sites of the chain that are {\it not} irreducible
representations of the corresponding Lie superalgebra. 
The structure of these spaces from the point of
view of the supersymmetry of the underlying physical problem is
a so far unresolved question.

A NLSM approach to the model studied by Dyson is not possible
as there is no diffusive regime in one dimension.
This difficulty can be remedied by considering
thick quantum wires, which are also often
referred to as quasi-one-dimensional wires. A thick quantum
wire with a large number of transverse
channels $N \gg 1$ appears in the limit where
both the bare conductance and the mean free path $\ell$ scale
with $N$. This limiting procedure justifies the derivation of a
one-dimensional NLSM with any one of the 10 classical symmetric
spaces as the target manifold. In particular, one can derive a
one-dimensional NLSM with the target manifold corresponding to
the chiral-unitary symmetry class. In the same way as a
two-dimensional NLSM can be regularized by a spin chain, a
one-dimensional NLSM can be regularized by a
zero-dimensional spin-Hubbard-like model with
one or two sites. Solving such a spin-Hubbard-like model is a
reasonable first step towards understanding the critical
properties of the two-dimensional models in
the chiral-unitary symmetry class.

The purpose of this paper is to calculate the density of states
in the close vicinity of the band center of a quantum wire
belonging to the chiral unitary symmetry class for an arbitrary
number $N$ of channels by using a representation of the problem
in terms of an effective zero-dimensional ``superspin'' model.
As we have pointed out earlier, in this case the space of
states is not obtained from irreducible representations of a
superalgebra. However, the Hamiltonian of the effective model
can be written as a bilinear form in the generators of a
superalgebra, and it is this construction that we refer to as
``the superspin approach''. The density of states has been
obtained by the superspin approach only for $N=1$
\cite{Balents97,Bocquet99,Gruzberg01}.

The density of states of
quasi-one-dimensional wires with arbitrary $N$ in the chiral
symmetry classes has been obtained
using an entirely different approach \cite{Brouwer00,Titov01}
based on the Dorokhov-Mello-Pereyra-Kumar (DMPK) equation.
This is an equation for the joint probability
distribution of the Lyapunov exponents that are related to the
eigenvalues of the transfer matrix \cite{DMPK} along the wire
viewed as a multi-channel scatterer. There it was shown that
the density of states depends sensitively on the parity of the
number of channels $N$. When $N$ is odd, the Dyson singularity
is recovered. When $N$ is even, the density of states is
controlled by random matrix theory up to multiplicative
logarithmic corrections. This paper is devoted to uncovering
this parity effect within the superspin formalism applied to
the chiral-unitary symmetry class, the simplest among all 10
symmetry classes from the point of view of the DMPK equation
\cite{Brouwer00,Mudry99}.
Whereas the approach based on the DMPK equation is geometric%
~\cite{Huffmann90,Brouwer00b}, the method to be presented below
is algebraic. However, the DMPK equation is limited to
quasi-one-dimensional wires, while our hope
is that the algebraic insights gained in this work can be of
use to solving spin-Hubbard-like models related to
two-dimensional NLSMs in the chiral symmetry classes.

The band center of a superconducting quantum wire with broken
spin-rotation symmetry is another example of a mobility edge
separating two insulating phases \cite{Titov01,Furusaki00}.
Remarkably, the density of states displays the same Dyson
singularity as a chiral quantum wire with an odd number of
channels \cite{Titov01,Furusaki00}. This suggests a deep
connection between the corresponding symmetric spaces that has
been partially explored from a geometric point of view in
Refs.\ \cite{Motrunich01,Gruzberg05}. Hereto we hope that the
algebraic method used in this paper could shed some light on
this connection.

As our computation of the density of states for a
quasi-one-dimensional wire in the chiral-unitary symmetry class
is quite involved, we now summarize the main steps leading to
the dependence
\begin{equation}
\nu(\varepsilon)\sim
\left\{
\begin{array}{ll}
(\varepsilon\tau)^{-1}\left|\ln^{-3}(\varepsilon\tau)\right|,
&
N
\hbox{ odd,}
\\
&
\\
(\varepsilon\tau)|\ln(\varepsilon\tau)|,
&
N
\hbox{ even,}
\end{array}
\right.
\label{eq: main result paper}
\end{equation}
of the density of states $\nu(\varepsilon)$ at the positive
energy $\varepsilon$ measured relative to the band center
whereby $0<\varepsilon\tau\ll 1$ with the mean scattering time
$\tau$ of order $N^2\ell/v^{\ }_{F}$ ($\ell$ is the mean free
path and $v^{\ }_{F}$ the Fermi velocity at the band center
without disorder). The divergence of the density of states when
$N$ is odd is nothing but the Dyson divergence for the single
chain random hopping problem. For even $N$, there is a
multiplicative logarithmic correction to the power law
predicted by random matrix theory.

In Sec.~\ref{sec: The quasi-one dimensional random hopping chain}, 
we define a quasi-one-dimensional random
hopping model on the square lattice
of length $L=Ma$ and width $Na$ with $a$ being the lattice
spacing and $L\gg Na$. We also provide a supersymmetric
representation for the global density of states. In
Sec.~\ref{sec: Mapping to a transfer Hamiltonian}, the density
of states in the thermodynamic limit $L\to\infty$, $L\gg Na$,
is recast as a certain expectation value in the ground state of
an operator $H$ acting on the direct product of a fermionic and
a bosonic Fock spaces with a positive definite scalar product.
The operator $H$ is an infinitesimal version of the transfer
matrix along the wire written in second quantized language,
and, therefore, we call $H$ the (effective) Hamiltonian, even
though it is not Hermitian so that we need to distinguish its
right and left eigenstates. Hamiltonian $H$ is quartic in terms
of the fermionic and bosonic operators spanning the Fock space
on which it acts. Moreover, Hamiltonian $H$, although not
Hermitian, possesses a certain degree of supersymmetry, which
allows us to construct its ground sate and thus to compute the
density of states of the underlying quasi-one-dimensional wire.
Finally, Hamiltonian $H$ depends parametrically on the energy
scale $\varepsilon$ at which the density of states of the wire
is to be evaluated and on the energy scale $1/\tau$
characterizing the strength of the static disorder in the wire.
The construction of the ground state of $H$ is done in Sec.%
~\ref{sec: Construction of the ground state sector}, 
where it is expanded in terms of certain basis states and
where we find a recursion relation for the expansion coefficients. 
We do not solve this recursion relation
exactly, but we solve it and calculate the corresponding
density of states approximately close to the band center
$\varepsilon \tau \ll 1$.
The approximate solution is first presented in Sec.%
~\ref{sec: Evaluation of the density of states when N=1 and
N=2} for $N=1$ and $N=2$, and then for arbitrary
even or odd $N$ in Sec.%
~\ref{sec: Evaluation of the density of states when N=1,2,3,...}.
We summarize the important steps of
our derivation and make a comparison with the DMPK derivation
in Sec.~\ref{sec: Discussion}.

\section{The quasi-one-dimensional random hopping model}
\label{sec: The quasi-one dimensional random hopping chain}

In this paper we reconsider the density of states for a single
quantum particle in a thick quantum wire.
We start with a quasi-one-dimensional tight-binding system of
spinless fermions with nearest-neighbor and
next-nearest-neighbor random hopping matrix elements. On the
square lattice the tight-binding Hamiltonian for $N$ coupled
random hopping chains takes the form
\begin{equation}
\mathcal{H}=
-
\sum_{i,j=1}^N
\sum_{n=1}^M
\left(
t^{\ }_{n;ij}
c^{\dagger}_{n,i}c^{\ }_{n+1,j}
+
t^{\prime }_{n;ij}
c^{\dagger}_{n,i}c^{\ }_{n+2, j}
+ \textrm{h.\ c.\ }
\right).
\label{eq: random hopping lattice hamiltonian}
\end{equation}
Here, the indices $i$ and $j$ label the $N$ chains
while $n$ labels the rows of sites in the
direction along the wire. We impose the periodic boundary
conditions along this direction, so that $M+1\equiv1$, etc.
The spinless fermions are represented by operators that satisfy
canonical anticommutation relations. The nearest-neighbor
hopping strength $t^{\ }_{n;ij}$ consists  of a large uniform
part $t \delta^{\ }_{ij}$ and a small random piece $\delta t^{\
}_{n;ij}$, whereas the next-nearest-hopping strength
$t^{\prime}_{n;ij}$ is assumed to be purely random and small.
That is, the chains are only weakly coupled by small random
hopping amplitudes. Perpendicular to the chains there is no
notion of distance, as every chain is coupled with equal
strength to every other one. We shall consider the
quasi-one-dimensional limit, where the length $L$ of the
disordered wire is much larger than its width $Na$, with $a$
the lattice constant.

For vanishing next-nearest-neighbor hopping, $t'_{n;ij}=0$, the
Hamiltonian~\re{eq: random hopping lattice hamiltonian} reduces
to the chiral random hopping model. This model is special
because its energy eigenvalue spectrum exhibits a symmetry
under charge conjugation, i.e., the Hamiltonian changes sign
under the unitary transformation
 \begin{equation}
\left( c^{\ }_{n,i}  ,   c^{\dag}_{n,i} \right)
 \rightarrow
 (-1)^{n}  \left( c^{\ }_{n,i}  ,   c^{\dag}_{n,i} \right).
 \end{equation}
Hence, the eigenvalue spectrum of the chiral random hopping
model has a reflection symmetry around the band center. As we
shall see below, because of this extra symmetry the states near
zero energy have localization properties that are dramatically
different from those of the states in other parts of the spectrum.

Under the assumption of weak disorder,
 \begin{equation}
 \delta t^{\ }_{n;ij}, t^{\prime}_{n;ij} \ll t,
 \end{equation}
it is legitimate to take the continuum limit in the direction
along the chains. In the absence of disorder the ground state
of each chain consists of a filled Fermi sea bounded by two 
distinct Fermi points. The low energy excitations, 
the plane waves with momenta near these two points, 
are commonly called left- and right-movers,
respectively. Linearizing the spectrum about the two Fermi
points yields a kinetic energy that is a first-order
differential operator. We thus model the dynamics of the single
quantum particle in a quasi-one-dimensional geometry by
\begin{equation} 
\begin{split}
&
\mathcal{H}^{\ }_{\mathrm{c}}=
\int\limits_{0}^{L} dx\,
\psi^{\dagger} (x) \,  h^{\ }_{\mathrm{c}}(x) \, \psi^{ } (x) \, ,
\\
&
h^{\ }_{ \mathrm{c} } (x)
=
-\mathrm{i}
\sigma^{\ }_3
\partial^{\ }_x
+
\sigma^{\ }_{0}
v^{\ }_{0}(x)
+
\sigma^{\ }_{1}
v^{\ }_{1}(x)
+
\sigma^{\ }_{2}
v^{\ }_{2}(x)
+
\sigma^{\ }_3
v^{\ }_{3}(x),
\end{split}
\label{eq: continuum Hamiltonian N chains}
\end{equation}
where we choose our units such that $\hbar$ and the Fermi
velocity $v^{\ }_{F}$ are one. The spinors $\psi(x)$ are
$2N$-component vectors, $v^{\ }_\mu (x)$ ($\mu=0,1,2,3$) are
$N\times N$ random Hermitian matrices, and $\sigma^{\ }_\mu$
denote the three Pauli matrices and the $2 \times 2$ unit
matrix acting on the left-right-mover degrees of freedom. Due
to gauge invariance the potential $v^{\ }_3(x)$ can be chosen
to be zero in a system with open boundary conditions. In the
continuum language the chiral symmetry is implemented by the
interchange of left and right movers and is represented by
\begin{equation}
\sigma^{\ }_{1} h^{\ }_{\mathrm{c}} (x) \sigma^{\ }_{1} =
-
h^{\ }_{\mathrm{c}} (x) .
\end{equation}
Chiral symmetry is thus only satisfied if the random matrices
$v^{\ }_{0}$ and $v^{\ }_{1}$ vanish, i.e., a chiral symmetric
disorder potential is here necessarily off-diagonal with
respect to the left-right-mover degrees of freedom. In the
presence of time-reversal symmetry, $ \sigma^{\ }_{1} h^{\ast
}_{\mathrm{c}} \sigma^{\ }_{1} = h^{\ }_{\mathrm{c}}, $ the
disorder potentials satisfy the additional symmetry constraints
\begin{equation}
\label{eq: time-reversal syms}
v^{*}_{0;ij} = v^{\ }_{0;ij},\quad
v^{*}_{1;ij} = v^{\ }_{1;ij},\quad
v^{*}_{2;ij} = v^{\ }_{2;ij},\quad
\end{equation}
where $i,j=1,\cdots,N$. The disorder potentials $v^{\ }_\mu$
are assumed to be independent and Gaussian distributed with
zero means and and with variances
\begin{equation}
\left[
v^{\ }_{\mu; ij}(x)
v^{* }_{\mu; kl}(x^\prime)
\right]^{\ }_{\mathrm{av}}
=
2g^{\ }_\mu \delta(x -x^\prime)
\left[
\delta^{\ }_{ik}\delta^{\ }_{jl}
+
(2/\beta -1)
\delta^{\ }_{il}\delta^{\ }_{jk}
\right],
\label{eq: gaussian disorder N chains}
\end{equation}
respectively, where $\mu =0,1,2$ and the chain index
$i,j,k,l=1,\cdots,N$. The disorder strength is denoted by
$g_\mu$ and $\beta$ is the Dyson index with $\beta=1(2)$ in the
presence (absence) of time-reversal symmetry.

We want to compute the Green function of the Hamiltonian
\re{eq: continuum Hamiltonian N chains} in order to obtain its
density of states at the energy $\varepsilon$, which we assume
to be positive, $\varepsilon>0$, without loss of generality.
The two-point single-particle Green function is given by
\begin{equation}\label{eq: def Green funtion for fixed disorder}
G(x,i,\alpha|x',i',\alpha';\varepsilon+\mathrm{i}\eta)
=
\left
\langle
x,i,\alpha
\left|
\left.
\frac{1}{h^{\ }_{\mathrm{c}}-\varepsilon-\mathrm{i} \eta}
\right.
\right|
x',i',\alpha'
\right
\rangle ,
\end{equation}
where $\alpha,\alpha'=L,R$ are left-right-mover indices,
$i,i'=1,\cdots,N$ are channel indices, the infinitesimal
regulator $\eta $ is strictly positive, $\eta>0$, and $\left|
x,i,\alpha \right\rangle$ denotes a position eigenstate. The global density of states
$\nu(\varepsilon)$ for a given realization of the disorder is
then obtained from
\begin{equation}\label{eq: def DOS}
\nu(\varepsilon)=
\lim_{\eta\to0}
\frac{1}{L}\int\limits_{0}^{L} dx\,
\frac{1}{N}
\sum_{i=1}^{N}
\sum_{\alpha=L,R}
\frac{1}{\pi}
\mathrm{Im}\,
G(x,i,\alpha|x,i,\alpha;\varepsilon+\mathrm{i}\eta).
\end{equation}
It is convenient to introduce the notation
\begin{equation}
\mathrm{i}\omega:=
\varepsilon+\mathrm{i}\eta
\
\Longleftrightarrow
\
\omega:=
-\mathrm{i}\varepsilon+\eta.
\end{equation}
The Green function~(\ref{eq: def Green funtion for fixed
disorder}) is an analytic function of $\mathrm{i}\omega$ in the
upper part of the complex plane for any realization of the
disorder. By analytical continuation of $\varepsilon>0$ to the
upper imaginary axis, $\mathrm{i}\omega$ is strictly imaginary
($\omega$ becomes strictly positive). If so, one has to
analytically continue $\omega$ to the lower imaginary axis in
order to obtain the density of states~(\ref{eq: def DOS}).
Translation invariance is restored in the disorder average of
the Green function, i.e., one can define
\begin{equation}\label{eq: def mean Green funtion}
G(x-x';\mathrm{i}\omega)
=
\frac{1}{N}
\left[
\sum_{i=1}^{N}
\sum_{\alpha=L,R}
G(x,i,\alpha|x',i,\alpha;\mathrm{i}\omega)
\right]^{\ }_{\mathrm{av}},
\end{equation}
where the square brackets on the right-hand side denote
disorder averaging. Correspondingly, 
the mean global density of states is
\begin{equation}\label{eq: mean DOS}
[\nu(\varepsilon)]^{\ }_{\mathrm{av}}=
\frac{1}{L}\int\limits_{0}^{L} dx\,
\lim_{\omega\to-\mathrm{i}\varepsilon}
\lim_{x'\to x}
\frac{1}{\pi}
\mathrm{Im}\,G(x-x';\mathrm{i}\omega).
\end{equation}
The global density of states becomes self-averaging in the
thermodynamic limit $L\to\infty$, so that the brackets on the
left-hand side of Eq.~(\ref{eq: mean DOS}) can be omitted.

The single-particle Green functions%
~(\ref{eq: def Green funtion for fixed disorder}) or%
~(\ref{eq: def mean Green funtion}) 
can be represented by functional path
integrals as long as $\mathrm{i}\omega$ belongs to the upper
part of the complex plane. Unless specified, we assume that
$\omega>0$. To perform the disorder average we make use of
supersymmetry, so that
\begin{subequations}
\begin{equation} 
G(x-x';\mathrm{i}\omega)=
\frac{\mathrm{i}}{N}
\left[
\frac{1}{Z}
\sum_{i=1}^{N}
\sum_{\alpha=L,R}
\int
D\psi^{* }D\psi\,
D\xi^{* } D\xi\;
\psi^{ }_{i,\alpha} (x)\psi^{*}_{i,\alpha} (x^{\prime})
\;
e^{-S}
\right]^{\ }_{\mathrm{av}},
\label{eq: susy G}
\end{equation}
where the action
\begin{equation} \label{eq: susy S}
S=
\int\limits_{0}^{L} d x \, \mathcal{L},
\end{equation}
is defined as the one-dimensions integral of the Lagrangian
\begin{equation} 
\mathcal{L}
=
\psi^{\dag}
\left(
\mathrm{i}h^{\ }_{\mathrm{c}}
+
\omega
\right)
\psi^{\ }
+
\xi^{\dag}
\left(
\mathrm{i}h^{\ }_{\mathrm{c}}
+
\omega
\right)
\xi^{\ },
\label{eq: susy L}
\end{equation}
and the partition function
\begin{equation} 
Z:=
\int
D\psi^{* }D\psi\,
D\xi^{* } D\xi\;
\;
e^{-S}.
\label{eq: partition fun 1}
\end{equation}
\end{subequations}
We have introduced, for any channel index $i=1,\cdots,N$ and
any left-right mover index $\alpha=L,R$, the complex-valued
integration variables $\xi^{* }_{i,\alpha}(x)$ and $\xi^{\
}_{i,\alpha}(x)$, here related by complex conjugation, together
with the pairs of independent Grassmann fields $\psi^{*
}_{i,\alpha}(x)$ and $\psi^{\ }_{i,\alpha}(x)$.
In the functional integral formulation, the
periodic boundary conditions imposed on the Hamiltonian%
~(\ref{eq: random hopping lattice hamiltonian}) 
translate as follows: 
the bosonic fields obey periodic boundary conditions,
but the fermionic fields obey antiperiodic boundary conditions
in the $x$ direction. The assumption $\omega>0$ guarantees the
convergence of the Gaussian integral over the complex-valued
integration variables $\xi^{* }_{i,\alpha}(x)$ and $\xi^{\
}_{i,\alpha}(x)$. The action and the measure
of integration in the field theory~\re{eq: susy G} contains
supersymmetries that rotate boson and fermion fields into each
other. These symmetries guarantee that the corresponding
partition function is unity in every disorder
realization: $Z = 1$,
which simplifies the computation of averaged Green functions
considerably. The ensemble average over the disorder
configurations in Eq.~\re{eq: susy G} can be performed
analytically by means of a cumulant expansion. With the
Gaussian distribution~(\ref{eq: gaussian disorder N chains}),
this gives the exact result
\begin{subequations}
\begin{equation} 
G(x-x';\mathrm{i}\omega)=
\frac{\mathrm{i}}{N}
\sum_{i=1}^{N}
\sum_{\alpha=L,R}
\int
D\psi^{* }D\psi\,
D\xi^{* } D\xi\;
\psi^{ }_{i,\alpha}(x)\psi^{*}_{i,\alpha}(x^{\prime})
\;
e^{-S^{\ }_{\mathrm{eff}}},
\label{eq: susy G eff}
\end{equation}
where the effective action
\begin{equation}\label{eq: S eff}
S^{\ }_{\mathrm{eff}}=
\int\limits_{0}^{L} d x \,
\mathcal{L}^{\ }_{\mathrm{eff}}(x)
\end{equation}
is defined from the effective Lagrangian
\begin{equation}\label{eq: mathcal L eff}
\begin{split}
\mathcal{L}^{\ }_{\mathrm{eff}}(x)=&\,
\mathrm{Tr}\,
\left(
-
\psi^{* }_{i,L}\partial^{\ }_x \psi^{\ }_{j,L}
+
\psi^{* }_{i,R}\partial^{\ }_x\psi^{\ }_{j,R}
-
\xi^{* }_{i,L} \partial^{\ }_x \xi^{\ }_{j,L}
+
\xi^{* }_{i,R} \partial^{\ }_x \xi^{\ }_{j,R}
+
\omega\mathcal{J}^{\ }_{0;ij}
\right)(x)
\\
&\,
+
\sum_{\mu=0}^{2}
g^{\ }_{\mu}
\Tr
\left[
\mathcal{J}^{\ }_{\mu}
\mathcal{J}^{\ }_{\mu}
+
(2/\beta-1)
\mathcal{J}^{\ }_{\mu}
\mathcal{J}^{T }_{\mu}
\right](x).
\end{split}
\end{equation}
Here, $\Tr\,(\cdots)$ refers to the trace over the channel
index, we made use of the Hermiticity of the three random
$N\times N$ matrix $v^{\ }_{0,1,2}$, and we also introduced the
$N\times N$ Hermitian matrices $\mathcal{J}^{\ }_{0,1,2}$ and
their transposed $\mathcal{J}^{T }_{0,1,2}$ through their
matrix elements
\begin{equation}
\begin{split}
&
\mathcal{J}^{\ }_{0;ij}
\equiv
\psi^{* }_{i,L}
\psi^{\ }_{j,L}
+
\psi^{* }_{i,R}
\psi^{\ }_{j,R}
+
\xi^{* }_{i,L}
\xi^{\ }_{j,L}
+
\xi^{* }_{i,R}
\xi^{\ }_{j,R},
\\
&
\mathcal{J}^{\ }_{1; ij}
\equiv
\psi^{* }_{i,R}
\psi^{\ }_{j,L}
+
\psi^{* }_{i,L}
\psi^{\ }_{j,R}
+
\xi^{* }_{i, R}
\xi^{\ }_{j,L}
+
\xi^{* }_{i,L}
\xi^{\ }_{j,R},
\\
&
\mathcal{J}^{\ }_{2; ij}
\equiv
\mathrm{i}
\left(
\psi^{* }_{i,R}\psi^{\ }_{j,L}
-
\psi^{* }_{i,L}\psi^{\ }_{j,R}
+
\xi^{* }_{i,R}\xi^{\ }_{j,L}
-
\xi^{* }_{i,L}\xi^{\ }_{j,R}
\right),
\end{split}
\label{eq: def J 0,1,2}
\end{equation}
\end{subequations}
with $i,j=1,\cdots,N$.

We are going to demonstrate how we can use Feynman's transfer
matrix method to trade the one-dimensional field theory encoded
by Eqs.~(\ref{eq: susy G eff}), (\ref{eq: S eff}), and
(\ref{eq: mathcal L eff}) with $\omega>0$ for a
zero-dimensional Schr\"odinger equation with a non-Hermitian
superspin Hamiltonian. We will show that the non-Hermitian
Hamiltonian possesses real eigenvalues and has manifestly
unitary ``imaginary-time evolution''. We will also show how the
density of states~(\ref{eq: mean DOS}) can be extracted from an
expectation value in the nondegenerate ground state annihilated
by this non-Hermitian Hamiltonian.

\section{
Mapping to a transfer Hamiltonian
        }
\label{sec: Mapping to a transfer Hamiltonian}

To perform the mapping onto a superspin Hamiltonian, we start
from the \textit{classical} Hamiltonian
\begin{subequations}
\label{eq: def H}
\begin{equation}
\mathcal{L}(p,q) -
\sum_{i=1}^{N}\sum_{\mathrm{deg}=0,1}\sum_{\alpha=L,R} p^{\
}_{i,\mathrm{deg},\alpha} \, \dot{q}^{\
}_{i,\mathrm{deg},\alpha}, \label{eq: def H a}
\end{equation}
that can be read off from the Lagrangian~\re{eq: susy L}
once $4N$ coordinates
\begin{equation}
q^{\ }_{i,\mathrm{deg},\alpha},
\qquad
i=1,\cdots,N,
\qquad
\mathrm{deg}=0,1,
\qquad
\alpha=1,2,
\end{equation}
have been chosen and by use of the $4N$ momenta
\begin{equation}\label{eq: def H b}
p^{\ }_{i,\alpha} :=
\frac{ \delta \mathcal{L}}{ \delta \dot{q}^{\ }_{i,\alpha} },
\quad
\dot{q}^{\ }_{i,\alpha}\equiv\partial^{\ }_{x}q^{\ }_{i,\alpha},
\quad
i=1,\cdots,N,
\quad
\mathrm{deg}=0,1,
\quad
\alpha=1,2,
\end{equation}
\end{subequations}
whereby $\delta / \delta \dot{q}_{i,\alpha}$ denotes the right
derivative. We are thus interpreting $x$ as the imaginary time.
There is considerable freedom in the choice of the $q$'s and
hence of the $p$'s. The most obvious choice is
\begin{subequations}
\begin{equation}
q^{\ }_i \equiv \left( q^{\ }_{i,0}, 	 q^{\ }_{i,1}\right),
\qquad
q^{\ }_{i,0} = \left(\xi^{\ }_{i,L} ,  \xi^{\ }_{i,R}\right),
\qquad
q^{\ }_{i,1} = \left(\psi^{\ }_{i,L}, \psi^{\ }_{i,R}\right),
\end{equation}
for which the momenta are
\begin{equation}
p^{\ }_i \equiv \left(p^{\ }_{i,0},        p^{\ }_{i,1}\right),
\qquad
p^{\ }_{i,0} = \left(-\xi^{\ast}_{i,L} , \xi^{\ast}_{i,R}\right),
\qquad
p^{\ }_{i,1} = \left(-\psi^{\ast}_{i,L},\psi^{\ast}_{i,R}\right),
\end{equation}
with the channel index running over $i=1,\cdots,N$.
With this choice,
\begin{eqnarray}
\mathcal{L} &=&
p\cdot\dot{q}
+
\Tr
\big(
\omega\mathcal{J}^{\ }_{0}
+
\mathrm{i}v^{\ast }_{0} \mathcal{J}^{\ }_{0}
+
\mathrm{i}v^{\ast }_{1} \mathcal{J}^{\ }_{1}
+
\mathrm{i}v^{\ast }_{2} \mathcal{J}^{\ }_{2}
\big) ,
\end{eqnarray}
where we have introduced the short-hand notation
\begin{equation}
p\cdot\dot{q}\equiv
\sum_{i=1}^{N}
\sum_{\mathrm{deg}=0,1}
\sum_{\alpha=L,R}
p^{\ }_{i,\mathrm{deg},\alpha}
\dot{q}^{\ }_{i,\mathrm{deg},\alpha}
\end{equation}
and, for any $i,j=1,\cdots,N$,
Eq.~(\ref{eq: def J 0,1,2})
now reads
\begin{equation}\label{eq: J 0,1,2 operator}
\begin{split}
&
\mathcal{J}^{\ }_{0;ij}
\equiv
-p^{\ }_{i,1,L} q^{\ }_{j,1,L}
+p^{\ }_{i,1,R} q^{\ }_{j,1,R}
-p^{\ }_{i,0,L} q^{\ }_{j,0,L}
+p^{\ }_{i,0,R} q^{\ }_{j,0,R},
\\
&
\mathcal{J}^{\ }_{1; ij}
\equiv
+p^{\ }_{i,1,R} q^{\ }_{j,1,L}
-p^{\ }_{i,1,L} q^{\ }_{j,1,R}
+p^{\ }_{i,0,R} q^{\ }_{j,0,L}
-p^{\ }_{i,0,L} q^{\ }_{j,0,R},
\\
&
\mathcal{J}^{\ }_{2; ij}
\equiv
\mathrm{i}
\left(
 p^{\ }_{i,1,R} q^{\ }_{j,1,L}
+p^{\ }_{i,1,L} q^{\ }_{j,1,R}
+p^{\ }_{i,0,R} q^{\ }_{j,0,L}
+p^{\ }_{i,0,L} q^{\ }_{j,0,R}
\right).
\end{split}
\end{equation}
\end{subequations}
Observe that the bilinear form $\mathcal{J}^{\
}_{0}\equiv(\mathcal{J}^{\ }_{0;ij})$ has the undesirable
property that it is not positive definite. Hence, the
\textit{quantum} operator corresponding to Eq.~(\ref{eq: def
H}) is seemingly ill defined as it has an unbounded spectrum of
eigenvalues from below upon quantization by which the classical
$p$'s and $q$'s are replaced by operators that obey the
canonical supercommutation relations
\begin{equation}\label{eq: canonical supercommutation relations}
\begin{split}
\left[
\hat q^{\ }_{i ,\mathrm{deg} ,\alpha },
\hat p^{\ }_{i',\mathrm{deg}',\alpha'}
\right]
:=&\,
\hat q^{\ }_{i ,\mathrm{deg} ,\alpha }
\hat p^{\ }_{i',\mathrm{deg}',\alpha'}
-
(-)^{\mathrm{deg}\,\mathrm{deg}'}
\hat p^{\ }_{i',\mathrm{deg}',\alpha'}
\hat q^{\ }_{i ,\mathrm{deg} ,\alpha }
=
\delta^{\ }_{i,i'}
\delta^{\ }_{\alpha,\alpha'} ,
\end{split}
\end{equation}
for any $i,i'=1,\cdots,N,$ $\mathrm{deg},\mathrm{deg}'=0,1$,
and $\alpha,\alpha'=L,R$. To cure this difficulty it is
convenient to redefine the vacuum in the fermionic sector of
the theory to be the Dirac sea. However, by the spin-statistics
theorem, this medicine is inoperative in the  bosonic sector
unless one is willing to either give up Hermiticity
\cite{Bocquet99} or the positive definite scalar product of the
Fock space on which the quantum operator corresponding to
Eq.~(\ref{eq: def H}) is defined \cite{Gruzberg01}.
In this paper, we choose the former route, so
our quantum Hamiltonian will act on a space with a
positive-definite scalar product, but will not be self-adjoint
with respect to this product. In this setting, one will have to
distinguish between the right and left eigenstates of the
Hamiltonian.

For example, consider the quantum theory defined by the assignments
\begin{subequations}
\label{eq: def quantum operators}
\begin{equation}
\begin{split}
&
\psi^{* }_{i,R} \to f^{\dag}_{i},
\quad
\psi^{\ }_{i,R} \to f^{\ }_{i},
\quad
\xi^{* }_{i,R} \to b^{\dag}_{i},
\quad
\xi^{\ }_{i,R} \to b^{\ }_{i},
\\
&
\psi^{* }_{i,L} \to  \bar{f}^{\ }_{i},
\quad
\psi^{\ }_{i,L} \to - \bar{f}^{\dag}_{i},
\quad
\xi^{* }_{i,L} \to  \bar{b}^{\ }_{i},
\quad
\xi^{\ }_{i,L} \to  \bar{b}^{\dag}_{i},
\end{split}
\end{equation}
with the only nonvanishing supercommutators
\begin{eqnarray}
&&
\left\{
f^{\   }_{i },
f^{\dag}_{j }
\right\}
=
\left\{
\bar{f}^{\   }_{i },
\bar{f}^{\dag}_{j }
\right\}=
\left[
b^{\   }_{i },
b^{\dag}_{j }
\right]=
\left[
\bar{b}^{\ }_i,
\bar{b}^{\dag}_j
\right]=
\delta^{\ }_{ij},
\qquad
i,j=1,\cdots,N.
\label{eq: f bar f b bar b super algebra}
\end{eqnarray}
\end{subequations}
The corresponding Fock space $\mathfrak{F}^{\ }_{\mathrm{R}}$
is obtained by action of the creation operators on the
vacuum state, i.e., the state $|0\rangle$ that is annihilated
by the right action of 
$f^{\ }_{i},\bar{f}^{\ }_{i},b^{\ }_{i},\bar{b}^{\ }_{i}$ 
with $i=1,2,\cdots,N$, i.e.,
$\mathfrak{F}^{\ }_{\mathrm{R}}$ 
is the linear span of the product states
\begin{subequations}
\label{eq: product states}
\begin{equation}
\prod_{i=1}^{N}
\left(     f ^{\dag}_{i}\right)^{n^{\ }_{     f^{\ }_{i}}}
\left(\bar{f}^{\dag}_{i}\right)^{n^{\ }_{\bar f^{\ }_{i}}}
\left(     b ^{\dag}_{i}\right)^{n^{\ }_{     b^{\ }_{i}}}
\left(\bar{b}^{\dag}_{i}\right)^{n^{\ }_{\bar b^{\ }_{i}}}
|0\rangle   ,
\end{equation}
where
\begin{equation}
n^{\ }_{     f^{\ }_{i}},n^{\ }_{\bar f^{\ }_{i}}=0,1,
\qquad
n^{\ }_{     b^{\ }_{i}},n^{\ }_{\bar b^{\ }_{i}}=0,1,2,\cdots.
\end{equation}
We assume that the vacuum state is normalized to one,
\begin{equation}
\left\langle0|0\right\rangle=1,
\end{equation}
\end{subequations}
whereby the dual vacuum state $\langle0|$ is the state
annihilated by the left action of
$f^{\dag}_{i},\bar{f}^{\dag}_{i},b^{\dag}_{i},\bar{b}^{\dag}_{i}$
with $i=1,2,\cdots,N$. The identification \re{eq: def quantum
operators} leads us to define the coherent states
\begin{subequations} \label{eq: coherent states}
\begin{equation}
\begin{split}
\left| \psi^{\ }_R \right\rangle^{\ }_{\mathrm{R}}
&=
e^{- \psi^{\ }_{i,R} f^{\dag}_i}  \left|  0 \right\rangle,
\;
{}^{\ }_{\mathrm{L}}\!\left\langle \psi^{\ }_R \right|
=
\left\langle 0 \right|
e^{- \psi^{\ast }_{i,R} f^{\ }_i },
\;
\left| \xi^{\ }_R \right\rangle^{\ }_{\mathrm{R}}
=
e^{+ \xi^{\ }_{i,R} b^{\dag}_i}  \left|  0 \right\rangle,
\;
{}^{\ }_{\mathrm{L}}\!\left\langle \xi^{\ }_R \right|
=
\left\langle 0 \right| e^{+ \xi^{\ast }_{i,R} b^{\ }_i } ,
\\
\left| \psi^{\ }_L \right\rangle^{\ }_{\mathrm{R}}
&=
e^{- \psi^{\ast }_{i,L} \bar{f}^{\dag}_i } \left|  0 \right\rangle,
\;
{}^{\ }_{\mathrm{L}}\!\left\langle \psi^{\ }_L \right|
=
\left\langle 0 \right| e^{+ \psi^{\  }_{i,L} \bar{f}^{\ }_i} ,
\;
\left| \xi^{\ }_L \right\rangle^{\ }_{\mathrm{R}}
=
e^{+ \xi^{\ast }_{i,L} \bar{b}^{\dag}_i } \left|  0 \right\rangle,
\;
{}^{\ }_{\mathrm{L}}\!\left\langle \xi^{\ }_L \right|
=
\left\langle 0 \right| e^{+ \xi^{\ }_{i,L} \bar{b}^{\ }_i } ,
\end{split}
\end{equation}
where summation over $i=1,\cdots,N$ is implied,
with the overlaps
\begin{equation}
\begin{split}
{}^{\ }_{\mathrm{L}}\!\left\langle \psi^{\ }_{R} \right|
\left. \psi^{\prime }_R \right\rangle^{\ }_{\mathrm{R}}
&=
e^{+\psi^{\ast}_{i,R} \psi^{\prime }_{i,R}} ,
\quad
{}^{\ }_{\mathrm{L}}\!\left\langle \xi^{\ }_{R} \right|
\left. \xi^{\prime }_R \right\rangle^{\ }_{\mathrm{R}}
=
e^{+\xi^{\ast}_{i,R} \xi^{\prime }_{i,R}},
\\
{}^{\ }_{\mathrm{L}}\!\left\langle \psi^{\ }_{L} \right|
\left. \psi^{\prime }_L \right\rangle^{\ }_{\mathrm{R}}
&=
e^{- \psi^{\ }_{i,L} \psi^{\prime \ast}_{i,L}} ,
\quad
{}^{\ }_{\mathrm{L}}\!\left\langle \xi^{\ }_{L} \right|
\left. \xi^{\prime }_L \right\rangle^{\ }_{\mathrm{R}}
=
e^{+\xi^{\ }_{i,L} \xi^{\prime \ast }_{i,L}} .
\end{split}
\end{equation}
\end{subequations}
We observe that $\left| \psi^{\ }_L \right\rangle^{\
}_{\mathrm{R}}$ and ${}^{\ }_{\mathrm{L}}\!\left\langle \psi^{\
}_L \right| $ are not related by the ``conventional'' adjoint
operation $(\cdots)^{\dag}$. Instead, the right coherent states
$\left| \psi^{\ }_L \right\rangle^{\ }_{\mathrm{R}}$ can be
mapped onto the left coherent states $\left| \psi^{\ }_L
\right\rangle^{\ }_{\mathrm{L}}$ by a unitary transformation
$U$,
\begin{equation}
\label{eq: def unitary trafo 1}
\left| \psi^{\ }_L \right\rangle^{\ }_{\mathrm{L}}
=
\left(
{}^{\ }_{\mathrm{L}}\!\left\langle \psi^{\ }_L \right|
\right)^{\dag}
=
U \left| \psi^{\ }_L \right\rangle^{\ }_{\mathrm{R}},
\qquad
U:=
\exp
\left(
\mathrm{i}\pi\bar{f}^{\dag}_i\bar{f}^{\ }_i
\right).
\end{equation}
Due to this fact, it is necessary to distinguish between
the right Fock space $\mathfrak{F}^{\ }_{\mathrm{R}}$ 
and its dual, the left Fock space 
$\mathfrak{F}^{* }_{\mathrm{L}}$, that are related by the
unitary transformation (\ref{eq: def unitary trafo 1}). That
is, for a given basis $\left| \Psi^{\ }_m \right\rangle^{\
}_{\mathrm{R}}$ of the right Fock space $\mathfrak{F}^{\
}_{\mathrm{R}}$, we define the corresponding basis $\left|
\Psi^{\ }_m \right\rangle^{\ }_{\mathrm{L}}$ of the left Fock
space  $\mathfrak{F}^{\ }_{\mathrm{L}}$ by acting with $U$ on
the right basis, i.e., $\left| \Psi^{\ }_m \right\rangle^{\
}_{\mathrm{L}}= U\left| \Psi^{\ }_m \right\rangle^{\
}_{\mathrm{R}}$. The dual Fock space $\mathfrak{F}^{*
}_{\mathrm{L}}$ is then the linear span of the basis $ {}{\
}_{\mathrm{L}}\! \left\langle \Psi^{\ }_m \right| \equiv
\left(\left| \Psi^{\ }_m \right\rangle^{\
}_{\mathrm{L}}\right)^{\dag} $. Similarly, we will need to
distinguish between left and right eigenstates of the transfer
Hamiltonian $H$.

Equipped with the definition~\re{eq: coherent states} for the
coherent states, we are now ready to derive the effective
transfer Hamiltonian. It is given by replacing $\mathcal{L}$ in
Eq.~(\ref{eq: def H a}) with $\mathcal{L}^{\ }_{\mathrm{eff}}$
from Eq.~(\ref{eq: mathcal L eff}). After disorder averaging,
the transfer Hamiltonian reads
\begin{subequations}
\label{eq: final hamiltonian n chains}
\begin{equation}
\begin{split}
&
H=
H^{\ }_{\omega}
+
H^{\ }_{\mathcal{V}},
\\
&
H^{\ }_{\omega}=
\omega \Tr J^{\ }_{0}
=
\omega
\left(
N^{\ }_{f^{\ }}
+
N^{\ }_{\bar{f}^{\ }}
+
N^{\ }_{b^{\ }}
+
N^{\ }_{\bar{b}^{\ }}
\right),
\\
&
H^{\ }_{\mathcal{V}}=
\sum_{\mu=0}^2
g^{\ }_{\mu}
\Tr
\left[
J^{\ }_{\mu}
J^{\ }_{\mu}
+
(2/\beta-1)
J^{\         }_{\mu}
J^{\mathrm{T}}_{\mu}
\right],
\end{split}
\label{eq: final hamiltonian n chains a}
\end{equation}
with the number operators and disorder-induced operators defined by
\begin{equation}
N^{\ }_{b^{\ }}:=
\sum_{i=1}^N b^{\dag}_i
b^{\   }_i,
\quad
N^{\ }_{\bar{b}^{\ }}:=
\sum_{i=1}^N
\bar{b}^{\dag}_i
\bar{b}^{\   }_i,
\quad
N^{\ }_{f^{\ }}:=
\sum_{i=1}^N  f^{\dag}_{i}
f^{\   }_{i},
\quad
N^{\ }_{\bar{f}^{\ }}:=
\sum_{i=1}^N  \bar{f}^{\dag}_i
\bar{f}^{\   }_i,
\label{eq: def number operators}
\end{equation}
and
\begin{equation} \label{def of Jij}
\begin{split}
&
J^{\ }_{0;ij}:=
+
f^{\dag}_{i}
f^{\ }_{j}
-
\bar{f}^{\   }_{i}
\bar{f}^{\dag}_{j}
+
b^{\dag}_{i}
b^{\   }_{j}
+
\bar{b}^{\   }_{i}
\bar{b}^{\dag}_{j},
\\
&
J^{\ }_{1; ij}:=
-
f^{\dag}_{i}
\bar{f}^{\dag}_{j}
+
\bar{f}^{\ }_{i}
f^{\ }_{j }
+
b^{\dag}_{i }
\bar{b}^{\dag}_{j}
+
\bar{b}^{\ }_{i}
b^{\ }_{j},
\\
&
J^{\ }_{2;ij}:=
-
\mathrm{i}
\left(
f^{\dag}_{i}
\bar{f}^{\dag}_{j}
+
\bar{f}^{\ }_{i}
f^{\ }_{j}
-
b^{\dag}_{i}
\bar{b}^{\dag}_{j}
+
\bar{b}^{\ }_{i }
b^{\ }_{j }
\right),
\end{split}
\end{equation}
\end{subequations}
respectively. In as much as the right Fock space
$\mathfrak{F}^{\ }_{\mathrm{R}}$ and its dual, the left Fock
space $\mathfrak{F}^{* }_{\mathrm{L}}$, that are associated
with $H$ are not related by the conventional adjoint operation,
the transfer Hamiltonian $H$ itself is not invariant under the
conventional adjoint operation. On the one hand,
$J^{\dag}_{0,ij} = J^{\ }_{0,ji}$. On the other hand, because
of the fermions, $J^{\dag}_{1,ij} \ne J^{\ }_{1,ji}$, and
$J^{\dag}_{2,ij} \ne J^{\ }_{2,ji}$. Consequently,
\begin{equation}
H^{\ }_{\mathcal{V}}\neq H^{\dag}_{\mathcal{V}}.
\end{equation}
Upon making the transition to the transfer Hamiltonian $H$
as encoded by Eq.~\re{eq: def quantum operators}
the partition function~\re{eq: partition fun 1}
becomes
\begin{equation} \label{eq: partition fun 2}
Z =
\sum_{m}\
{}^{\ }_{\mathrm{L}}\!\big\langle \Psi^{\ }_m \big|
(-1)^{N^{\ }_{F}} e^{-L H }
\big| \Psi^{\ }_m \big\rangle^{\ }_{\mathrm{R}},
\qquad
N^{\ }_{F}:=
N^{\ }_{f}
+
N^{\ }_{\bar{f}},
\end{equation}
where $N^{\ }_{F}$ is the total fermion number operator, and
the sum runs over some basis of the Fock space pair $(
\mathfrak{F}^{\ }_{\mathrm{L}}, \mathfrak{F}^{\ }_{\mathrm{R}}
)$, with $\big|\Psi^{\ }_m\big\rangle^{\ }_{\mathrm{L}}=
U\big|\Psi^{\ }_m\big\rangle^{\ }_{\mathrm{R}}$ for all $m$.
The factor $(-1)^{N^{\ }_{F}}$ is a consequence of the
antiperiodic boundary conditions obeyed by the original
Grassmann integration variables at the ends of the wire.

Finally, we note that for a single channel, $N=1$,
alternative representations of Hamiltonian%
~\re{eq: final hamiltonian n chains a} have been studied in
Refs.~\cite{Balents97}, \cite{Bocquet99}, and \cite{Bunder01}
as well as in connection with the random network problem
introduced in Ref.~\cite{Gruzberg01}. We are now going to show
that the eigenvalues of $H$ for the chiral unitary symmetry
class are manifestly real and that the matrix elements of
$\exp(-LH)$ between left and right eigenvectors of $H$ define
transition amplitudes of a unitary evolution operator, in spite
of the fact that $H$ is non-Hermitian.

\subsection{Properties of the transfer Hamiltonian}

In this section and thereafter we shall limit ourselves
to the chiral-unitary symmetry class defined by the conditions
\begin{equation}
g^{\ }_{0}=g^{\ }_{1}=0,
\qquad
\beta=2,
\end{equation}
in Eq.~(\ref{eq: gaussian disorder N chains}).
Hamiltonian~(\ref{eq: final hamiltonian n chains a})
thus reduces to
\begin{equation}
H=
H^{\ }_{\omega}
+
H^{\ }_{2},
\qquad
H^{\ }_{\omega}
= \omega \Tr J^{\ }_{0} ,
\qquad
H^{\ }_{2}
=
g^{\ }_{2} \Tr J^{\ }_{2} J^{\ }_{2},
\label{eq: final hamiltonian n chains chiral-unitary}
\end{equation}
in the chiral-unitary symmetry class. The original
supersymmetry of the action~(\ref{eq: S eff}) is realized by
the invariance of Hamiltonian~\re{eq: final hamiltonian n
chains a} under the interchanges
\begin{equation}\label{eq: SUSYa}
\left(
     f^{\dag}_i ,      f^{\   }_i,
+\bar f^{\dag}_i , \bar f^{\   }_i
\right)
\leftrightarrow
\left(
     b^{\dag}_i ,      b^{\   }_i,
-\bar b^{\dag}_i , \bar b^{\   }_i
\right)
\qquad
i=1,\cdots,N.
\end{equation}
In addition, for the chiral classes  ($g^{\ }_{0} = g^{\
}_{1}=0$), $H^{\ }_{2}$ is unchanged by the interchanges
\begin{equation}
\left(
f^{\ }_i , f^{\dag}_i,  +b^{\ }_i , b^{\dag}_i
\right)
\leftrightarrow
\left(
\bar{f}^{\dag}_i , \bar{f}^{\ }_i,  -\bar{b}^{\dag}_i , \bar{b}^{\ }_i
\right),
\qquad
i=1,\cdots,N ,
\end{equation}
whereas $H^{\ }_{\omega}$ changes its sign under this
transformation. This symmetry has its origin in the chiral
symmetry. We also note that $H^{\ }_{2}$ commutes with the
eight generators of the Lie superalgebra
$gl(1|1)\oplus gl(1|1)$, while $H^{\ }_{\mathcal{\omega}}$
commutes with the four generators of the diagonal sub-algebra
$gl(1|1)\subset gl(1|1)\oplus gl(1|1)$ 
(see Appendix~\ref{app sec: Symmetries of the Hamiltonian}).

The invariance of $H$ under the supersymmetric transformation%
~(\ref{eq: SUSYa}) has two consequences. First, it implies that
an eigenvalue $E^{\ }_{l}$ of $H$ can only be nondegenerate if
its right eigenvector belongs to a singlet, i.e., if the right
eigenvector is annihilated by the supersymmetric
transformation~(\ref{eq: SUSYa}). Second, any nonvanishing
eigenvalue $E^{\ }_l$ must be at least twofold degenerate with
a pair of right eigenstates that are simultaneous eigenstates
of the total fermion occupation number $N^{\ }_{F}$ that differ
by an odd number of fermions. That is, all the eigenstates with
nonvanishing eigenenergy $E^{\ }_l$ can be grouped into
supersymmetric  multiplets of degenerate eigenstates. The only
exception to this rule, is the nondegenerate ground state, a
singlet state, with $E^{\ }_l = 0$. Since the transfer
Hamiltonian $H$ is non-Hermitian we need to distinguish left
eigenstates from right eigenstates. In order to do so, we
define right eigenstates, $\big|\varphi^{\
}_{l,\iota}\big\rangle^{\ }_{\mathrm{R}} \in \mathfrak{F}^{\
}_{\mathrm{R}}$, with eigenvalues $E^{\ }_{l}$ by
\begin{subequations}
\label{eq: eigenval eq}
\begin{equation}
H\left|\varphi^{\ }_{l,\iota} \right\rangle^{\ }_{\mathrm{R}}=
E^{\ }_{l}\left|\varphi^{\ }_{l,\iota} \right\rangle^{\ }_{\mathrm{R}}
\quad \Longleftrightarrow \quad
{}^{\ }_{\mathrm{R}}\!\left\langle\varphi^{\ }_{l,\iota}\right|H^{\dag}=
{}^{\ }_{\mathrm{R}}\!\left\langle\varphi^{\ }_{l,\iota}\right|E^{\ast}_{l}.
\label{eq: def right eigenstates}
\end{equation}
and left eigenstates, ${}^{\ }_{\mathrm{L}}\!\big\langle
\varphi^{\ }_{l,\iota} \big| \in \mathfrak{F}^{*
}_{\mathrm{L}}$, of $H$ with eigenvalues $E^{\ }_{l}$ by
\begin{equation}
{}^{\ }_{\mathrm{L}}\!\left\langle\varphi^{\
}_{l,\iota}\right|H= {}^{\
}_{\mathrm{L}}\!\left\langle\varphi^{\
}_{l,\iota}\right|E^{\ast }_{l} \quad \Longleftrightarrow \quad
H^{\dag} \left|\varphi^{\ }_{l,\iota} \right\rangle^{\
}_{\mathrm{L}}= E^{\ }_{l}\left|\varphi^{\ }_{l,\iota}
\right\rangle^{\ }_{\mathrm{L}}, \label{eq: def left
eigenstates}
\end{equation}
\end{subequations}
where the index $\iota$ labels the different elements of the
supersymmetric multiplet with eigenenergy $E^{\ }_l$. Since the
right and left eigenstates are elements of the Fock space
$\mathfrak{F}^{\ }_{\mathrm{R}}$ and $\mathfrak{F}^{*
}_{\mathrm{L}}$, respectively, it is possible to map
$\big|\varphi^{\ }_{l,\iota}\big\rangle^{\ }_{\mathrm{R}}$ onto
$\big|\varphi^{\ }_{l,\iota}\big\rangle^{\ }_{\mathrm{L}}$ by
use of the unitary transformation $U$ as defined in Eq.~\re{eq:
def unitary trafo 1},
\begin{equation}
\left| \varphi^{\ }_{l,\iota} \right\rangle^{\ }_{\mathrm{L}} =
U
\left| \varphi^{\ }_{l,\iota} \right\rangle^{\ }_{\mathrm{R}}
 \quad \Longleftrightarrow \quad
{}^{\vphantom{\dag}}_{\mathrm{L}}\!\left\langle \varphi^{\ }_{l,\iota}
\right|
=
{}^{\vphantom{\dag}}_{\mathrm{R}}\!\left\langle \varphi^{\ }_{l,\iota}
\right|
U^{\dag}.
\label{eq: mapping btw RL ev}
\end{equation}
Conversely, the unitary transformation $U$ can be used to
compute the adjoint of both $J^{\ }_{2}$ and the transfer
Hamiltonian $H$. Namely, we find that
\begin{equation} \label{eq: adjoint of Ha}
J^{\dag}_{2} = U \, J^{T}_{2}  \, U^{\dag},
\qquad
H^{\dag}  =  U H U^{\dag},
\end{equation}
where $( \cdots )^{T}$ denotes the transpose operating on the $N$
scattering channels.

As for the single channel problem of Ref.~\cite{Balents97}, 
we can now use the left and right eigenstates~(\ref{eq: eigenval eq}) 
to construct a normalized eigenbasis and
the corresponding resolution of identity.
Thereto, we argue that the left and right eigenstates of $H$ in
any given supersymmetric multiplet with eigenenergy $E^{\ }_l$
can be grouped into pairs 
$ 
\big(
\big|
\varphi^{\ }_{l,\iota}
\big\rangle^{\ }_{\mathrm{R}/\mathrm{L}}, 
\big|
\varphi^{\ }_{l,\sigma[\iota]}
\big\rangle^{\ }_{\mathrm{R}/\mathrm{L}} 
\big)
$
normalized by the condition
\begin{subequations}
\label{eq: orhonormality}
\begin{equation}
{}^{\vphantom{\dag}}_{\mathrm{L}}\!\big\langle \varphi^{\ }_{l,\iota}\big|
\varphi^{\ }_{l,\sigma[\iota]}\big\rangle^{\ }_{\mathrm{R}}
=1,
\label{eq: normalization of ev}
\end{equation}
and such that the left and right eigenstates form
a well defined biorthogonal system with
\begin{equation}
{}^{\ }_{\mathrm{L}}\!\big\langle\varphi^{\ }_{l,\iota}\big|
\varphi^{\ }_{l^{\prime},\iota^{\prime}}\big\rangle^{\ }_{\mathrm{R}}
=
\delta^{\ }_{l,l'} \delta^{\ }_{\sigma[\iota],\iota'}
\end{equation}
\end{subequations}
for all energy eigenvalue indices $l,l'$ and for all
supersymmetric multiplet index $\iota,\iota'$. Here, we have
introduced the permutation function $\sigma[\iota]$ of the
index $\iota$ that parametrizes the pairing of the eigenstates.
The permutation function is involutive, i.e., it satisfies
$\sigma^{2} = \mathrm{1}$. The resolution of identity is then
given by
\begin{equation} \label{eq: res identity}
\sum_{l,\iota}
\big|\varphi^{\ }_{l,\sigma[\iota]}\big\rangle^{\ }_{\mathrm{R}}
{
\vphantom{\big\rangle^{\ }_{\mathrm{R}}}
}_{\mathrm{L}}\!\big\langle\varphi^{\ }_{l,\iota}\big|
=1.
\end{equation}

These properties of the eigenstates can now be used to show
that the transfer Hamiltonian $H$ with its associated Fock
space pair $(\mathfrak{F}^{\ }_{\mathrm{L}}, \mathfrak{F}^{\
}_{\mathrm{R}})$ possesses a real eigenvalue spectrum, and that
the operator $\exp (-L H)$ defines a unitary time evolution
with respect to the Fock space pair $(\mathfrak{F}^{\
}_{\mathrm{L}}, \mathfrak{F}^{\ }_{\mathrm{R}})$. First, for
any given eigenenergy $E^{\ }_l$ with left eigenvector $ \big|
\varphi^{\ }_{l,\iota} \big\rangle^{\ }_{\mathrm{L}}$ and right
eigenvector $ \big| \varphi^{\ }_{l,\sigma[\iota]}
\big\rangle^{\ }_{\mathrm{R}} $ we find
\begin{equation}
E^{\ast}_{l}=
{}^{\vphantom{\dag}}_{\mathrm{R}}\!
\big\langle\varphi^{\ }_{l,\sigma[\iota]}\big|
H^{\dag}
\big|\varphi^{\ }_{l,\iota}\big\rangle^{\ }_{\mathrm{L}}
=
{}^{\vphantom{\dag}}_{\mathrm{R}}\big\langle
\varphi^{\ }_{l,\sigma[\iota]}
\big|U^{\ }H^{\ }U^{\dag}\big|
\varphi^{\ }_{l,\iota}\big\rangle^{\ }_{\mathrm{L}}
=
{}^{\vphantom{\dag}}_{\mathrm{L}}
\big\langle\varphi^{\ }_{l,\sigma[\iota]}
\big|H\big|
\varphi^{\ }_{l,\iota}\big\rangle^{\ }_{\mathrm{R}}
=
E^{\ }_{l} ,
\end{equation}
where we have used the mapping between left and right eigenstates%
~\re{eq: mapping btw RL ev}, the Hermitian adjoint of
$H$~\re{eq: adjoint of Ha}, and the normalization of the
states~\re{eq: orhonormality}. Second, the transition amplitude
from the right energy eigenstate $\big|\varphi^{\ }_{l,\iota}
\big\rangle^{\ }_{\mathrm{R}}$ to the left energy eigenstate
$\big|\varphi^{\ }_{l^{\prime},\iota^{\prime}}\big\rangle^{\
}_{\mathrm{L}}$ is defined by
\begin{equation}
W^{\ }_{l\iota,l^{\prime}\iota^{\prime}}(t):=
{}^{\vphantom{\dag}}_{\mathrm{L}}\!
\big\langle\varphi^{\ }_{l^{\prime},\iota^{\prime}} \big|
e^{-\mathrm{i}tH} \big|\varphi^{\ }_{l,\iota}
\big\rangle^{\ }_{\mathrm{R}}
\label{eq: def Wml}
\end{equation}
after the analytic continuation
$L\to\mathrm{i}t$.
Using Eqs.~\re{eq: eigenval eq},
Eqs.~\re{eq: orhonormality},
and the involutive property of $\sigma[\iota]$,
one verifies that, for all $t$,
\begin{equation}
\begin{split}
\sum_{l^{\prime},\iota^{\prime}}
\left|W^{\ }_{l\iota,l^{\prime}\iota^{\prime}}\right|^2(t)
=\,&
\sum_{l',\iota'}
\vphantom{\big\rangle}^{\ }_{\mathrm{R}}\!
\big\langle\varphi^{\ }_{l,\iota}\big|
e^{+\mathrm{i}tH^{\dag}}
\big|\varphi^{\ }_{l^{\prime},\iota^{\prime}}\big\rangle
^{\ }_{\mathrm{L}}\,
{\vphantom{\big\rangle}}^{\ }_{\mathrm{L}}\!
\big\langle\varphi^{\ }_{l^{\prime},\iota^{\prime}}\big|
e^{-\mathrm{i}tH}
\big|\varphi^{\ }_{l,\iota}\big\rangle
^{\ }_{\mathrm{R}}
\\
=&\,
\sum_{l',\iota'}
{}^{\ }_{\mathrm{R}}
\big\langle\varphi^{\ }_{l,\iota}\big|
e^{+\mathrm{i}tE^{\ }_{l'}}
\big|\varphi^{\   }_{l^{\prime},\iota^{\prime}}\big\rangle^{\ }_{\mathrm{L}}
e^{-\mathrm{i}tE^{\ }_{l}}
\delta^{\ }_{l',l}
\delta^{\ }_{\sigma[\iota'],\iota}
\\
=&\,
1.
\end{split}
\label{eq: proof unitarity quantum evolution}
\end{equation}
Hence, the expression~(\ref{eq: def Wml}) constitutes a
well-defined transition amplitude, since it satisfies the
condition of probability conservation associated to a unitary
time evolution.

Summarizing, we have found that Eq.~\re{eq: adjoint of Ha}
implements the adjoint operation for the non-Hermitian
Hamiltonian $H$. The operator $U$ maps the right eigenstates
from $\mathfrak{F}^{\ }_{\mathrm{R}}$ into the dual space
$\mathfrak{F}^{\ }_{\mathrm{L}}$ (i.e., the space of left
eigenstates) in such a way that the transfer Hamiltonian
becomes Hermitian within the Fock space pair $(\mathfrak{F}^{\
}_{\mathrm{L}}, \mathfrak{F}^{\ }_{\mathrm{R}})$. That is, in
order to reinstate unitarity of the evolution operator,
$\exp(-\mathrm{i}tH)$, it is necessary to include the action of
$U$ in the scalar product.

\subsection{
Quantum representation of the density of states
           }

We are going to give a quantum representation of the density of states%
~(\ref{eq: mean DOS}) in the long wire limit $L \to \infty$ and
show how the original supersymmetry and a well-defined density
of states~(\ref{eq: mean DOS}) imply that the spectrum of $H$
is positive definite.

We start by reexpressing the partition function \re{eq:
partition fun 1} [see also Eq.~\re{eq: partition fun 2}] in
terms of the eigenbasis \re{eq: eigenval eq}
\begin{equation}
Z =
\sum_{l,\iota}\
{}^{\ }_{\mathrm{L}}\!
\left\langle \varphi^{\ }_{l,\iota}\right|
(-1)^{N^{\ }_{F}}
e^{-LE^{\ }_l}
\left|\varphi^{\ }_{l,\iota}\right\rangle^{\ }_{\mathrm{R}}
=
1.
\label{eq: part func 3}
\end{equation}
By the construction of Section%
~\ref{sec: The quasi-one dimensional random hopping chain},
the supersymmetry implies that the partition function is exactly one
whatever the length $L$ of the wire.
In the transfer Hamiltonian representation of the partition function%
~\re{eq: part func 3}, this is born out by the fact that each
supersymmetric eigenmultiplet of $H$
contains equal numbers of fermionic ($N^{\ }_{F}$ odd)
and bosonic ($N^{\ }_{F}$ even) eigenstates, which
thereby cancel in the supertrace due to the factor
$(-1)^{N^{\ }_{F}}$. Hence, the sum in Eq.~\re{eq: part func 3}
must then reduce to the expectation value in the zero-energy eigenenergy
sector $E^{\ }_{0}=0$ of the pair
$(\mathfrak{F}^{\ }_{\mathrm{L}}, \mathfrak{F}^{\ }_{\mathrm{R}})$.
This sum then gives unity if and only if the eigenspace
$E^{\ }_{0}=0$ in
$(\mathfrak{F}^{\ }_{\mathrm{L}}, \mathfrak{F}^{\ }_{\mathrm{R}})$
is of dimension one.
For the partition function to be independent of the wire length $L$,
the spectrum of $H$ must contain a nondegenerate eigenenergy
$E^{\ }_{0}=0$.

In the quantum representation, the density of states%
~(\ref{eq: mean DOS})
is expressed by
\begin{subequations}
\label{eq: QM DOS}
\begin{equation}
\begin{split}
\nu(\varepsilon)
=&\,
\lim_{\omega\to-\mathrm{i}\varepsilon}
\lim_{L\to\infty}
\pi^{-1}
\mathrm{Re} \,
\sum_{l,\iota}
{}^{\vphantom{\dag}}_L \! \left\langle \varphi^{\ }_{l,\iota} \right|
(-1)^{N^{\ }_{F}}
\left(\widebar{B}-B \right)\,e^{-LH}
\left| \varphi^{\ }_{l,\iota} \right\rangle^{\ }_{\mathrm{R}}
\\
=&\,
\lim_{\omega\to-\mathrm{i}\varepsilon}
\lim_{L\to\infty}
\pi^{-1}
\mathrm{Re} \,
\sum_{l,\iota}
{}^{\ }_{\mathrm{L}}\!\left\langle \varphi^{\ }_{l,\iota} \right|
(-1)^{N^{\ }_{F}}
 \left( Q - \widebar{Q} \right)\,e^{-LH}
 \left| \varphi^{\ }_{l,\iota} \right\rangle^{\ }_{\mathrm{R}} ,
\end{split}
\end{equation}
where either
\begin{equation}
B^{}_{}:=
+
\sum_{i=1}^{N}
f^{\dag}_{i}
f^{\ }_{i}
-\frac{1}{2}N,
\qquad
\widebar{B}^{}_{}:=
+
\sum_{i=1}^{N}
\bar{f}^{\ }_{i}
\bar{f}^{\dag}_{i}
-
\frac{1}{2}N,
\end{equation}
or
\begin{equation}
Q^{}_{}:=
+
\sum_{i=1}^{N}
b^{\dag}_{i}
b^{\ }_{i}
+
\frac{1}{2}N,
\qquad
\widebar{Q}^{}_{}:=
-
\sum_{i=1}^{N}
\bar{b}^{\ }_{i}
\bar{b}^{\dag}_{i}
+
\frac{1}{2}N.
\end{equation}
\end{subequations}
As the system becomes infinite in length, $L \to \infty$,
any exponential term in Eq.~\re{eq: QM DOS},
vanishes for all positive energy eigenstates.
In order for the density of states to be well defined
in the long wire limit,
the eigenvalues of $H$ need to be positive.
If so, the density of states in the limit $L \to \infty$ is dominated
by the ground state expectation value with $E^{\ }_{0}=0$
\begin{equation}
\begin{split}
\nu(\varepsilon)
=&\,
\lim_{\omega\to-\mathrm{i}\varepsilon}
\pi^{-1}
\mathrm{Re} \,
{}^{\vphantom{\dag}}_{\mathrm{L}}\!\langle\varphi^{\ }_{0}|
\left(\widebar{B}-B \right)\
| \varphi^{\ }_{0}\rangle^{\ }_{\mathrm{R}}
\\
=&\,
\lim_{\omega\to-\mathrm{i}\varepsilon}
\pi^{-1}
\mathrm{Re} \,
{}^{\vphantom{\dag}}_{\mathrm{L}}\!\langle\varphi^{\ }_{0}|
\left( Q - \widebar{Q} \right)
|\varphi^{\ }_{0}\rangle^{\ }_{\mathrm{R}} ,
\end{split}
\label{eq: dos in ground state}
\end{equation}
where
$\left| \varphi^{\ }_{0} \right\rangle^{\ }_{\mathrm{L}}$
and
$\left| \varphi^{\ }_{0} \right\rangle^{\ }_{\mathrm{R}}$
denote the nondegenerate left and right ground state wave functions,
respectively.  In Eq.~\re{eq: dos in ground state}
we have dropped the factor
$(-1)^{N^{\ }_{F}}$
since the left and right ground state wave functions
contain an even number of fermions, as we will explicitly verify in
Section~\ref{sec: Construction of the ground state sector}.

\section{Construction of the ground state sector}
\label{sec: Construction of the ground state sector}

In order to compute the density of states in the long wire limit%
~\re{eq: dos in ground state},
we need to make an appropriate Ansatz for the solutions
$\left| \varphi^{\ }_{0} \right\rangle^{\ }_{\mathrm{R}}$
and
$\left| \varphi^{\ }_{0} \right\rangle^{\ }_{\mathrm{L}}$
of the right and left Schr\"odinger equations
\begin{equation}
H \left|\varphi^{\ }_{0}\right\rangle^{\ }_{\mathrm{R}}
=
\left(
H^{\ }_{\omega}
+
H^{\ }_{2}
\right)
\left|\varphi^{\ }_{0}\right\rangle^{\ }_{\mathrm{R}}
=
0,
\qquad
H^{\dag} \left|\varphi^{\ }_{0}\right\rangle^{\ }_{\mathrm{L}}
=
\left(
H^{\dag }_{\omega}
+
H^{\dag }_{2}
\right)
\left|\varphi^{\ }_{0}\right\rangle^{\ }_{\mathrm{L}}
=
0 ,
\label{eq: Schroedinger eq 1}
\end{equation}
respectively.
The left and right ground state wave functions are related by
$\left| \varphi^{\ }_{0} \right\rangle^{\ }_{\mathrm{L}} =
U\left| \varphi^{\ }_{0} \right\rangle^{\ }_{\mathrm{R}}$.
As dictated by supersymmetry,
the ground state wave function is nondegenerate and constitutes a
supersymmetric singlet, i.e., it is annihilated
by the supersymmetric transformation~\re{eq: SUSYa}.
Therefore, the ground state sector can be expanded in terms of states
that transform as singlets under the supersymmetry~\re{eq: SUSYa}.
One such state is the vacuum $\left| 0 \right\rangle$.
Other states that transform as singlets under the symmetry%
~\re{eq: SUSYa} can be generated
by repeated action of $H$ on the vacuum $\left| 0 \right\rangle$.
All the states that are generated by repeated action of $H$
on the vacuum $\left|0\right\rangle$
form a vector space,
which we call the right ground state sector $\mathfrak{G}^{\ }_{\mathrm{R}}$.
The dual ground state sector  $\mathfrak{G}^{* }_{\mathrm{L}}$ is generated
by repeated action of $H^{\dag}$ on the vacuum $\left\langle0\right|$.
Alternatively, $\mathfrak{G}^{\ }_{\mathrm{L}}$
can be obtained by the application of the
unitary transformation $U$ defined in Eq.~\re{eq: def unitary trafo 1}
on the right ground state sector $\mathfrak{G}^{\ }_{\mathrm{R}}$.
In turn, $\mathfrak{G}^{* }_{\mathrm{L}}$ follows from
$\mathfrak{G}^{\ }_{\mathrm{L}}$ with the help of the adjoint operation.
The ground state sectors are subsets of the corresponding Fock spaces,
i.e.,
$\mathfrak{G}^{\ }_{\mathrm{L}}\subsetneq \mathfrak{F}^{\ }_{\mathrm{L}}$
and
$\mathfrak{G}^{\  }_{\mathrm{R}}\subsetneq \mathfrak{F}^{\ }_{\mathrm{R}}$.
We shall thus expand the right (left) ground state
$\left| \varphi^{\ }_{0} \right\rangle^{\ }_{\mathrm{R}}$
($\left| \varphi^{\ }_{0} \right\rangle^{\ }_{\mathrm{L}}$)
of $H$ in a basis of the right (left) ground state sector
$\mathfrak{G}_{\mathrm{R}}$ ($\mathfrak{G}_{\mathrm{L}}$).

Thereto we construct in what follows a right basis
\begin{subequations} \label{eq: GS sector basis}
\begin{equation} \label{eq: GS sector basisa}
\phantom{AAA}
\left\{ \left| m \right\rangle^{(n)}_{\mathrm{R}} \right\}_{m,n}
\; \; \cup  \; \;
\left| 0 \right\rangle,
\qquad
m,n= 1,2, \cdots,
\end{equation}
and a left basis
\begin{equation} \label{eq: GS sector basisb}
\phantom{AAA}
\left\{
{}^{(n)}_{\phantom{w}\mathrm{L}}
\left\langle m \right|  \right\}_{m,n} \; \; \cup  \; \;
\left\langle 0 \right| ,
\qquad
m,n=1,2, \cdots,
\end{equation}
\end{subequations}
of the ground state sectors $\mathfrak{G}^{\ }_{\mathrm{R}}$
and  $\mathfrak{G}^{\ }_{\mathrm{L}}$, respectively.
The left orthogonal set
$\left| m \right\rangle^{(n)}_{\mathrm{L}}$
and the right orthogonal set
$\left| m \right\rangle^{(n)}_{\mathrm{R}}$
are here generated with the help of the auxiliary orthogonal states
\begin{subequations}
\label{auxiliary states}
\begin{equation}
\begin{split}
&
\left| 2m+2, 2n; \downarrow \right\rangle
:=
\left[
 \left( A^{\ }_{-} \right)^{2n} \left( D^{\ }_{-} \right)^{2m+2}
-
2n  S^{\ }_{-} \widebar{S}^{\ }_{+}  \left( A^{\ }_{-} \right)^{2n-1}
\left( D^{\ }_{-} \right)^{2m+1}
\right] \left| 0 \right\rangle,
\\
&
\left| 2m+1, 2n+1; \uparrow \right\rangle
:=
\left[
\left( A^{\ }_{-} \right)^{2n+1} \left( D^{\ }_{-} \right)^{2m+1}
+(2m+1)
S^{\ }_{-} \widebar{S}^{\ }_{+}
\left( A^{\ }_{-} \right)^{2n} \left( D^{\ }_{-} \right)^{2m}
\right] \left| 0 \right\rangle,
\\
&
\left| 2m, 2n+2; \uparrow \right\rangle
:=
\left[
\left( A^{\ }_{-} \right)^{2n+2} \left( D^{\ }_{-} \right)^{2m}
+
2m
S^{\ }_{-} \widebar{S}^{\ }_{+}
\left( A^{\ }_{-} \right)^{2n+1} \left( D^{\ }_{-} \right)^{2m-1}
\right]
 \left| 0 \right\rangle,
\\
&
\left| 2m+1, 2n+1; \downarrow \right\rangle
:=
\left[
\left( A^{\ }_{-} \right)^{2n+1} \left( D^{\ }_{-} \right)^{2m+1}
-
(2n+1)
S^{\ }_{-} \widebar{S}^{\ }_{+}
\left( A^{\ }_{-} \right)^{2n} \left( D^{\ }_{-} \right)^{2m}
\right] \left| 0 \right\rangle,
\end{split}
\label{eq: states for N arbitrary a orthonormal}
\end{equation}
defined by taking arbitrary integer powers of the raising operators
\begin{equation}
A^{\ }_{-}:=
\sum_{a=1}^N
f^{\dag}_a \bar{f}^{\dag}_a ,
\quad
D^{\ }_{-}:=
\sum_{a=1}^N
b^{\dag}_a \bar{b}^{\dag}_a ,
\quad
S^{\ }_{-}:=
\sum_{a=1}^N
f^{\dag}_a \bar{b}^{\dag}_a ,
\quad
\widebar{S}^{\ }_{+}
:=
\sum_{a=1}^N
b^{\dag}_a \bar{f}^{\dag}_a ,
\end{equation}
\end{subequations}
acting on $|0\rangle$.
The basis sets $\left| m \right\rangle^{(n)}_{\mathrm{L}}$
and $\left| m \right\rangle^{(n)}_{\mathrm{R}}$
are countably infinite due to the presence of boson creation operators.
The basis set $\left|m\right\rangle^{(n)}_{\mathrm{R}}$  is given by
\begin{subequations} \label{eq: right and left basis}
\begin{equation}
\label{eq: right and left basisa}
\begin{split}
&
\left|m \right\rangle^{(2n+1)}_{\mathrm{R}}
:=
\frac{1}{\sqrt{2}} \frac{ (N-2n-1)!}{(N+2m-1)!}
\left(\vphantom{\sqrt{A^A}}
\left|2m,2n  ;\downarrow\right\rangle
+
\left|2m-1,2n+1;\uparrow  \right\rangle
\right),
\\
&
\left|m \right\rangle^{(2n+2)}_{\mathrm{R}}:=
\frac{1}{\sqrt{2}} \frac{(N-2n-2)!}{(N+2m-2)!}
\left(\vphantom{\sqrt{A^A}}
\left| 2m -2, 2n+2; \uparrow \right\rangle
+
\left| 2m-1, 2n+1; \downarrow \right\rangle
\right),
\end{split}
\end{equation}
where $m = 1,2,3,\cdots$.
The basis $\left|m \right\rangle^{(n)}_{\mathrm{L}}$ is given by
\begin{equation}
\label{eq: right and left basisb}
\begin{split}
&
\left|m\right\rangle^{(2n+1)}_{\mathrm{L}}:=
\frac{1}{\sqrt{2}} \frac{(N-2n-1)!}{(N+2m-1)!}
\left(\vphantom{\sqrt{A^A}}
\left|2m,2n  ;\downarrow\right\rangle
-
\left|2m-1,2n+1;\uparrow  \right\rangle
\right),
\\
&
\left|m\right\rangle^{(2n+2)}_{\mathrm{L}}:=
\frac{1}{\sqrt{2}} \frac{ (N-2n-2)!}{(N+2m-2)!}
\left(\vphantom{\sqrt{A^A}}
\left| 2m-2, 2n+2; \uparrow \right\rangle
-
\left| 2m-1, 2n+1; \downarrow \right\rangle
\right),
\end{split}
\end{equation}
\end{subequations}
where $m=1,2,3, \cdots$.

\begin{figure}[t!]
\begin{center}
\includegraphics[width=0.485\textwidth]{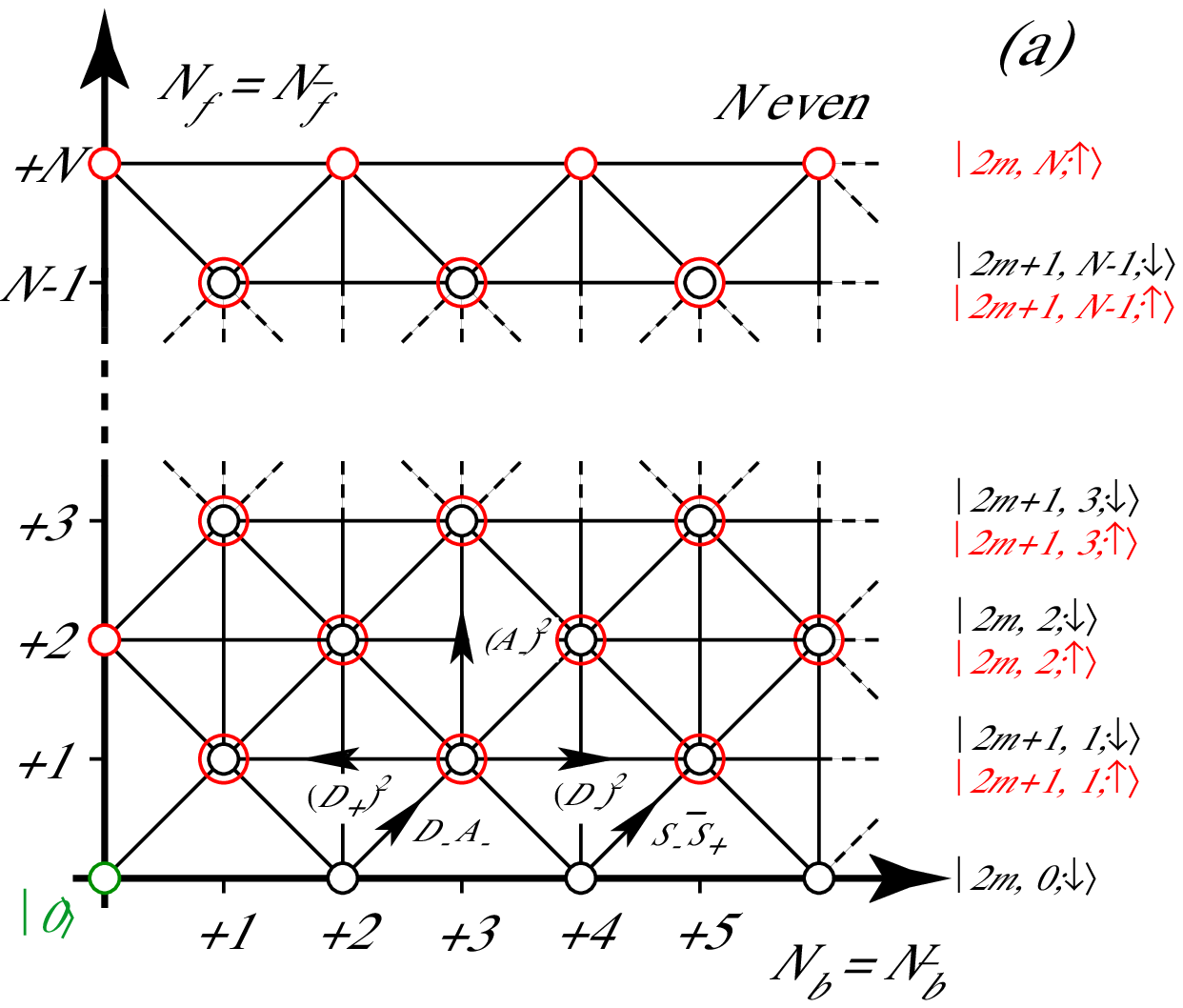}
\hfill
\includegraphics[width=0.485\textwidth]{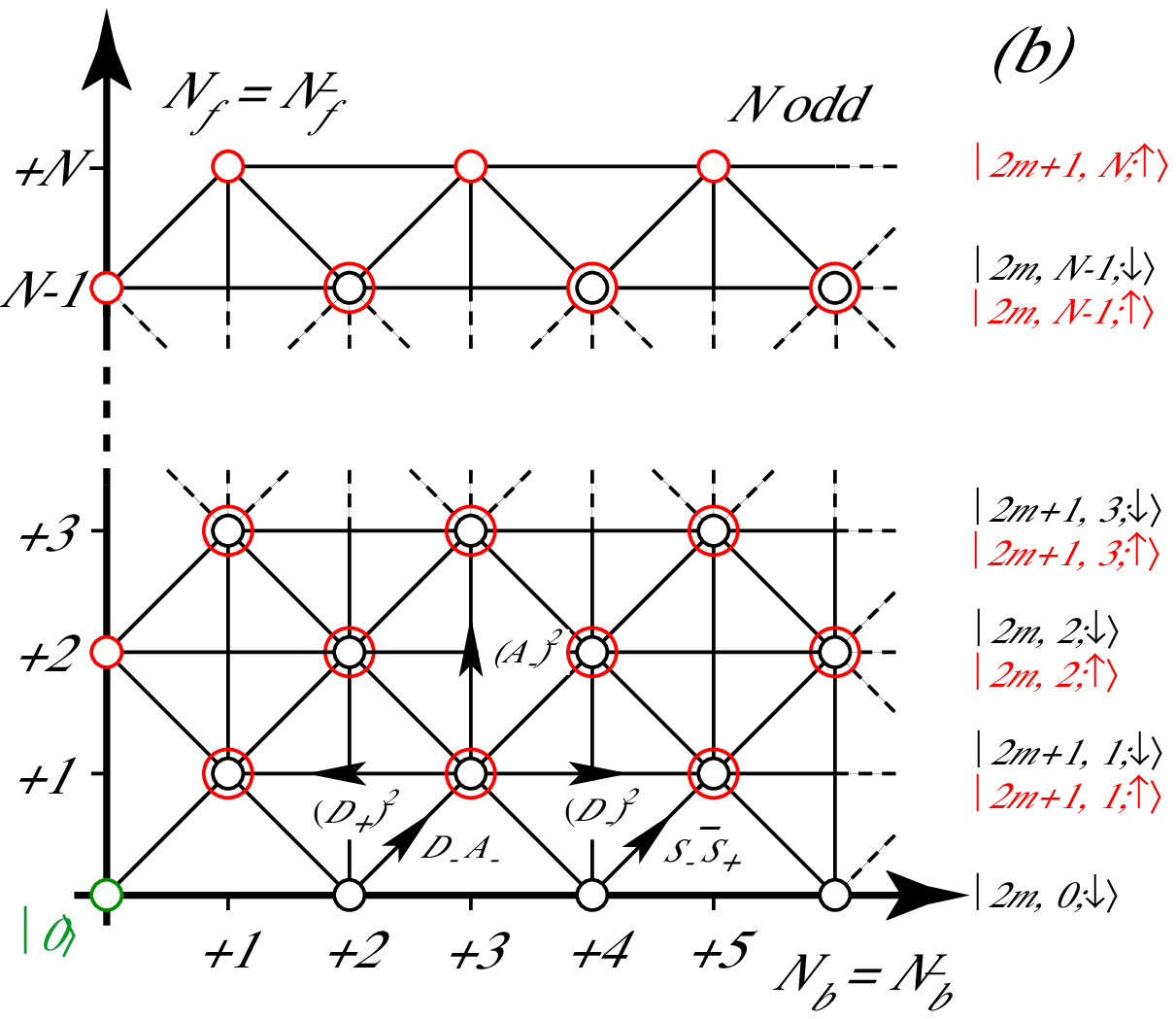}
\end{center}
\caption{
(Color online)
These weight diagrams depict states
\re{eq: states for N arbitrary a orthonormal}
in the $(N^{\ }_{b}, N^{\ }_{f})$ plane.
The left panel (a) displays the case
of an even channel number $N$, whereas the right panel (b) shows
the case of an odd channel number.
}
\label{fig: states for N even}
\end{figure}

The states \re{eq: states for N arbitrary a orthonormal}
are eigenstates of
the operators $N^{\ }_{f}$, $N^{\ }_{\bar{f}}$, $N^{\ }_{b}$,
and $N^{\ }_{\bar{b}}$ defined in Eq.~(\ref{eq: def number operators}),
whereby the eigenvalues
of $N^{\ }_{f} - N^{\ }_{\bar{f}}$ and $N^{\ }_{b} - N^{\ }_{\bar{b}}$
are vanishing for all the states in
\re{eq: states for N arbitrary a orthonormal}.
It is therefore convenient to depict these
states in the $(N^{\ }_{b}, N^{\ }_{f})$-plane
(see Fig.~\ref{fig: states for N even}).
As a consequence of Pauli's principle, the fermion number $N^{\ }_{f}$ is
restricted to a finite range. That is,
any power larger than one of any of the
fermionic operators
$S^{\ }_{-}$ and $\widebar{S}^{\ }_{+}$ vanishes,
and for a given channel number $N$ we have
\begin{equation} \label{nilpotency}
\left(
A^{\ }_{-}
\right)^{N+1}=0,
\end{equation}
and
\begin{equation}
\left( A_- \right)^N D_-
=
N S_- \widebar{S}_+ \left( A_- \right)^{N-1} .
\end{equation}
Condition \re{nilpotency} determines the range over which the index $n$ runs
in Eqs.~(\ref{auxiliary states})
and (\ref{eq: right and left basis}).
We note that there is, as we shall see,
a fundamental difference between even and odd channel numbers.
For $N$ even the states with the highest fermion number $N^{\ }_{f}$ are
labeled by even boson numbers $N^{\ }_{b}$,
whereas for $N$ odd they are labeled by odd boson numbers $N^{\ }_{b}$.
The right and left bases \re{eq: right and left basis}
are related by the unitary transformation $U$ of
Eq.~\re{eq: def unitary trafo 1}, through
\begin{equation}
|m\rangle^{(n)}_{\mathrm{L}}=
U
|m\rangle^{(n)}_{\mathrm{R}},
\qquad
m=1,2,\cdots,
\qquad
n=1,2,\cdots,N,
\label{eq: m(n)R and m(n) left unitary related}
\end{equation}
as it should be.
The norms in the right sets~(\ref{eq: right and left basisa})
are
\begin{equation}
\begin{split}
{ }^{(2n+1)}_{\phantom{AAA}\mathrm{R}}\!
\left\langle m | m \right\rangle^{(2n+1)}_{\mathrm{R}}
=
(2m+2n)(2m-1)!(2n)! \frac{ (N-2n-1)!}{(N+2m-1)!}
=:
\mathcal{N}^{(1)}_{m,n},
\\
{ }^{(2n+2)}_{\phantom{AAA}\mathrm{R}}\!
\left\langle
m |  m
\right\rangle^{(2n+2)}_{\mathrm{R}}
=
(2m+2n)(2m-2)!(2n+1)! \frac{ (N-2n-2)!}{(N+2m-2)!}
=:
\mathcal{N}^{(2)}_{m,n},
\end{split}
\label{eq: norm of basis states}
\end{equation}
with $m = 1,2,3, \cdots$.
It follows from \re{eq: m(n)R and m(n) left unitary related},
that the norm for a given state in the left set is
equal to the norm of the corresponding state in the right set.
Moreover, the left and right sets~\re{eq: right and left basisa}
and~\re{eq: right and left basisb}
are biorthogonal in the sense that
\begin{equation}
{}^{(n)}_{\phantom{A}\mathrm{L}}\!
\left\langle m|m^{\prime}\right\rangle^{(n^{\prime})}_{\mathrm{R}}
= 0,
\qquad
m,m^{\prime}=1,2,\cdots,
\qquad
n,n^{\prime}=1,2,\cdots,N,
\label{eq: zero overlap}
\end{equation}
as follows from
Eq.~\re{eq: norm of basis states}
and the orthogonality of the states
Eq.~\re{eq: states for N arbitrary a orthonormal}.

Equipped with a basis for the right and left ground state sectors,
we are now in a position to expand the right ground state
\begin{subequations}
\label{eq: expansion of eigenstates}
\begin{equation}
\left| \varphi^{\ }_{0} \right\rangle^{\ }_{\mathrm{R}}
\equiv
\left( {}^{\vphantom{\dag}}_{\mathrm{R}}\!
\left\langle \varphi^{\ }_{0} \right| \, \right)^{\dag}
=
a^{(0)}   \left| 0 \right\rangle
+
\sum_{n=1}^N \sum_{m=1}^{\infty}
a^{(n)}_{m}  \left| m \right\rangle^{(n)}_{\mathrm{R}},
\label{eq: expansion of eigenstatesa}
\end{equation}
in terms of the basis Eq.~\re{eq: GS sector basisa} and
the left ground state
\begin{equation}
\left| \varphi^{\ }_{0} \right\rangle^{\ }_{\mathrm{L}}
=
\left( {}^{\vphantom{\dag}}_{\mathrm{L}}\!
\left\langle \varphi^{\ }_{0} \right| \, \right)^{\dag}
=
a^{(0)}   \left| 0 \right\rangle
+
\sum_{n=1}^N \sum_{m=1}^{\infty}
a^{(n)}_{m}  \left| m \right\rangle^{(n)}_{\mathrm{L}},
\label{eq: expansion of eigenstatesb}
\end{equation}
\end{subequations}
in terms of the basis Eq.~\re{eq: GS sector basisb}.
The coefficients of these two expansions are identical and
are determined by solving the Schr\"odinger equations
\re{eq: Schroedinger eq 1}.
The overlap between the expansions~\re{eq: expansion of eigenstates}
for the right and left ground states
follows form Eq.~\re{eq: zero overlap},
\begin{subequations}
\label{eq: normalization a(0)}
\begin{equation}
{}^{\ }_{\mathrm{L}}\!
\langle \varphi^{\ }_{0}|
\varphi^{\ }_{0}\rangle^{\ }_{\mathrm{R}}
=
\big|a^{(0)}\big|^2.
\end{equation}
Consequently, in order to normalize the ground state wave function,
we set
\begin{equation}
a^{(0)}=1
\end{equation}
\end{subequations}
from now on.

\subsection{Ground state Schr\"odinger equation}

Hamiltonian~(\ref{eq: final hamiltonian n chains chiral-unitary})
depends on two energy scales: the (imaginary) energy
$\omega$ at which the density of states in the thermodynamic limit
is to be evaluated and the chiral disorder strength $g^{\ }_{2}$
defined in Eq.~(\ref{eq: gaussian disorder N chains}).
Since the density of states in the thermodynamic limit
is controlled solely by the right
$\left|\varphi^{\ }_{0}\right\rangle^{\ }_{\mathrm{R}}$
and left
$\left|\varphi^{\ }_{0}\right\rangle^{\ }_{\mathrm{L}}$
eigenstates annihilated by
Hamiltonian~(\ref{eq: final hamiltonian n chains chiral-unitary}),
it must be a scaling function of the dimensionless variable
\begin{equation}
\frac{\omega}{g^{\ }_{2}}\to \omega,
\label{eq: setting g2 to unity}
\end{equation}
provided we set $g^{\ }_{2}$ to be unity,
as we shall do from now on unless stated otherwise.
The right (left) ground state
$\left|\varphi^{\ }_{0}\right\rangle^{\ }_{\mathrm{R}}$
$( \left|\varphi^{\ }_{0}\right\rangle^{\ }_{\mathrm{L}})$
satisfies the eigenvalue equation \re{eq: Schroedinger eq 1}.
In the basis~\re{eq: GS sector basis},
the eigenvalue problem~\re{eq: Schroedinger eq 1} yields
the recursion relation for the coefficients $a^{(n)}_m$ of
Eq.~\re{eq: expansion of eigenstates} given by
\begin{subequations}
\label{eq: recursion relations gen N}
\begin{equation} \label{eq: rec rel gen N a}
\begin{split}
2 (2m+2n)\,\omega \,a^{(2n+1)}_{m}
=&\,
- 2(2m+2n) (2m-1-2n+N) a^{(2n+1)}_{m}
\\
&\,
+ 2m(2m+1) a^{(2n+1)}_{m+1}
\\
&\,
+ (N+2m-1)(N+2m-2) a^{(2n+1)}_{m-1}
\\
&\,
- (N-2n)(N-2n+1)a^{(2n-1)}_{m}
\\
&\,
- (2n+1)(2n+2)a^{(2n+3)}_{m},
\end{split}
\end{equation}
when $n=0,1, \cdots, [(N-1)/2]$ and
\begin{equation}
\label{eq: rec rel gen N b}
\begin{split}
2(2m+2n)\, \omega \,a^{(2n+2)}_{m}
=&\,
-2 (2m+2n)(2m-3-2n+N) a^{(2n+2)}_{m}
\\
&\,
+2m(2m-1) a^{(2n+2)}_{m+1}
\\
&\,
+(N+2m-3)(N+2m-2) a^{(2n+2)}_{m-1}
\\
&\,
- (N-2n)(N-2n-1)a^{(2n)}_{m}
\\
&\,
- (2n+2)(2n+3) a^{(2n+4)}_{m},
\end{split}
\end{equation}
when $n=0,1, \cdots, [N/2]-1$ and where $m=2,3,4,\cdots$.
Here, it is understood that $a^{(n)}_m = 0$ when $n \leq 0$ or $n>N$.
The initial conditions are
\begin{equation}
\begin{split}
&
4\, \omega \,a^{(1)}_{1}
=
+ \sqrt{2} N(N+1)
- 4(N+1) a^{(1)}_1
+ 6 a^{(1)}_2
-2 a^{(3)}_1,
\\
&
4\, \omega \,a_{1}^{(2)}
=
- \sqrt{2} N (N-1)
- 4(N-1) a^{(2)}_1
+ 2 a^{(2)}_2
- 6 a^{(4)}_1.
\end{split}
\end{equation}
\end{subequations}
Equations~(\ref{eq: recursion relations gen N})
are a generalization to the case of an arbitrary number $N$ of channels
of the recursion relations obtained by Refs.%
~\cite{Balents97,Bocquet99,Bunder01} when $N=1$.
The rest of this paper is devoted to solving them and evaluating the density of states using Eq.~\re{eq: dos in ground state}.

\section{Evaluation of the density of states when $N=1$ and $N=2$}
\label{sec: Evaluation of the density of states when N=1 and N=2}

Before we solve the recursion relations%
~\re{eq: recursion relations gen N}
for arbitrary $N$,
we first specialize to the case $N=1$ and $N=2$
for pedagogical reasons. Only the case $N=1$
using the supersymmetric quantum representation
(see Refs.~\cite{Balents97,Bocquet99,Bunder01})
was treated in the literature.
The case $N=2$ was solved with the help of the DMPK
equation in Refs.\ \cite{Brouwer00} and \cite{Titov01}.

\subsection{Density of states for a single channel $N=1$}
\label{subsec: Density of states for a single channel N=1}

For a single channel, $N=1$,
the recursion relations Eq.~\re{eq: recursion relations gen N}
for the coefficients  $a^{(1)}_m$ simplify to
\begin{subequations}
\label{eq: recursion relation N=1}
\begin{eqnarray}
2 \omega  a^{(1)}_{1}
&=&
+\sqrt{2}
-4 a^{(1)}_{1} + 3 a^{(1)}_{2},
\label{eq: recursion relation N=1b}
\\
2 \omega  a^{(1)}_{m+1}
&=&
- 2 (2m+2) a^{(1)}_{m+1}
+ (2m+3) a^{(1)}_{m+2}
+ (2m+1) a^{(1)}_{m} ,
\label{eq: recursion relation N=1a}
\end{eqnarray}
\end{subequations}
with $m=1,2,3,\cdots$.
This recursion relation is identical
to the one obtained by Gogolin \textit{et al.}~\cite{Gogolin77}
by means of the Berezinskii diagram technique.
Hence, one can identify the coefficients $a^{(1)}_{m}$
as the so-called right-hand side Berezinskii block.
The recursion relation \re{eq: recursion relation N=1a} can also be written
in terms of the finite differences
\begin{subequations}
\begin{equation}
\label{eq: finite differences}
\Delta a^{(1)}_{m}
:=
a^{(1)}_{m+1}
-
a^{(1)}_{m},
\qquad
\Delta^{2} a^{(1)}_{m} :=
a^{(1)}_{m+2}
-
2
a^{(1)}_{m+1}
+
a^{(1)}_{m}.
\end{equation}
That is, Eqs.~\re{eq: recursion relation N=1a}
and~\re{eq: finite differences}
yield
\begin{eqnarray}
\label{eq: recursion relation N=1d}
 2\omega   a^{(1)}_{m}
&=&
+ (2m+3) \Delta^2 a^{(1)}_{m}
+ \left(2- 2 \omega  \right) \Delta a^{(1)}_{m}
\label{eq: difference eqn case N=1 if mathcal{E}=0} ,
\end{eqnarray}
\end{subequations}
where $m=1,2,3 \cdots$.

The strategy to analyze equations \re{eq: recursion relation N=1}
in the limit of asymptotically small frequency, $ \omega  \ll 1$,
is to derive two approximate solutions that are valid
in two distinct asymptotic limits. For large
$m \gg 1$, the variable $m$ can be taken
as continuous. This reduces the recursion relation
\re{eq: recursion relation N=1a}
to a Bessel differential equation.
For small $m\omega\ll1$,
we can neglect the $\omega$ term in Eq. \re{eq: recursion relation N=1a}.
With this approximation,
it becomes possible to solve the resulting recursion relation exactly,
for example, by means of the generating function technique.

For $\omega$ asymptotically small (the limit we are interested in),
the overlapping region $1 \ll m \ll \omega^{-1}$
becomes arbitrarily large.
We can then derive matching conditions between
the two approximate solutions to fix the
coefficients of the Bessel equation
and the initial values of the zero-frequency recursion relation.
Hence, in the limit of asymptotically small frequency,
this procedure gives a complete solution for the ground state of $H$.
This solution can be used to compute the expectation value of
$(\widebar{B}-B)$ in the ground state, which yields the
asymptotic form of the density of states~(\ref{eq: dos in ground state})
as $\varepsilon\to0$.

We first consider the limit $m\gg1$.
In this limit, we can neglect terms of order one compared to $m$
in Eq.~(\ref{eq: difference eqn case N=1 if mathcal{E}=0})
and replace finite differences by derivatives with respect to $m$,
i.e.,
the discrete index $m$ is now treated as a continuous one with
$a^{(1)}_m \to a^{(1)}_{\vphantom{m}}(m)$.
This gives
\begin{equation}
2\omega
a^{(1)}
=
2
\left(
m \frac{d^{2}a^{(1)}}{dm^{2}}
+
\frac{d a^{(1)}}{dm}
\right).
\label{eq: continuum equation n=1}
\end{equation}
Here, we have assumed that $a^{(1)}$ varies slowly
so that $d a^{(1)}/dm$ can be neglected
relative to $a^{(1)}$.
After the substitution
\begin{subequations}
\begin{equation}
x^2:=4\omega  m,
\end{equation}
whereby
\begin{equation}
\frac{x^2}{4\omega}\gg1,
\end{equation}
\end{subequations}
this differential equation can be transformed
into a modified Bessel equation
and we obtain the solutions
\begin{equation}
a^{(1)}(x)=
c^{(1)}_{\textrm{o},1} K^{\ }_{0} (x)
+
c^{(1)}_{\textrm{o},2} I^{\ }_{0} (x),
\label{eq: solution to Bessel eq n=1}
\end{equation}
where $I^{\ }_{0}(x)$ is the modified Bessel function of the first kind and
$K^{\ }_{0}(x)$ the modified Bessel function of the second kind
(see chapter~9.6.~in~\cite{Abramowitz65}).
We shall demand that $a^{(1)}_{m}$
decays to zero as $m\to\infty$,
i.e., we must set
\begin{equation}
c^{(1)}_{\textrm{o},2} = 0
\end{equation}
in Eq.~(\ref{eq: solution to Bessel eq n=1}).
The remaining
coefficient $c^{(1)}_{\textrm{o},1}$
of the modified Bessel function $K^{\ }_{0}$
is determined by matching
the solution~\re{eq: solution to Bessel eq n=1} to the one for
$m \ll \omega^{-1}$. The coefficient
$c^{(1)}_{\textrm{o},1}$
of the modified Bessel function $K^{\ }_{0}$
will thereby acquire an $\omega$ dependence.
In order to carry out this matching procedure, we will need the
small $x$ behavior of the continuum solution
\begin{equation}
a^{(1)}(x)\sim
-
c^{(1)}_{\textrm{o},1}
\left[\ln(x/2)+\gamma\right]
=
-
\frac{c^{(1)}_{\textrm{o},1}}{2}
\left[\ln(m\omega )+2\gamma\right] ,
\label{eq: small x limit case N=1}
\end{equation}
where $\gamma$ is the Euler's constant.

Second, we consider the limit $m \ll \omega^{-1}$,
in which case we can safely drop the terms containing $\omega$ in
Eqs.~\re{eq: recursion relation N=1}.
Equation~\re{eq: recursion relation N=1d} can then be written as
\begin{subequations}
\begin{equation}
\begin{split}
0
=&\,
(2m+3) \Delta a^{(1)}_{m+1}
-
(2m+1) \Delta a^{(1)}_{m}
\\
=&\,
\Delta
\left(
(2m+1) \Delta a^{(1)}_{m}
\right)
\end{split}
\label{eq: rec rel for small m N=1}
\end{equation}
with $m=1,2,3,\cdots,$
together with the initial condition
\begin{equation}
3 a^{(1)}_{2}
=
4 a^{(1)}_{1}
-
\sqrt{2}.
\label{eq: initial conditions N=1}
\end{equation}
\end{subequations}
Integrating Eq.%
~\re{eq: rec rel for small m N=1}
and determining the integration constant
by use of Eq.~\re{eq: initial conditions N=1},
we find the first order difference equation
\begin{equation}
(2m+1)
\left(
a^{(1)}_{m+1} - a^{(1)}_{m}
\right) =
a^{(1)}_{1} - \sqrt{2}
\end{equation}
with $m=1,2,3,\cdots,$
whose solution is given in terms of the Digamma function $\Psi$
(for references on the Digamma function $\Psi$,
see chapters~6.3 or 8.36 in Refs.~\cite{Abramowitz65}
or \cite{Gradshteyn94}, respectively)
\begin{equation}
a^{(1)}_m =
\sqrt{2}
+
\frac{1}{2} \left(a^{(1)}_{1}
-
\sqrt{2}  \right)
\Big(
\Psi(m+1/2)
-
\Psi(1/2)
\Big),
\label{eq: psi in terms of digamma N=1}
\end{equation}
with the limiting form
\begin{equation}
a^{(1)}_{m}\sim
\sqrt{2}
+
\frac{1}{2}
\left(
a^{(1)}_{1}
-
\sqrt{2}
\right)
\Big(\ln m-\Psi(1/2)\Big)
\label{eq: large m limit for m w << 1 case N=1}
\end{equation}
for large $m\in\mathbb{N}$.

Having solved the recursion relation in the two limits
$1 \ll m$
and
$m \ll  \omega^{-1}$ ,
we now match solutions%
~(\ref{eq: small x limit case N=1})
and%
~(\ref{eq: large m limit for m w << 1 case N=1})
in the overlapping regime $1 \ll m \ll \omega^{-1}$
to fix the coefficient
$c^{(1)}_{\textrm{o},1}$ of the modified Bessel function $K^{\ }_{0}$
and the initial value $a^{(1)}_{1}$
of the $\omega=0$ recursion relation.
Equating Eq.~\re{eq: small x limit case N=1}
with Eq.~\re{eq: large m limit for m w << 1 case N=1}
yields the two conditions
\begin{equation}
\begin{split}
c^{(1)}_{\textrm{o},1}
=&\,
-
a^{(1)}_{1}
+
\sqrt{2} ,
\\
-
c^{(1)}_{\textrm{o},1}
\ln\omega
-
2
c^{(1)}_{\textrm{o},1}
\gamma
=&\,
2 \sqrt{2}
-
\left( a^{(1)}_{1}
-
\sqrt{2}  \right) \Psi(1/2),
\end{split}
\end{equation}
for the two unknowns
$c^{(1)}_{\textrm{o},1}$
and
$a^{(1)}_{1}$.
Solving for $c^{(1)}_{\textrm{o},1}$
and
$a^{(1)}_{1}$, we get
\begin{equation}
a^{(1)}_{1}=
\sqrt{2}\
\left(
\frac{
2
     }
     {
\ln (C^{\ }_1  \omega )
     }
+
1
\right),
\quad
c^{(1)}_{\textrm{o},1}
=
\frac{
-2\sqrt{2}
     }
     {
\ln\left(C^{\ }_1 \omega \right)
     },
\quad
C^{\ }_1:=\exp\Big(2\gamma+\Psi(1/2)\Big).
\label{eq: def constant C}
\end{equation}
We have thus derived two limiting solutions
to the recursion relations \re{eq: recursion relation N=1}
\begin{eqnarray}
a^{(1)}_{m}=
\left\{
\begin{array}{l l}
-\frac{2\sqrt{2} }{\ln(C^{\ }_1 \omega  )}
K^{\ }_{0}\left(2\sqrt{m\omega }\right),
&
\qquad \qquad  1 \ll  m,
\\
&
\\
\sqrt{2}
\left(
1
+
\frac{\Psi(m+1/2)-\Psi(1/2)}
     {\ln(C^{\ }_1\omega)}
\right),
&
\qquad  \qquad   m \ll \omega^{-1}.
\end{array}
\right.
\label{eq: limiting solutions to rec rel N=1}
\end{eqnarray}

Knowledge of the  asymptotics%
~(\ref{eq: limiting solutions to rec rel N=1})
is sufficient to determine the leading behavior of
the density of states $\nu (\varepsilon)$,
Eq.~(\ref{eq: dos in ground state}),
for asymptotically small energies $\varepsilon$.
The density of states in the long wire limit
Eq.~\re{eq: QM DOS} is given by the expectation value of the operator
$(\widebar{B}-B)$  in the ground state
\begin{subequations}
\begin{equation}
\begin{split}
\nu(\varepsilon)
=&\,
\lim_{\omega\to-\mathrm{i}\varepsilon}
\pi^{-1}
\mathrm{Re} \,
{}^{\ }_{\mathrm{L}}\!\langle\varphi^{\ }_{0}|
\left(\widebar{B}-B \right)\
| \varphi^{\ }_{0}\rangle^{\ }_{\mathrm{R}}
\\
=&\,
\lim_{ \omega\to -i \varepsilon }
\pi^{-1}
\mathrm{Re}
\left(
1
+
\sum_{m=1}^{\infty}
\left( a^{(1)}_{m} \right)^2
\right),
\end{split}
\label{eq: dos n=1 first}
\end{equation}
where we have used
Eq.~\re{eq: norm of basis states}
and the identity
\begin{equation}
(\widebar{B} - B) \left| m \right\rangle^{(1)}_{\mathrm{R}}=
\left| m \right\rangle^{(1)}_{\mathrm{L}}.
\end{equation}
\end{subequations}
In view of Eq.~\re{eq: limiting solutions to rec rel N=1}, it is necessary
to break the  sum over $m$
in Eq.~\re{eq: dos n=1 first}
into two parts separated by the integer $m^{\ }_{0}$, with
$1\ll m^{\ }_{0} \ll \omega^{-1}$.
We choose $m^{\ }_{0}$ to be some fixed number, independent
of $\omega$.
With this choice and for sufficiently small $\omega$,
the sum on the right-hand side of Eq.~(\ref{eq: dos n=1 first})
is dominated by the part $m>m^{\ }_{0}$, i.e.,
by the contribution from the first line in Eq.%
~\re{eq: limiting solutions to rec rel N=1}.
(In the limit $\omega \to 0$, the  sum over  $m \leq m^{\ }_{0}$,
i.e., the contribution from the second line in Eq.%
~\re{eq: limiting solutions to rec rel N=1},
 is less divergent than the sum over $m >m^{\ }_{0}$,
and can therefore be neglected.)
We then find the estimate
\begin{equation}
\begin{split}
{}^{\ }_{\mathrm{L}}\!\left\langle\varphi^{\ }_{0} \right|
\left(\widebar{B}-B\right)
\left|\varphi^{\ }_{0} \right\rangle^{\ }_{\mathrm{R}}
\approx&\,
\frac{8}
     {\ln^2( C^{\ }_1\omega)}
\int\limits_{m_{0}}^{\infty}dm\,
K^2_{0}\left(2\sqrt{m\omega}\right)
\\
=&\,
\frac{4 \omega^{-1}   }{    \ln^2 ( C^{\ }_1 \omega   )}
\int\limits_{x_{0}}^{\infty} dx \,
x K^2_{0} \left(x \right).
\end{split}
\end{equation}
This gives the asymptotic behavior of the density of states
\begin{equation}
\begin{split}
\nu(\varepsilon)
\propto&\,
\lim_{\omega\to -  \mathrm{i} \varepsilon }\textrm{Re} \,
\frac{\mathrm{1} }{\omega\ln^2 ( C_1 \omega ) }
\sim \,
\frac{1 }{\varepsilon \left| \ln^3  \varepsilon \right| },
\end{split}
\end{equation}
which has precisely the form of Dyson's singularity.

\subsection{Density of states for two channels $N=2$}
\label{subsec: Density of states for two channels N=2}

For the case of two channels, $N=2$, the recursion relations%
~\re{eq: recursion relations gen N}
for the coefficients $a^{(1)}_m$ and $a^{(2)}_m$ reduce to
\begin{subequations}
\label{eq: rec rel n=2}
\begin{equation}
\begin{split}
2\omega  a^{(1)}_{m+1}
=&\,
(2m+3)
\left(
-
2a^{(1)}_{m+1}
+
a^{(1)}_{m+2}
+
a^{(1)}_{m}
\right),
\\
2\omega  a^{(2)}_{m+1}
=&\,
(2m+1)
\left(
-
2a^{(2)}_{m+1}
+
a^{(2)}_{m+2}
+
a^{(2)}_{m}
\right),
\end{split}
\label{eq: rec rel n=2a}
\end{equation}
with $m=1,2,3,\cdots$ and with the initial conditions
\begin{equation}
\begin{split}
2\omega  a^{(1)}_{1}
=&\,
+
\sqrt{2}\,3
-
6a^{(1)}_{1}
+
3a^{(1)}_{2},
\\
2 \omega a^{(2)}_{1}
=&\,
-
\sqrt{2}
-
2a^{(2)}_{1}
+
a^{(2)}_{2}.
\end{split}
\label{eq: recursion rel for N=2 mathcal{E}=0 b}
\end{equation}
Alternatively, we can rewrite the two decoupled recursion
relations  \re{eq: rec rel n=2a}
in terms of the finite differences
\begin{equation}
\Delta a^{(i)}_{m}
:=
a^{(i)}_{m+1}
-
a^{(i)}_{m},
\qquad
\Delta^{2} a^{(i)}_{m} :=
a^{(i)}_{m+2}
-
2
a^{(i)}_{m+1}
+
a^{(i)}_{m},
\qquad
i=1,2.
\end{equation}
With this,  Eq.~\re{eq: rec rel n=2a} becomes
\begin{equation}
\begin{split}
2\omega  a^{(1)}_{m}
=&\,
(2m+3)
\Delta^2 a^{(1)}_m
-  2\omega  \Delta a^{(1)}_{m},
\\
2\omega  a^{(2)}_{m}
=&\,
(2m+1)
\Delta^2 a^{(2)}_m
-
2\omega  \Delta a^{(2)}_{m} .
\end{split}
\label{eq: rec rel n=2d}
\end{equation}
\end{subequations}
To analyze these equations,
we follow the same strategy
as in the previous section for the $N=1$ case and solve the
Eqs.~\re{eq: rec rel n=2} first in the $m \gg 1$ limit and then in the
$m \ll \omega^{-1}$ limit.

We begin with the $m \gg 1$ limit.
Treating $m$ as a continuous index, we replace
finite differences by derivatives in Eq.~\re{eq: rec rel n=2d},
i.e., we let $a^{(i)}_m \to a^{(i)}_{\vphantom{m}}(m)$
with $i=1,2$.
This gives
\label{eq: N=2, large, ODE}
\begin{equation}
2\omega a^{(i)}=
2m \frac{d^{2}a^{(i)}}{d m^{2}}.
\qquad i=1,2.
\label{eq: N=2, large, ODE a}
\end{equation}
Here, $a^{(i)}$ is assumed to be slowly varying so that
we can neglect $d a^{(i)}/dm$ compared to
$a^{(i)}$.
After the substitution
\begin{subequations}
\begin{equation}
x^2:=4  \omega m,
\end{equation}
whereby
\begin{equation}
\frac{x^2}{4\omega}\gg1,
\end{equation}
\end{subequations}
Eq.~\re{eq: N=2, large, ODE a} reduces to
a Bessel-type equation whose solution is given in terms of the modified
Bessel functions $K^{\ }_{1}(x)$ and $I^{\ }_{1}(x)$,
\begin{equation}
a^{(i)} (x)=
c^{(i)}_{\textrm{e},1}
\frac{x}{2}\,
K^{\ }_{1}
\left( x \right)
+
c^{(i)}_{\textrm{e},2}
\frac{x}{2}\,
I^{\ }_{1}
( x ),
\qquad
i=1,2,
\label{eq: N=2, large, ODE b}
\end{equation}
with the
coefficients
$c^{(i)}_{\textrm{e},j}$ with $i=1,2$
of the modified Bessel functions
$K^{\ }_{1}$ when $j=1$ and $I^{\ }_{1}$ when $j=2$.
We shall demand that $a^{(i)}_{m}$
with $i=1,2$ decay to zero as $m\to\infty$,
i.e., we must set
\begin{equation}
c^{(i)}_{\textrm{e},2}=0.
\end{equation}
in Eq.~(\ref{eq: N=2, large, ODE b}).
The remaining coefficients
$c^{(i)}_{\textrm{e},1}$ with $i=1,2$
of the modified Bessel function $K^{\ }_{1}$
will be fixed by matching solutions%
~(\ref{eq: N=2, large, ODE b}) to
the small $m\omega$ solution.
In order to do so we will need the small $x$
behavior of Eq.~(\ref{eq: N=2, large, ODE b})
\begin{subequations}
\label{eq: small omega asymptotic to large m solution N=2}
\begin{equation}
\begin{split}
a^{(i)}(x)
\sim&\,
\frac{1}{2}
c^{(i)}_{\textrm{e},1}
+
c^{(i)}_{\textrm{e},1}
\frac{x^2}{8}
\left[
  \ln ( x^2 /4 ) + 2 \gamma -1
\right]
\\
=&\,
\frac{1}{2}
c^{(i)}_{\textrm{e},1}
+
c^{(i)}_{\textrm{e},1}
\frac{m\omega}{2}\,
\ln(C^{\textrm{(e)}}_0m\omega),
\end{split}
\end{equation}
with $i=1,2$ and where
\begin{equation}
C^{\textrm{(e)}}_0:=\exp(2\gamma-1).
\end{equation}
\end{subequations}

Next, we turn to the limit $m\omega\ll1$,
in which we can neglect the $\omega$ term in
Eq.~\re{eq: rec rel n=2d}. If so,
Eq.~\re{eq: rec rel n=2d}
becomes
\begin{subequations}
\begin{equation}
\label{eq: simplified rec rel n=2a}
\begin{split}
0=&\,
(2m+3)
\Delta^2 a^{(1)}_m  ,
\\
0=&\,
(2m+1)
\Delta^2 a^{(2)}_m,
\end{split}
\end{equation}
together with the initial conditions
\begin{equation}
\label{eq: simplified rec rel n=2b}
\begin{split}
2a^{(1)}_{1}
=&\,
+
\sqrt{2}\,
+
 a^{(1)}_{2},
\\
2a^{(2)}_{1}
=&\,
-
\sqrt{2}
+
a^{(2)}_{2}.
\end{split}
\end{equation}
\end{subequations}
The solution of Eqs.~\re{eq: simplified rec rel n=2a} is given by
\begin{equation}
\begin{split}
a^{(1)}_{m}
=&\,
m \, a^{(1)}_{1}
-
(m-1)\sqrt{2}\, ,
\\
a^{(2)}_{m}
=&\,
m \,
a^{(2)}_{1}
+
(m-1)
\sqrt{2}\, ,
\end{split}
\label{eq: omega<<1 solution N=2}
\end{equation}
where we have fixed two initial values by use of
Eq.~\re{eq: simplified rec rel n=2b}.

Having solved the two decoupled recursion relations
\re{eq: rec rel n=2} in the two limits
$m \gg1$  and $m  \ll  \omega^{-1}$,
we now match solutions~\re{eq: N=2, large, ODE b}
and~\re{eq: omega<<1 solution N=2}
in the overlapping region
in order to fix the
coefficients
$c^{(i)}_{\textrm{e},1}$
of the modified Bessel functions
$K^{\ }_{1}$
and initial values $a^{(i)}_{1}$ with $i=1,2$
of the $\omega=0$ recursion relation.
Equating~(\ref{eq: small omega asymptotic to large m solution N=2})
with~(\ref{eq: omega<<1 solution N=2}) and matching equal powers
of $m$ (whereby we neglect $\ln m$ compared to $\ln \omega$)
gives the two equations
\begin{subequations}
\begin{equation}
\begin{split}
+ \sqrt{2}
=&\,
\frac{1}{2}
c^{(1)}_{\textrm{e},1},
\\
a^{(1)}_{1}
-
\sqrt{2}\,
=&\,
c^{(1)}_{\textrm{e},1}
\frac{\omega}{2}
\ln\left(C^{(\textrm{e})}_0\omega\right),
\end{split}
\end{equation}
for the two unknowns
$c^{(1)}_{\textrm{e},1}$
and $a^{(1)}_{1}$
and the two equations
\begin{equation}
\begin{split}
- \sqrt{2}
=&\,
\frac{1}{2}
c^{(2)}_{\textrm{e},1},
\\
a^{(2)}_{1}
+
\sqrt{2}
=&\,
c^{(2)}_{\textrm{e},1}
\frac{\omega}{2}
\ln\left(C^{(\textrm{e})}_0\omega\right),
\end{split}
\end{equation}
\end{subequations}
for the two unknowns $c^{(2)}_{\textrm{e},1}$ and $a^{(2)}_{1}$.
Solving for $a^{(1)}_{1}$ and $a^{(2)}_{1}$, respectively, we obtain
\begin{equation}
\begin{split}
a^{(1)}_{1}
=&\,
+
\sqrt{2}\,
\left[
1
+
\omega
\ln\left(C^{(\textrm{e})}_0\omega\right)
\right],
\\
a^{(2)}_{1}
=&\,
-
\sqrt{2}\,
\left[
1
+
\omega
\ln\left(C^{(\textrm{e})}_0\omega\right)
\right].
\end{split}
\label{eq: omega dependence seeds rec rel N=2}
\end{equation}
We have thus derived two limiting solutions
to the recursion relations \re{eq: rec rel n=2}
\begin{equation}
\label{eq: asymptotic solutions n=2}
\begin{split}
a^{(1)}_{m}
=&\,
\left\{
\begin{array}{l l}
+
 \,
2 \sqrt{2 m\omega } \,
K^{\ }_{1}
\left(
2 \sqrt{ m\omega }
\right),
&\qquad
1\ll m,
\\
+
m \,
\sqrt{2}\,
\left[
\omega
\ln\left(C^{( \textrm{e}) }_0\omega\right)
\right]
+
\sqrt{2}\, ,
&
\qquad m \ll \omega^{-1},
\end{array}
\right.
\\
 &
\\
a^{(2)}_{m}
=&\,
\left\{
\begin{array}{l l}
-
 \,
2 \sqrt{2 m\omega } \,
K^{\ }_{1}
\left(
2 \sqrt{m\omega }
\right),
&\qquad
1\ll m,
\\
-
m \,
\sqrt{2}\,
\left[
\omega
\ln\left(C^{(\textrm{e})}_0\omega\right)
\right]
-
\sqrt{2}\, ,
&
\qquad m \ll \omega^{-1}.
\end{array}
\right.
\end{split}
\end{equation}

In the long wire limit, the density of states
is given by the expectation value of the operator $(\widebar{B}-B)$
in the ground state [see Eq.~\re{eq: QM DOS}]
\begin{equation}
\label{eq: dos n=2 first}
\begin{split}
\nu(\varepsilon)
=&\,
\lim_{  \omega \to -i \varepsilon }
\pi^{-1}
\textrm{Re} \,
{}^{\ }_{\mathrm{L}}\!\left\langle \varphi^{\ }_{0} \right|
(\widebar{B}-B)
\left|Ê\varphi^{\ }_{0} \right\rangle^{\ }_{\mathrm{R}}
\\
=&\,
\lim_{  \omega \to -i \varepsilon }
\pi^{-1}
\textrm{Re} \,
\left[
2
+
\sum_{m=1}^{\infty}
\left(
\frac{ \left( a^{(1)}_{m}\right)^2 }{2m+1}
-
\frac{ \left( a^{(2)}_{m}\right )^2 }{2m-1}
\right)
\right].
\end{split}
\end{equation}
Here, we have used Eq.~\re{eq: norm of basis states},
the identity
$( \widebar{B} - B ) \left| 0 \right\rangle = 2 \left| 0 \right\rangle$
and
\begin{equation}
\begin{split}
&
\left(
\widebar{B}-B
\right)
\left|m\right\rangle^{(1)}_{\mathrm{R}}
=
+
\left| m \right\rangle^{(1)}_{\mathrm{R}}
+
\left| m \right\rangle^{(1)}_{\mathrm{L}},
\\
&
\left(
\widebar{B}-B
\right)
\left| m \right\rangle^{(2)}_{\mathrm{R}}
=
-
\left| m \right\rangle^{(2)}_{\mathrm{R}}
-
\left| m \right\rangle^{(2)}_{\mathrm{L}}.
\end{split}
\end{equation}

We want to estimate the leading behavior
of the density of states for two scattering channels, $N=2$,
and for asymptotically small energies $\varepsilon$,
i.e. $\omega \ll1$.
Thereto, we break the sum over $m$ in
Eq.~\re{eq: dos n=2 first}
into two parts separated by the integer
$1\ll m^{\ }_{0} \sim  \omega^{-1}$. With
the choice
$m_{0}=\omega^{-1}$, we find that the sum in
Eq.~\re{eq: dos n=2 first}
is dominated by the small $m\ll 1/\omega$ solutions in
Eq.~\re{eq: asymptotic solutions n=2} [the rest of the sum
is easily shown to be $\mathcal{O}(\omega)$],
\begin{equation}
\begin{split}
 {}^{\ }_{\mathrm{L}}\!\left\langle \varphi^{\ }_{0} \right|
(\widebar{B}-B)
\left|Ê\varphi^{\ }_{0} \right\rangle^{\ }_{\mathrm{R}}
\approx&\,
2
+
2 \sum_{m=1}^{1/\omega}
\left(
\frac{1 }{2m+1}
-
\frac{1 }{2m-1}
\right)
\Big[
1
+
 m \,
\omega
\ln\left(C^{(\textrm{e}) }_0\omega\right)
\Big]^2.
\end{split}
\label{eq: dropping S2 N=2}
\end{equation}
This sum over $m$ can be computed exactly
by means of a telescopic expansion. We find
\begin{equation}
\begin{split}
 {}^{\ }_{\mathrm{L}}\!\left\langle \varphi^{\ }_{0} \right|
(\widebar{B}-B)
\left|Ê\varphi^{\ }_{0} \right\rangle^{\ }_{\mathrm{R}}
\approx&\,
2
- \frac{4}{2 + \omega}
- 2 \omega
\left(\frac{1 + \omega}{2+ \omega }\right)
\ln^2\left(C^{(\textrm{e})}_0\omega\right)
+
\omega
\Big[
2
-
2 \gamma
\\
&\,
-
2 \ln 4
-
\Psi \left(\omega^{-1}+1/2\right)-\Psi\left(\omega^{-1}+3/2\right)
\Big]
\ln\left(C^{(\textrm{e})}_0\omega\right).
\end{split}
\label{eq: telescopic expansion N=2}
\end{equation}
To compute the leading behavior of the density of states
it is sufficient to retain only the lowest order in $\omega$.
This gives
\begin{eqnarray}
&&
{}^{\ }_{\mathrm{L}}\!\left\langle \varphi^{\ }_{0} \right|
(\widebar{B}-B)
\left|Ê\varphi^{\ }_{0} \right\rangle^{\ }_{\mathrm{R}}
\propto
-
\omega \ln^2 \omega + \textrm{higher order terms}.
\end{eqnarray}
We thus find the estimate
\begin{equation}
\begin{split}
\nu(\varepsilon)
\propto&\,
\lim_{\omega\to-\mathrm{i}\varepsilon}
\pi^{-1}
\mathrm{Re}\,
\left(
-
\omega
\ln^{2} \omega
\right)
\propto 
 \varepsilon
\left| \ln  \varepsilon \right|
\end{split}
\end{equation}
for the density of states in the thermodynamic limit.

\section{
Evaluation of the density of states when $N=1,2,3,\cdots$
        }
\label{sec: Evaluation of the density of states when N=1,2,3,...}

We are going to compute the density of states
$\nu(\varepsilon)$ from Eq.~(\ref{eq: dos in ground state})
to leading order in the positive
dimensionless energy $\varepsilon\ll1$
by solving the recursion relation~(\ref{eq: recursion relations gen N})
for an arbitrary number $N$ of channels.
We begin with the case of $N$ odd and proceed with the
case of $N$ even. In doing so, we are going to reproduce
the parity effect~(\ref{eq: main result paper})
in the density of states that was
obtained for the first time using the DMPK approach
in Refs.\ \cite{Brouwer00} and \cite{Titov01}.

\subsection{
Density of states for an odd number $N$ of channels
           }

The calculation of the density of states for an odd number
$N$ of channels follows along the lines of the $N=1$ case
with the caveat that for $N>1$ the recursion
relations~\re{eq: recursion relations gen N} no longer decouple.
This difference complicates the calculation considerably.
To overcome this difficulty we introduce a linear transformation of
the coefficients $a^{(n)}_m$ that
approximately decouples the recursion relations%
~\re{eq: recursion relations gen N} in the two regimes
$m \gg N$ and $m \ll \omega^{-1}$. That is, we introduce

\begin{subequations}
\label{eq: transf N odd}
\begin{equation}
\begin{split}
&
b^{(2k+1)}_{m}
=
\sum_{n=0}^{(N-1)/2}
\left[\mathcal{M}^{\ }_{\mathrm{o};N,1}\right]^{k}_{n}
a^{(2n+1)}_m ,
\qquad
k=0, 1, \cdots, \frac{N-1}{2},
\\
&
b^{(2k+2)}_{m}
=
\sum_{n=0}^{(N-3)/2}
\left[\mathcal{M}^{\ }_{\mathrm{o};N,2}\right]^{k}_{n}
a^{(2n+2)}_m,
\qquad
k=0, 1, \cdots, \frac{N-3}{2},
\end{split}
\label{eq: transf N odd a}
\end{equation}
with $m=1,2,3,\cdots$
and the transformation matrices
\begin{equation}
\begin{split}
\left[\mathcal{M}^{\ }_{\mathrm{o};N,1}\right]^{k}_{n}
:=&\,
(-1)^n
\frac{ (2n)!  ( N-2n-1)!}
{ n! \left( \frac{N-1}{2} -n \right)! }
\;
{}_3F_{2}
\left[
\begin{array}{c}
-k, k, -n  \\
\frac{1}{2} ,  \frac{1-N}{2}   \end{array}; 1
\right],
\\
\left[\mathcal{M}^{\ }_{\mathrm{o};N,2}\right]^{k}_{n}
:=&\,
(-1)^n
\frac{  (2n+1)!  ( N-2n-2)!}
{ n! \left( \frac{ N-3}{2} -n \right)!}
\;
{}_3F_{2}
\left[
\begin{array}{c}
-k, k+2, -n   \\
\frac{3}{2} ,  \frac{3-N}{2}  \end{array}; 1
\right].
\end{split}
\label{eq: transf N odd b}
\end{equation}
Here, the symbol ${}_3 F_2$ denotes
a generalized hypergeometric function
(see for example Ref.~\cite{Gradshteyn94}).
It is also possible to express the $a^{(n)}_m$'s
in terms of the $b^{(k)}_{m}$'s,
\begin{equation}
\begin{split}
&
a^{(2n+1)}_{m}
=
\sum_{k=0}^{(N-1)/2}
\left[\mathcal{M}^{-1}_{\mathrm{o};N,1}\right]^{n}_{k}
b^{(2k+1)}_m ,
\qquad
n=0, 1, \cdots, \frac{N-1}{2},
\\
&
a^{(2n+2)}_{m}
=
\sum_{k=0}^{(N-3)/2}
\left[\mathcal{M}^{-1}_{\mathrm{o};N,2}\right]^{n}_{k}
b^{(2k+2)}_m,
\qquad
n=0, 1, \cdots, \frac{N-3}{2}.
\end{split}
\label{eq: transf N odd a bis}
\end{equation}
\end{subequations}
With these definitions, we are going to
rewrite the recursion relations%
~\re{eq: recursion relations gen N}
in a form that decouples both in the limit $\omega=0$ and
in the large $m$ limit.

To see this, we first make use of
identities~(\ref{prop: HYP GEO 1b}) and~(\ref{prop: HYP GEO 2b}) to combine
Eqs.~(\ref{eq: recursion relations gen N}) and
(\ref{eq: transf N odd})
into the recursion relations
\begin{subequations}
\label{eq: rc rel for N odd}
\begin{equation}
\begin{split}
\label{eq: rec rel for b1 a}
4 m \omega b^{(2k+1)}_m
=&\,
\left[
-4m(2m-1+N)
+(N-1-4k^2)
\right]
b^{(2k+1)}_m
\\
&\,
+ 2m(2m+1) b^{(2k+1)}_{m+1}
+ (2m-2+N)(2m+N-1) b^{(2k+1)}_{m-1}
\\
&\,
-
4 \omega
\sum_{n=0}^{(N-1)/2}
n
\left[\mathcal{M}^{\ }_{\mathrm{o};N,1}\right]^{k}_{n}
a_m^{(2n+1)}
\end{split}
\end{equation}
with $k=0,1,\cdots,(N-1)/2$ and
\begin{equation}
\label{eq: N odd rec rel for b(2n+2) b}
\begin{split}
4m \omega b^{(2k+2)}_{m}
=&\,
\big\{
-4m(2m-3 +N)
+\big[3N-5-4(k+1)^2 \big]
\big\}
b^{(2k+2)}_m
\\
&\,
+2m(2m-1)b^{(2k+2)}_{m+1}
+(2m-3+N)(2m-2+N)b^{(2k+2)}_{m-1}
\\
&\,
-
4 \omega
\sum_{n=0}^{(N-3)/2}
n
\left[\mathcal{M}^{\ }_{\mathrm{o};N,2}\right]^{k}_{n}
a^{(2n+2)}_m
\end{split}
\end{equation}
with $k=0,1,\cdots,(N-3)/2$ and $m=2,3,\cdots$.
Here, the initial conditions are
\begin{equation}
\begin{split}
4 \omega  b^{(2k+1)}_1
=&\,
+ \sqrt{2} N (N+1)
- (3N+5+4k^2) b^{(2k+1)}_1
+ 6 b^{(2k+1)}_2
\\
&\,
-
4 \omega
\sum_{n=0}^{(N-1)/2}
n
\left[\mathcal{M}^{\ }_{\mathrm{o};N,1}\right]^{k}_{n}
a_1^{(2n+1)}
\end{split}
\end{equation}
with $k=0,1,\cdots,(N-1)/2$ and
\begin{equation}
\begin{split}
4 \omega b^{(2k+2)}_1
=&\,
- \sqrt{2} N (N-1)
-  \big[ N+1+ 4(k+1)^2 \big]
b^{(2k+2)}_1
+ 2 b^{(2k+2)}_2
\\
&\,
-
4
\omega
\sum_{n=0}^{(N-3)/2}
n
\left[\mathcal{M}^{\ }_{\mathrm{o};N,2}\right]^{k}_{n}
a_1^{(2n+2)}
\end{split}
\end{equation}
\end{subequations}
with $k=0,1,\cdots,(N-3)/2$.
We aim at a solution of Eq.~\re{eq: rc rel for N odd}
for asymptotically small frequency~$\omega$.
As in Sec.~\ref{subsec: Density of states for a single channel N=1},
our strategy is to solve Eq.~\re{eq: rc rel for N odd}
in the two limits $N \ll m$ and $m \ll \omega^{-1}$.
In the former limit, $m \gg N$, we treat $m$ as a continuous variable.
Then, Eq.~\re{eq: rc rel for N odd}
decouples, provided we assume that $a^{(n)}_m$ decays
rapidly for $m \gg N$.
In the other limit, $m \ll \omega^{-1}$,
we can neglect $\omega$ in Eq.~\re{eq: rc rel for N odd},
which again decouples.
In this way,
it is possible to find approximate solutions in the two regions
$N \ll m$ and $m \ll \omega^{-1}$
that are uniquely fixed up to
some multiplicative coefficients and initial values,
respectively.
For asymptotically small $\omega$,
the overlap between these two regions
$N \ll m \ll \omega^{-1}$ is large.
We can then match the two approximate
solutions in the overlapping region.
This gives a unique and approximate solution for
the ground state wave function,
which in turn determines the density
of states~\re{eq: dos in ground state} in the long wire limit.

\subsubsection{
Solution when $m\gg N$
              }

First, we treat the limit $m \gg N$.
If we assume that $a^{(n)}_m$ decays rapidly for large $m$,
we can drop the last line on the right-hand side
of Eqs.~\re{eq: rec rel for b1 a}
and~\re{eq: N odd rec rel for b(2n+2) b}, respectively.
Consequently, in terms of the finite differences
\begin{equation}
\Delta b^{(i)}_m :=
b^{(i)}_{m+1} - b^{(i)}_m,
\qquad
\qquad
\Delta^2 b^{(i)}_m :=
b^{(i)}_{m+2} - 2 b^{(i)}_{m+1} + b^{(i)}_m,
\end{equation}
with $i=1,\cdots,N$,
the recursion relations Eqs.~\re{eq: rec rel for b1 a}
and
and~\re{eq: N odd rec rel for b(2n+2) b}
read
\begin{subequations}
\label{eq: decoupled rec rel N odd large omega limit}
\begin{equation}
\begin{split}
(m+1) 4 \omega
\left(
\Delta b^{(2k+1)}_{m} + b^{(2k+1)}_{m}
\right)
=&\,
-2(2m+2)(N-2)
\Delta b^{(2k+1)}_m
\\
&\,
+(2m+2)(2m+3)
\Delta^2 b^{(2k+1)}_m
\\
&\,
+(N-1)(N-2)b^{(2k+1)}_m
\\
&\,
+(N-1-4k^2)
\left(
\Delta b^{(2k+1)}_m+b^{(2k+1)}_m
\right)
\end{split}
\label{eq: N odd m>>N rec a}
\end{equation}
with $k=0,1,\cdots,(N-1)/2$ and
\begin{equation}
\begin{split}
(m+1)  4 \omega
\left(
\Delta b^{(2k+2)}_m + b^{(2k+2)}_m
\right)
=&\,
-2(2m+2)(N-2) \Delta b^{(2k+2)}_m
\\
&\,
+ (2m+2)(2m+1) \Delta^2 b^{(2k+2)}_m
\\
&\,
+ (N-2)(N-3) b^{(2k+2)}_m
\\
&\,
+ \big[ 3N -5 - 4(k+1)^2 \big]
\left(
\Delta b^{(2k+2)}_m + b^{(2k+2)}_m
\right)
\end{split}
\label{eq: N odd m>>N rec b}
\end{equation}
\end{subequations}
with $k=0,1,\cdots,(N-3)/2$.
In the limit  $m\gg N$,
we can neglect terms of order $N$
compared to $m$ and replace finite differences by derivatives.
In place of
Eq.~(\ref{eq: decoupled rec rel N odd large omega limit})
and if we assume that
$b^{(i)}_{m}\to b^{(i)}_{\vphantom{m}}(m)$
with $i=1,2,\cdots,N$ is slowly varying,
we get
\begin{subequations}
\label{eq: diff eq for lage m N odd}
\begin{equation}
4 m \omega b^{(2k+1)}
=
- 4m (N-2) \frac{d   b^{(2k+1)}}{d m}
+ 4 m^2    \frac{d^2 b^{(2k+1)}}{d m^2}
+ \left[ (N-1)^2 -4 k^2 \right]  b^{(2k+1)}
\end{equation}
with $k=0,1,\cdots,(N-1)/2$ and
\begin{equation}
4 m \omega b^{(2k+2)}
=
-4m(N-2) \frac{d  b^{(2k+2)}}{d m  }
+4m^2    \frac{d^2b^{(2k+2)}}{d m^2}
+ \big[ (N-1)^2-4(k+1)^2 \big]
b^{(2k+2)}
\end{equation}
\end{subequations}
with $k=0,1,\cdots,(N-3)/2$.
By use of the substitution
\begin{subequations}
\label{eq: def x as fct m and omega if large m N odd}
\begin{equation}
x^2:=4\omega m,
\label{eq: def x as fct m and omega N odd}
\end{equation}
whereby
\begin{equation}
\frac{x^{2}}{4\omega}\gg N,
\label{eq: large m condition on x N odd}
\end{equation}
\end{subequations}
we find that the solutions
to Eq.~\re{eq: diff eq for lage m N odd}
are given by the linear combinations of
\begin{subequations}
\label{eq: sol large m N odd}
\begin{equation} \label{eq: sol large m N odd a}
b^{(2k+1)}(x)=
c^{(2k+1)}_{\textrm{o},1}
\left( \frac{x}{2}\right)^{N-1}
K^{\ }_{2k} (x)
+
c^{(2k+1)}_{\textrm{o},2}
\left(\frac{x}{2}\right)^{N-1}
I^{\ }_{2k}(x)
\end{equation}
with $k=0,1,\cdots,(N-1)/2$ and
\begin{equation} \label{eq: sol large m N odd b}
b^{(2k+2)}(x)=
c^{(2k+2)}_{\textrm{o},1}
\left(\frac{x}{2}\right)^{N-1}
K^{\ }_{2k+2}(x)
+
c^{(2k+2)}_{\textrm{o},2}
\left(\frac{x}{2}\right)^{N-1}
I^{\ }_{2k+2} (x)
\end{equation}
\end{subequations}
with $k=0,1,\cdots,(N-3)/2$,
of modified Bessel functions $K^{\ }_{2k}$ and $I^{\ }_{2k}$.
We shall demand that $b^{(i)}_{m}$
with $i=1,\cdots,N$ decay to zero as $m\to\infty$,
i.e., we must set
\begin{equation}
c^{(i)}_{\textrm{o},2}=0
\label{eq: c(i) 0,2=0}
\end{equation}
with $i=1,\cdots,N$
in Eq.~\re{eq: sol large m N odd}.
The remaining $N$
coefficients
$c^{(i)}_{\textrm{o},1}$
with $i=1,\cdots,N$
of the modified Bessel functions
$K^{\ }_{0}, K^{\ }_{2}, K^{\ }_{4}, \cdots,K^{\ }_{N-1}$
are fixed by matching
solutions \re{eq: sol large m N odd}
to the solutions in the $m \ll \omega^{-1}$  region.
Thereto, we need to extract the terms that are of order
$x^{N+2k-1}$
and
$x^{N+2k-3}$
from the expansion of Eq.~\re{eq: sol large m N odd a}
and the terms that  are of order
$x^{N+2k+1}$
and
$x^{N+2k-1}$
from the expansion of Eq.~\re{eq: sol large m N odd b},
when $x\ll1$.
This gives
\begin{subequations}
\label{eq: final m>>N if N odd}
\begin{equation}
\begin{split}
b^{(1)}(x)
\sim&\,
c^{(1)}_{\textrm{o},1}
\left(\frac{x}{2}\right)^{N-1}
\frac{1}{2}
\left(-2\gamma-\ln\frac{x^2}{4}\right)
\\
=&\,
-
c^{(1)}_{\textrm{o},1}
(m\omega )^{\frac{N-1}{2}}
\frac{1}{2}
\ln \left(C^{(\textrm{o})}_{0}m\omega\right),
\end{split}
\label{eq: sol N odd large m}
\end{equation}
\begin{equation}
\label{eq: sol N odd large mBB}
\begin{split}
b^{(2k+1)}(x)
\sim& \,
c^{(2k+1)}_{\textrm{o},1}
\left[
\frac{(x/2)^{N+2k-1}}{2(2k)!}
\left(
-\gamma-\ln\frac{x^2}{4}+\Psi(2k+1)
\right)
-
\frac{(x/2)^{N+2k-3}}{2(2k-1)!}
\right]
\\
=&\,
c^{(2k+1)}_{\textrm{o},1}
\left[
-
\frac{(m\omega)^{\frac{N-1}{2}+k}}{2(2k)!}
\ln\left(C^{(\textrm{o})}_{k}m\omega\right)
-
\frac{(m\omega)^{\frac{N-1}{2}+k-1}}{2(2k-1)!}
\right],
\end{split}
\end{equation}
with $k=1,2,\cdots,(N-1)/2$,
and
\begin{equation}
\label{eq: sol N odd large mCC}
\begin{split}
b^{(2k+2)} (x)
\sim&\,
c^{(2k+2)}_{\textrm{o},1}
\left[
\frac{(x/2)^{N+2k+1}}
{2(2k+2)!}
\left(
-\gamma-\ln\frac{x^2}{4}
+\Psi(2k+3)
\right)
-
\frac{(x/2)^{N+2k-1}}
     {2(2k+1)!}
\right]
\\
=&\,
c^{(2k+2)}_{\textrm{o},1}
\left[
-\frac{(m\omega)^{\frac{N+1}{2}+k}}
{2(2k+2)!}
\ln\left(C^{(\textrm{o})}_{k+1}m\omega\right)
-
\frac{(m\omega )^{\frac{N+1}{2}+k-1}}
{2(2k+1)!}
\right],
\end{split}
\end{equation}
with $k=0,1,\cdots, (N-3)/2$, and where
\begin{equation}
C^{(\textrm{o})}_k:=
\exp\Big(\gamma - \Psi (2k+1)\Big).
\end{equation}
\end{subequations}
(For references on the Digamma function $\Psi$,
see chapters~6.3 or 8.36 in Refs.~\cite{Abramowitz65}
or \cite{Gradshteyn94}, respectively.)

\subsubsection{
Solution when $1\leq m\ll\omega^{-1}$
              }

Second, we treat the limit $1\leq m\ll\omega^{-1}$,
in which case we can neglect the $\omega$ terms
in Eq.~\re{eq: rc rel for N odd}.
In doing so, Eq.~\re{eq: rc rel for N odd}
becomes
\begin{subequations}
\label{eq: rc rel for N odd w=0}
\begin{equation}
\begin{split}
\label{eq: rec rel for b1 w=0 a}
0
=&\,
\left[
- 4 m(2m-1+N)
+ (N-1-4k^2)
\right]
b^{(2k+1)}_m
+ 2m(2m+1) b^{(2k+1)}_{m+1}
\\
&\,
+ (2m-2+N)(2m-1+N) b^{(2k+1)}_{m-1}
\end{split}
\end{equation}
with $k=0,1,\cdots,(N-1)/2$ and
\begin{equation}
\label{eq: N odd rec rel for b(2n+2) w=0 b}
\begin{split}
0
=&\,
\big\{
- 4m(2m-3+N)
+ \big[ 3N-5-4(k+1)^2 \big]
\big\}
b^{(2k+2)}_m
+ 2m (2m-1) b^{(2k+2)}_{m+1}
\\
&\,
+ (2m-3 +N)(2m-2+N) b^{(2k+2)}_{m-1}
\end{split}
\end{equation}
with $k=0,1,\cdots,(N-3)/2$, $m=2,3,\cdots$,
together with the initial conditions
\begin{equation}
\begin{split}  \label{eq: first initial cond ODD}
0
=&\,
+ \sqrt{2} N (N+1)
- (5+ 3N  + 4 k^2) b^{(2k+1)}_1
+ 6 b^{(2k+1)}_2
\end{split}
\end{equation}
with $k=0,1,\cdots,(N-1)/2$ and
\begin{equation}
\begin{split} \label{eq: second initial cond ODD}
0
=&\,
- \sqrt{2} N (N-1)
-  \big[ N+1+ 4(k+1)^2 \big]
b^{(2k+2)}_1
+ 2 b^{(2k+2)}_2
\end{split}
\end{equation}
\end{subequations}
with $k=0,1,\cdots,(N-3)/2$.
The solution to the recursion relation~(\ref{eq: rc rel for N odd w=0})
can be expressed in terms of generalized hypergeometric functions
\begin{subequations}
\label{eq: sol rec rel odd N w=0}
\begin{equation}
\begin{split}
b_m^{(1)}
= & \,
\mathcal{C}^{(\mathrm{o},1)}_0
\sum_{l=0}^{m-1}
\frac{ \left(m-1-l + \frac{N-1}{2} \right)! }
{ (2l+1)(m-1-l)!}
+
\mathcal{C}^{(\mathrm{o},2)}_0
\frac{ \left( m -1 + \frac{N}{2} \right)! }
{ \left( m - \frac{1}{2} \right) ! },
\end{split}
\label{solution N ODD a}
\end{equation}
\begin{equation}
\begin{split}
b_m^{(2k+1)}
=&\,
\mathcal{C}^{(\mathrm{o},1)}_k
\frac{ \left( m+k- \frac{3}{2} + \frac{N}{2} \right)! }
{ ( m -1)!}
 {}_3 F_2
\left[
\begin{array}{c}
 \frac{1}{2} -k,  1 -k, 1-m \\
 \frac{3}{2} , \frac{3}{2}-k-m- \frac{N}{2}
\end{array}; 1
\right]
\\
&\,
+
\mathcal{C}^{(\mathrm{o},2)}_k
\frac{ \left( m + k -1 + \frac{N}{2} \right)! }
{ \left( m - \frac{1}{2} \right) ! }
 {}_3 F_2
\left[
\begin{array}{c}
 \frac{1}{2} -k,   -k, \frac{1}{2}-m \\
 \frac{1}{2} , 1-k-m- \frac{N}{2}
\end{array}; 1
\right]
\end{split}
\label{solution N ODDb}
\end{equation}
with $k=1,2,\cdots,(N-1)/2$, and
\begin{equation}
\begin{split}
b_m^{(2k+2)}
=&\,
\mathcal{C}^{(\mathrm{o},3)}_k
\frac{ \left( m+k-\frac{1}{2} + \frac{N}{2} \right)! }
{(m-1)!}
{}_3 F_2
\left[
\begin{array}{c}
 -\frac{1}{2} -k, -1  -k, 1-m \\
 \frac{1}{2} , \frac{1}{2}-k-m- \frac{N}{2}
\end{array}; 1
\right]
\\
&\,
+
\mathcal{C}^{(\mathrm{o},4)}_k
\frac{ \left(m+k-1 +\frac{N}{2} \right)! }
{ \left( m - \frac{3}{2} \right)! }
{}_3 F_2
\left[
\begin{array}{c}
 -\frac{1}{2} -k,   -k, \frac{3}{2}-m \\
 \frac{3}{2} , 1-k-m - \frac{N}{2}
\end{array}; 1
\right]
\end{split}
\label{solution N ODDc}
\end{equation}
\end{subequations}
with $k=0,1,\cdots,(N-3)/2$ and $m=1,2,3,\cdots$.
The prefactors $\mathcal{C}^{(\mathrm{o},i)}_k$
$i=1,2,3,4$
to the hypergeometric function ${}_3 F_2$
are determined by the initial conditions \re{eq: first initial cond ODD}
and \re{eq: second initial cond ODD},
\begin{subequations}
\label{eq: sol rec rel odd N w=0 coefficients}
\begin{equation}
\begin{split}
&
\mathcal{C}^{(\mathrm{o},1)}_0
=
\frac{b^{(1)}_1-\sqrt{2}N }
     {\left(\frac{N-1}{2}\right)!},
\qquad
\mathcal{C}^{(\mathrm{o},2)}_0
=
\frac{\sqrt{2\pi}}{\left(\frac{N-2}{2}\right)!},
\end{split}
\label{eq: sol rec rel odd N w=0 coefficients a}
\end{equation}
\begin{equation}
\begin{split}
&
\mathcal{C}^{(\mathrm{o},1)}_k
=
\frac{1}
     {\left( k + \frac{N-1}{2} \right)!}
\left\{
b_1^{(2k+1)}
-
\mathcal{C}^{(\mathrm{o},2)}_k
\frac{2}{\sqrt{\pi}} \left( k + \frac{N}{2} \right)!
{}_3 F_2
\left[
\begin{array}{c}
  \frac{1}{2} -k,   -k, -\frac{1}{2} \\
 \frac{1}{2} ,  -k - \frac{N}{2}
\end{array}; 1
\right]
\right\},
\\
&
\mathcal{C}^{(\mathrm{o},2)}_k
=
\sqrt{\pi} 2^{k + \frac{1}{2} }
\frac{ \left( N -2k -1 \right)!! }
{ (N-1)!! \left( \frac{N-2}{2} \right)! },
\end{split}
\label{eq: sol rec rel odd N w=0 coefficients b}
\end{equation}
with $k=1,2, \cdots, (N-1)/2$, and
\begin{equation}
\begin{split}
&
\mathcal{C}^{(\mathrm{o},3)}_k
=
\frac{1}{ \left( k + \frac{N+1}{2} \right)! }
\left\{
b_1^{(2k+2)}
-
\mathcal{C}^{(\mathrm{o},4)}_k
\frac{1}{\sqrt{\pi}}
\left( k + \frac{N}{2} \right)!
{}_3 F_2
\left[
\begin{array}{c}
 -\frac{1}{2} -k,   -k,  \frac{1}{2} \\
 \frac{3}{2} ,  -k - \frac{N}{2}
\end{array}; 1
\right]
\right\},
\\
&
\mathcal{C}^{(\mathrm{o},4)}_k=
\sqrt{\pi} 2^{k+\frac{3}{2} }
\frac{(N-2k-3)!!}
{ (N-3)!! \left( \frac{N-2}{2}  \right)! } ,
\end{split}
\label{eq: sol rec rel odd N w=0 coefficients c}
\end{equation}
\end{subequations}
with $k=0,1,\cdots, (N-3)/2$.

In order to match this solution
when $1\leq m\ll 1/\omega$
to the solution when $N\ll m$,
we need the leading and subleading terms of the large $m$
behavior of
$b^{(2k+1)}_{m}$
and
$b^{(2k+2)}_{m}$.
They are
\begin{subequations}
\label{eq: sol N odd large m w=0}
\begin{equation}  \label{eq: sol N odd large m w=0a}
\begin{split}
b^{(1)}_m
\simeq &\,
\frac{ \mathcal{C}^{(\mathrm{o},1)}_0}{2}
m^{\frac{N-1}{2}} \ln m
+
\mathcal{C}^{(\mathrm{o})}_0
m^{\frac{N-1}{2}},
\end{split}
\end{equation}
with $\mathcal{C}^{(\mathrm{o})}_0$
some number that depends on $N$,
\begin{equation}
\begin{split}
b^{(2k+1)}_m
\simeq&\,
4^{k - \frac{1}{2} }
\left(
\frac{\mathcal{C}^{(\mathrm{o},1)}_k }{2k}
+
\mathcal{C}^{(\mathrm{o},2)}_k
\right)
m^{\frac{N-1}{2}+k}
\\
&\,
+
4^{k-2} (N+2k-1)(N-1)
\left(
\frac{\mathcal{C}^{(\mathrm{o},1)}_k }{2k}
+
\mathcal{C}^{(\mathrm{o},2)}_k
\right)
m^{\frac{N-1}{2}+k-1} ,
\end{split} 
\label{asymptotics N ODDb}
\end{equation}
with $k=1,2,\cdots,(N-1)/2$, and
\begin{equation}
\begin{split}
b^{(2k+2)}_m
\simeq&\,
4^{k+\frac{1}{2} }
\left(
\mathcal{C}^{(\mathrm{o},3)}_k
+
\frac{  \mathcal{C}^{(\mathrm{o},4)}_k }{2k+2}
\right)
m^{\frac{N+1}{2}+k}
\\
&\,
+
 4^{k-1} (N+2k+1)(N-3)
 \left(
\mathcal{C}^{(\mathrm{o},3)}_k
+ \frac{  \mathcal{C}^{(\mathrm{o},4)}_k }{2k+2}
 \right)
m^{\frac{N+1}{2}+k-1} ,
\end{split}
\label{asymptotics N ODDc}
\end{equation}
\end{subequations}
with $k=0,1,\cdots, (N-3)/2$,
to leading and subleading order in $m$.

\subsubsection{
Matching solutions when $N\ll m\ll\omega^{-1}$
              }

Having solved the recursion relations%
~\re{eq: rc rel for N odd} in the
two limits $N\ll m$ and $1\ll m\ll\omega^{-1}$,
we are going to match
the $m=x^{2}/(4\omega)$ dependences of the solutions
\re{eq: final m>>N if N odd}
and
\re{eq: sol N odd large m w=0}
in the overlapping
region $N\ll m\ll\omega^{-1}$,
where both solutions are valid.
We start with the coefficient
$b^{(1)}_m$.
Matching
Eq.~(\ref{eq: sol N odd large m w=0a})
with Eq.~(\ref{eq: sol N odd large m})
gives the two equations
\begin{equation}
\begin{split}
\mathcal{C}^{(\textrm{o})}_{0}
=&\,
-
\frac{c^{(1)}_{\textrm{o},1}}{2}
\omega^{\frac{N-1}{2}}
\ln
\left(
C^{(\textrm{o})}_0
\omega
 \right),
\\
\mathcal{C}^{(\textrm{o},1)}_0
=&\,
-
c^{(1)}_{\textrm{o},1} \omega^{\frac{N-1}{2}},
\end{split}
\end{equation}
for the two unknowns $b^{(1)}_1$ and $c^{(1)}_{\textrm{o},1}$
with the solutions
\begin{equation}
\begin{split}
c^{(1)}_{\mathrm{o},1}
=&\,
\frac{ -2 \mathcal{C}^{(\textrm{o})}_0 }
{\omega^{\frac{N-1}{2}} \ln \left( C^{(\textrm{o})}_0 \omega \right) } ,
\\
b^{(1)}_1
=&\,
\frac{ 2 \mathcal{C}^{(\textrm{o})}_0 \left( \frac{N-1}{2} \right)! }
{ \ln \left( C^{(\textrm{o})}_0 \omega \right) }
+ \sqrt{2} N.
\end{split}
\label{eq: N odd c 1}
\end{equation}
Matching Eq.~\re{asymptotics N ODDb}
with  Eq.~\re{eq: sol N odd large mBB}
gives two equations
(whereby we neglect $\ln m$ and $\ln C^{(\textrm{o})}_k$ compared
to $\ln\omega$)
\begin{subequations}
\label{eq: N odd 2k+1 or 2k+2}
\begin{equation}
\begin{split}
-c^{(2k+1)}_{\textrm{o},1}
\frac{\omega^{\frac{N-1}{2}+k}}
{2(2k)!}\ln\omega
=&\,
4^{k-\frac{1}{2}}
\left(
\frac{\mathcal{C}^{(\textrm{o},1)}_k}{2k}
+
\mathcal{C}^{( \textrm{o},2)}_k
\right),
\\
- c^{(2k+1)}_{\textrm{o},1}
\frac{\omega^{\frac{N-1}{2}+k-1}}
{2(2k-1)!}
=&\,
4^{k-2}
(N+2k-1)(N-1)
\left(
\frac{\mathcal{C}^{(\textrm{o},1)}_k}{2k}
+
\mathcal{C}^{(\textrm{o},2)}_k
\right),
\end{split}
\label{eq: N odd 2k+1 a}
\end{equation}
for the two unknowns $b^{(2k+1)}_1$ and $c^{(2k+1)}_{\textrm{o},1}$,
while matching  Eq.~\re{asymptotics N ODDc}
with  Eq.~\re{eq: sol N odd large mCC}
gives two equations
\begin{equation}
\begin{split}
-c^{(2k+2)}_{\textrm{o},1}
\frac{\omega^{\frac{N+1}{2}+k}}{2(2k+2)!}\ln\omega
=&\,
4^{k+\frac{1}{2}}
\left(
\mathcal{C}^{(\textrm{o},3)}_k
+
\frac{\mathcal{C}^{(\textrm{o},4)}_k}{2k+2}
\right),
\\
-c^{(2k+2)}_{\textrm{o},1}
\frac{\omega^{\frac{N+1}{2}+k-1}}{2(2k+1)!}
=&\,
4^{k-1}
(N+2k+1)(N-3)
\left(
\mathcal{C}^{(\textrm{o},3)}_k
+
\frac{\mathcal{C}^{(\textrm{o},4)}_k }{2k+2}
\right),
\end{split}
\label{eq: N odd 2k+2 b}
\end{equation}
\end{subequations}
for the two unknowns $b^{(2k+2)}_1$ and $c^{(2k+2)}_{\textrm{o},1}$.
Solving Eqs.~(\ref{eq: N odd 2k+1 a})
and~(\ref{eq: N odd 2k+2 b})
for $c^{(i)}_{\textrm{o},1}$ and $b^{(i)}_1$ gives
\begin{subequations}
\label{sol c1 and b1 for N odd}
\begin{equation}
c^{(i)}_{\textrm{o},1}=0
\label{eq: c(i) 0,1=0 i=2,3...}
\end{equation}
with $i=2,3,\cdots,N$,
\begin{equation}
\begin{split}
b^{(2k+1)}_1
=&\,
\mathcal{C}^{(\textrm{o},2)}_k
\left\{
-2k\left(\frac{N-1}{2}+k\right)!
+
\frac{2}{\sqrt{\pi}}
\left(\frac{N}{2}+k\right)!
{}_3 F_2
\left[
\begin{array}{c}
\frac{1}{2} -k, -k ,  - \frac{1}{2}\\
\frac{1}{2}, -k - \frac{N}{2}
\end{array}; 1 \right]
\right\}
\end{split}
\label{eq: N odd 2k+1 b}
\end{equation}
with $k=1,2,\cdots, (N-1)/2$, and
\begin{equation}
\begin{split}
b^{(2k+2)}_1
=&\,
\mathcal{C}^{(\textrm{o},4)}_k
\left\{
\frac{-1}{2k+2}
\left(\frac{N+1}{2}+k\right)!
+
\frac{1}{\sqrt{\pi}}
\left(\frac{N}{2}+k\right)!
{}_3 F_2
\left[
\begin{array}{c}
 -\frac{1}{2} -k, -k ,   \frac{1}{2}\\
\frac{3}{2}, -k - \frac{N}{2}
\end{array}; 1
\right]
\right\}
\end{split}
\label{eq: N odd 2k b}
\end{equation}
\end{subequations}
with $k=0,1,\cdots, (N-3)/2$.
Observe that
the coefficients~(\ref{eq: c(i) 0,1=0 i=2,3...})
of the modified Bessel functions
$K^{\ }_{2}, K^{\ }_{4},\cdots,K^{\ }_{N-1}$
are trivially independent of $\omega$,
while the initial values%
~(\ref{eq: N odd 2k+1 b})
and~(\ref{eq: N odd 2k b})
of the $\omega=0$ recursion relation
are independent of $\omega$
in view of Eqs.~(\ref{eq: sol rec rel odd N w=0 coefficients b})
and~(\ref{eq: sol rec rel odd N w=0 coefficients c}),
respectively.

\subsubsection{
$\omega$ dependence of
the recursion relation~(\ref{eq: rc rel for N odd})
              }

Equipped with the solutions
\re{eq: N odd c 1}
and
\re{sol c1 and b1 for N odd}
for the $\omega$ dependence of
$b^{(i)}_{1}$
and
$c^{(i)}_{\textrm{o},1}$
with $i=1,2,\cdots, N$,
we can derive the
$\omega$ dependence
of the coefficients
$b^{(i)}_m$
with $m=2,3,\cdots$
in the limit $1\leq m\ll\omega^{-1}$
from Eq.~(\ref{eq: sol rec rel odd N w=0}).
Insertion of Eq.~(\ref{eq: N odd c 1})
into the prefactors%
~(\ref{eq: sol rec rel odd N w=0 coefficients a})
appearing in Eq.~(\ref{solution N ODD a})
gives
\begin{subequations}
\label{final solution m small N odd}
\begin{equation}
b^{(1)}_m=
\frac{2\mathcal{C}^{(\textrm{o})}_0}
     {\ln\left(C^{(\textrm{o})}_0\omega\right)}
\sum_{l=0}^{m-1}
\frac{\left(m-1-l+\frac{N-1}{2}\right)!}
     {(2l+1)(m-1-l)!}
+
\mathcal{C}^{(\textrm{o},2)}_0
\frac{\left(m-1+\frac{N}{2}\right)!}
     {\left(m-\frac{1}{2}\right)!}
\label{eq: rec rel sol N odd small m a}
\end{equation}
with $\mathcal{C}^{(\textrm{o})}_0$ some number that
depends on $N$.
Insertion of Eq.~(\ref{eq: N odd 2k+1 b})
into the prefactor%
~(\ref{eq: sol rec rel odd N w=0 coefficients b})
appearing in Eq.~(\ref{solution N ODDb})
gives
\begin{equation}
\begin{split}
b^{(2k+1)}_m
=&\,
\mathcal{C}^{(\textrm{o},2)}_k
\frac{\left(m+k-1+\frac{N}{2}\right)!}
     {\left(m-\frac{1}{2}\right)!}
{}_3 F_2
\left[
\begin{array}{c}
\frac{1}{2}-k,-k,\frac{1}{2}-m\\
\frac{1}{2},1-k-m-\frac{N}{2}
\end{array};1
\right]
\\
&\,
-
\mathcal{C}^{(\textrm{o},2)}_k
2k
\frac{\left(m+k-\frac{3}{2}+\frac{N}{2}\right)!}
     {( m-1)!}
{}_3 F_2
\left[
\begin{array}{c}
\frac{1}{2} -k, 1 -k ,   1 - m\\
\frac{3}{2}, \frac{3}{2} -k - m- \frac{N}{2}
\end{array}; 1
\right]
\end{split}
\label{eq: rec rel sol N odd small m b}
\end{equation}
with $k=1,2, \cdots, (N-1)/2$.
Insertion of Eq.~(\ref{eq: N odd 2k b})
into the prefactor%
~(\ref{eq: sol rec rel odd N w=0 coefficients c})
appearing in Eq.~(\ref{solution N ODDc})
gives
\begin{equation}
\begin{split}
b^{(2k+2)}_m
=&\,
\mathcal{C}^{(\textrm{o},4)}_k
\frac{\left(m+k-1+\frac{N}{2}\right)!}
     {\left(m-\frac{3}{2}\right)!}
{}_3 F_2
\left[
\begin{array}{c}
-\frac{1}{2}-k,-k ,\frac{3}{2}-m\\
\frac{3}{2},1-k-m-\frac{N}{2}
\end{array};1
\right]
\\
&\,
-
\mathcal{C}^{(\textrm{o},4)}_k
\frac{1}{2k+2}
\frac{\left(m+k-\frac{1}{2}+\frac{N}{2}\right)!}
     {(m-1)!}
{}_3 F_2
\left[
\begin{array}{c}
-\frac{1}{2}-k,-1-k,1- m\\
\frac{1}{2},\frac{1}{2}-k-m-\frac{N}{2}
\end{array};1
\right]
\end{split}
\label{eq: rec rel sol N odd small m c}
\end{equation}
\end{subequations}
with $k=0,1,\cdots,(N-3)/2$.

In conclusion,
the $\omega$ dependence of
the solution to the recursion relation%
~(\ref{eq: rc rel for N odd})
follows from combining
Eqs.~\re{eq: sol large m N odd},
(\ref{eq: c(i) 0,2=0}),
(\ref{eq: N odd c 1}),
(\ref{eq: c(i) 0,1=0 i=2,3...})
when $m\gg N$
together with
Eq.~(\ref{final solution m small N odd})
when $m\omega\ll1$.
This gives
\begin{equation}
\begin{split}
b^{(1)}_m
=&\,
\left\{
\begin{array}{ll}
-\mathcal{C}^{(\textrm{o})}_{0}
\frac{2}
     {\ln\left(C^{(\textrm{o})}_0\omega\right)}
m^{\frac{N-1}{2}}K^{\ }_0
\left(2\sqrt{m\omega}\right),
&\quad
N\ll m,
\\
\frac{2\mathcal{C}^{(\textrm{o})}_0}
     {\ln\left(C^{(\textrm{o})}_0\omega\right)}
\sum\limits_{l=0}^{m-1}
\frac{\left(m-1-l+\frac{N-1}{2}\right)!}
     {(2l+1)(m-1-l)!}
+
\mathcal{C}^{(\textrm{o},2)}_0
\frac{\left(m-1+\frac{N}{2}\right)!}
     {\left(m-\frac{1}{2}\right)!},
&\quad
m\ll\omega^{-1},
\end{array}
\right.
\\
b^{(2k+1)}_m
=&\,
\left\{
\begin{array}{l l}
0,
&\quad
N\ll m,
\\
\textrm{ \re{eq: rec rel sol N odd small m b},}
&
\quad
m\ll\omega^{-1},
\end{array}
\right.
\\
b^{(2k^{\prime}+2)}_m
=&\,
\left\{
\begin{array}{l l}
0,
&\quad
N\ll m,
\\
\textrm{ \re{eq: rec rel sol N odd small m c}},
&
\quad
m\ll\omega^{-1},
\end{array}
\right.
\end{split}
\label{eq: limiting solutions to rec rel N odd}
\end{equation}
with $k=1,2,\cdots,(N-1)/2$ and $k^{\prime}=0,1,\cdots,(N-3)/2$.
Equation~(\ref{eq: limiting solutions to rec rel N odd})
should be compared to Eq.~(\ref{eq: limiting solutions to rec rel N=1}).
Observe that
Eqs.~(\ref{eq: rec rel sol N odd small m b})
and~(\ref{eq: rec rel sol N odd small m c})
are independent of $\omega$.

\subsubsection{
Leading energy dependence of the density of states
              }

We are ready to extract the leading behavior
of the average density of
states $\nu(\varepsilon)$, Eq.~(\ref{eq: dos in ground state}),
for asymptotically small energies $\varepsilon$.
The density of states in the long wire limit Eq.~\re{eq: QM DOS} is given by
\begin{equation} \label{DOS odd start}
\nu ( \varepsilon )
=
\lim_{ \omega \to -i \varepsilon  }
\pi^{-1} \mathrm{Re}\
{\vphantom{A}}^{\ }_{\mathrm{L}}
\left\langle \varphi^{\ }_{0} \right|
\left(\widebar{B} - B\right)
\left| \varphi^{\ }_{0} \right\rangle^{\ }_{\mathrm{R}}.
 \end{equation}
with  the left and right
ground states%
~(\ref{eq: expansion of eigenstatesa})
and~(\ref{eq: expansion of eigenstatesb}).
The normalizations%
~(\ref{eq: norm of basis states})
and%
~(\ref{eq: normalization a(0)}),
the biorthogonal relations%
~(\ref{eq: zero overlap}),
and the identities
\begin{equation}
\begin{split}
&
\left(\widebar{B}-B\right)\left|0\right\rangle
=
N
\left|0\right\rangle,
\\
&
\left(\widebar{B}-B\right)
\left|m\right\rangle^{(2n+1)}_{\mathrm{R}}=
(N-4n-1)\left|m\right\rangle^{(2n+1)}_{\mathrm{R}}
+
\left|m\right\rangle^{(2n+1)}_{\mathrm{L}},
\\
&
\left(\widebar{B}-B\right)
\left|m\right\rangle^{(2n+2)}_{\mathrm{R}}=
(N-4n-3)\left|m\right\rangle^{(2n+2)}_{\mathrm{R}}
-
\left|m\right\rangle^{(2n+2)}_{\mathrm{L}},
\end{split}
\end{equation}
deliver the expectation value
\begin{equation}
\label{dos in terms of bs NODD}
\begin{split}
{\vphantom{A}}^{\ }_{\mathrm{L}}
\left\langle \varphi^{\ }_{0} \right|
\left(\widebar{B} - B\right)
\left| \varphi^{\ }_{0} \right\rangle^{\ }_{\mathrm{R}}
=&\,
N
+
\sum_{m=1}^{\infty}
\sum_{n=0}^{(N-1)/2}
\mathcal{N}^{(1)}_{m,n}
\left( a^{(2n+1)}_m \right)^2
-
\sum_{m=1}^{\infty}
\sum_{n=0}^{(N-3)/2}
\mathcal{N}^{(2)}_{m,n}
\left( a^{(2n+2)}_m \right)^2
\\
=&\,
N
+
\sum\limits_{m=1}^{\infty}
\sum\limits_{n=0}^{(N-1)/2}
\mathcal{N}^{(1)}_{m,n}
\left(
\sum\limits_{k=0}^{(N-1)/2}
\left[\mathcal{M}^{-1}_{\mathrm{o};N,1}\right]^{n}_{k}
b^{(2k+1)}_{m}
\right)^2
\\
&
-
\sum\limits_{m=1}^{\infty}
\sum\limits_{n=0}^{(N-3)/2}
\mathcal{N}^{(2)}_{m,n}
\left(
\sum\limits_{k=0}^{(N-3)/2}
\left[\mathcal{M}^{-1}_{\mathrm{o};N,2}\right]^{n}_{k}
b^{(2k+2)}_{m}
\right)^2.
\end{split}
\end{equation}
The matrices
$\mathcal{M}^{\ }_{\mathrm{o};N,1}$
and
$\mathcal{M}^{\ }_{\mathrm{o};N,2}$
are given in Eq.~\re{eq: transf N odd},
while
the normalization factors $\mathcal{N}^{(1)}_{m,n}$
and
$\mathcal{N}^{(2)}_{m,n}$
are given by Eq.~(\ref{eq: norm of basis states}).
In view of Eq.~\re{eq: limiting solutions to rec rel N odd},
it is necessary to break the sum over $m$
in Eq.~\re{dos in terms of bs NODD}
into two parts separated by the integer $m^{\ }_{0}$,
with $N \ll m^{\ }_{0} \ll \omega^{-1}$.
We choose $m^{\ }_{0}$ to be some number independent of $\omega$.
With this choice and for sufficiently small $\omega$, we find that
the sum over $m$ in Eq.~\re{dos in terms of bs NODD}
is dominated by the contributions
from $b^{(1)}_{m}$ with
$m > m^{\ }_{0}$, i.e., by the contributions
from the first line in Eq.%
~(\ref{eq: limiting solutions to rec rel N odd}),
\begin{equation}
{\vphantom{A}}^{\ }_{\mathrm{L}}
\left\langle \varphi^{\ }_{0} \right|
\left(\widebar{B} - B\right)
\left| \varphi^{\ }_{0} \right\rangle^{\ }_{\mathrm{R}}
\approx
\int\limits_{m^{\ }_{0}}^{\infty} dm
\sum\limits_{n=0}^{(N-1)/2}
\mathcal{N}^{(1)}_{m,n}
\left(
\left[\mathcal{M}^{-1}_{\mathrm{o};N,1}\right]^{n}_{0}
b^{(1)}_{m} \right)^2.
\end{equation}
By inserting the solution for $b^{(1)}_m$
from Eq.~\re{eq: limiting solutions to rec rel N odd}
we obtain
\begin{subequations}
\begin{equation}
{\vphantom{A}}^{\ }_{\mathrm{L}}
\left\langle \varphi^{\ }_{0} \right|
\left(\widebar{B} - B\right)
\left| \varphi^{\ }_{0} \right\rangle^{\ }_{\mathrm{R}}
\approx
\frac{
4\big(\mathcal{C}^{(\textrm{o})}_0\big)^2
     }
     {
\ln^2\big(C^{(\textrm{o})}_0\omega\big)
     }
\int\limits_{m^{\ }_{0}}^{\infty}
d m \,
\mathcal{S}^{(\textrm{o})}_{N}(m)\,
m^{N-1}
K^2_0\left(2\sqrt{m\omega}\right) ,
\end{equation}
with the combinatorial factor
\begin{equation}
\mathcal{S}^{(\textrm{o})}_{N}(m)
:=
\sum_{n=0}^{(N-1)/2}
\mathcal{N}^{(1)}_{m,n}
\left(
\left[\mathcal{M}^{-1}_{\mathrm{o};N,1}\right]^{n}_{0}
\right)^2 .
\end{equation}
\end{subequations}
Using the substitution $x^2=4 m \omega$,
the integral transforms into
\begin{eqnarray}
{\vphantom{A}}^{\ }_{\mathrm{L}}
\left\langle \varphi^{\ }_{0} \right|
\left(\widebar{B} - B\right)
\left| \varphi^{\ }_{0} \right\rangle^{\ }_{\mathrm{R}}
\approx
\frac{
4 \big( \mathcal{C}^{(\textrm{o})}_0 \big)^2  \omega^{-N}
}
{ \ln^2 \big( C^{(\textrm{o})}_0 \omega \big) }
\int\limits_{x^{\ }_{0}}^{\infty}
d x \,
\mathcal{S}^{(\textrm{o})}_{N}
\left(\frac{x^2}{4\omega}\right)\,
\left(\frac{x}{2}\right)^{2N-1}
K^2_0\left(x\right) ,
\end{eqnarray}
with $x^{\ }_{0}:=2\sqrt{m^{\ }_{0}\omega}$.
Let us now expand
$\mathcal{S}^{(\textrm{o})}_{N}$
in small $\omega$ (i.e., large argument)
\begin{eqnarray}
\label{eq: exp of comb number N odd}
\mathcal{S}^{(\textrm{o})}_{N}
\left(  \frac{x^2}{4 \omega}  \right)
\approx
\mathcal{S}^{(\textrm{o},1)}_{N} \times
\left(
\frac{2 \omega}{x^2}
\right)^{N-1}
+
\mathcal{O}
\left[ \left(
\frac{2 \omega}{x^2}
\right)^{N}
\right],
\end{eqnarray}
where $\mathcal{S}^{(\textrm{o},1)}_{N} $
is some number that only depends on $N$.
In order to derive the asymptotic  density of states,
it is sufficient to keep the leading $\omega$ dependence
in Eq.~\re{eq: exp of comb number N odd}. This gives
\begin{equation}
{\vphantom{A}}^{\ }_{\mathrm{L}}
\left\langle \varphi^{\ }_{0} \right|
\left(\widebar{B} - B\right)
\left| \varphi^{\ }_{0} \right\rangle^{\ }_{\mathrm{R}}
\approx
\frac{
4\big(\mathcal{C}^{(\textrm{o})}_0\big)^22^{-N}\omega^{-1}
     }
     {
\ln^2\big(C^{(\textrm{o})}_0\omega\big)
     } \,
\mathcal{S}^{(\textrm{o},1)}_{N}
\int\limits_{x_0}^{\infty}
dx\,
x
K^{2}_{0}\left(x\right).
\label{eq: exp val 10}
\end{equation}
Combining
Eq.~\re{eq: exp val 10}
and
\re{DOS odd start},
we find that the density of states,
for asymptotically small energies,
is given by
 \begin{equation}
\begin{split}
\nu (\varepsilon)
\propto&\,
\lim\limits_{ \omega  \to -i \varepsilon }
\mathrm{Re}\,
\frac{1}
     {\omega   \ln^2 \big( C^{(\textrm{o})}_0 \omega \big) }
\sim
\frac{1}
     {\varepsilon \left| \ln^3\varepsilon \right| }.
\end{split}
\end{equation}
Recalling that $\varepsilon$ is measured in units
of the disorder strength, we have recovered
Eq.~(\ref{eq: main result paper})
when the number of channels $N$ is odd.

\subsection{
Density of states for an even number $N$ of channels
           }

The calculation of the density of states for an even number $N$
of channels follows along the lines of the $N=2$ case
with the caveat that for $N>2$ the recursion
relations~\re{eq: recursion relations gen N} no longer decouple.
This difference complicates the calculation considerably.
To overcome this difficulty we introduce a linear transformation of
the coefficients $a^{(n)}_m$ that
approximately decouples the recursion relations%
~\re{eq: recursion relations gen N} in the two regimes
$m \gg N$ and $m \ll \omega^{-1}$. That is, we introduce

\begin{subequations}
\label{eq: transf N even}
\begin{equation}
\begin{split}
&
b^{(2k+1)}_{m}=
\sum_{n=0}^{N/2-1}
\left[\mathcal{M}^{\ }_{\mathrm{e};N,1}\right]^{k}_{n}
a^{(2n+1)}_m ,
\qquad
k=0,1,\cdots,\frac{N}{2}-1,
\\
&
b^{(2k+2)}_{m}=
\sum_{n=0}^{N/2-1}
\left[\mathcal{M}^{\ }_{\mathrm{e};N,2}\right]^{k}_{n}
a^{(2n+2)}_m ,
\qquad
k=0,1,\cdots,\frac{N}{2}-1,
\end{split}
\label{eq: transf N even a}
\end{equation}
with $m=1,2,3,\cdots$ and
the transformation matrices
\begin{equation}
\begin{split}
\left[\mathcal{M}^{\ }_{\mathrm{e};N,1}\right]^{k}_{n}
:=&\,
(-1)^n
\frac{  (2n)! (N-1-2n)! \left( \frac{ N-2}{2} \right)! }
{ n! \left( \frac{N-2}{2}-n \right)! (N-1)! }
{}_3 F_2 \left[
\begin{array}{c}
-k, k+1, -n \\
\frac{1}{2},  \frac{2-N}{2}
\end{array} ;
1
\right],
\\
\left[\mathcal{M}^{\ }_{\mathrm{e};N,2}\right]^{k}_{n}
:=&\,
(-1)^n
\frac{ (2n+1)! (N-2 -2n)! \left( \frac{N-2}{2} \right)! }
{ n! \left( \frac{N-2}{2} -n \right)! (N-2)! }
{}_3 F_2 \left[
\begin{array}{c}
-k, k+1, -n \\
\frac{3}{2}, \frac{2-N}{2}
\end{array};
1
\right].
\end{split}
\label{eq: transf N even b}
\end{equation}
Here, the symbol ${}_3 F_2$ denotes
a generalized hypergeometric function
(see for example Ref.~\cite{Gradshteyn94}).
It is also possible to express the $a^{(n)}_m$'s
in terms of the $b^{(k)}_{m}$'s.
Indeed,
with the help of identities%
~\re{identity for inverse of F3}
and \re{identity for inverse of F4},
inverting the linear relation~(\ref{eq: transf N even a}), i.e.,
for any $m=1,2,3,\cdots$,
\begin{equation}
\begin{split}
&
a^{(2n+1)}_{m}=
\sum_{k=0}^{N/2-1}
\left[\mathcal{M}^{-1 }_{\mathrm{e};N,1}\right]^{n}_{k}
b^{(2k+1)}_m ,
\qquad
n=0,1,\cdots,\frac{N}{2}-1,
\\
&
a^{(2n+2)}_{m}=
\sum_{k=0}^{N/2-1}
\left[\mathcal{M}^{-1 }_{\mathrm{e};N,2}\right]^{n}_{k}
b^{(2k+2)}_m ,
\qquad
n=0,1,\cdots,\frac{N}{2}-1,
\end{split}
\label{eq: transf N even a bis}
\end{equation}
can be done with the inverses
\begin{equation}
\label{eq: a's in terms b's N even}
\begin{split}
\left[\mathcal{M}^{-1}_{\mathrm{e};N,1}\right]^{n}_{k}
=&\,
\frac{ \frac{ 2^{2- N }   }{ (2n)!  } \left[ (N-1)! \right]^2}{(N-1-2n)!
\left( \frac{N-2}{2}  -  k \right)! \left( \frac{N}{2}+k \right)! }
{}_3 F_2 \left[
\begin{array}{c}
-k-\frac{N}{2}, k+\frac{2-N}{2}, -n \\
\frac{1}{2}-\frac{N}{2} , \frac{2-N}{2}
\end{array};
1
\right],
\\
\left[\mathcal{M}^{-1}_{\mathrm{e};N,2}\right]^{n}_{k}
=&\,
\frac{
\frac{2^{2-N} }{ (2n+1)! }
\left[ (N-2)! \right]^2  (2k+1)^2
}
{ (N -2 -2n)!  \left( \frac{N-2}{2} -k \right)! \left( \frac{N}{2} + k \right)! }
{}_3 F_2 \left[
\begin{array}{c}
-k-\frac{N}{2}, k+\frac{2-N}{2}, -n \\
1 -\frac{N}{2} , \frac{3-N}{2}
\end{array};
1
\right],
\end{split}
\end{equation}
\end{subequations}
of the transformation matrices%
~(\ref{eq: transf N even b}).
With these definitions, we are going to
rewrite the recursion relations%
~\re{eq: recursion relations gen N}
in a form that decouples both in the limit $\omega=0$ and
in the large $m$ limit.

To see this, we first make use of
identities~(\ref{prop: HYP GEO 3b}) and~(\ref{prop: HYP GEO 4b})
to combine
Eqs.~(\ref{eq: recursion relations gen N}) and
(\ref{eq: transf N even})
into the recursion relations
\begin{subequations}
\label{eq: rc rel for N even}
\begin{equation}
\label{eq: rec rel decop even N 1}
\begin{split}
4m  \omega  b^{(2k+1)}_{m }
=&\,
\left\{
-4m(2m-1+N)
+\left[N-2-4k(k+1)\right]
\right\}
b^{(2k+1)}_m
\\
&\,
+2m(2m+1)
b^{(2k+1)}_{m+1}
\\
&\,
+(2m-2+N)(2m-1+N)
b^{(2k+1)}_{m-1}
\\
&\,
-4\omega
\sum_{n=0}^{N/2-1}
n\left[\mathcal{M}^{\ }_{\mathrm{e};N,1}\right]^{k}_{n}
a^{(2n+1)}_{m}
\end{split}
\end{equation}
and
\begin{equation}
\label{eq: rec rel decop even N 2}
\begin{split}
4m   \omega  b^{(2k+2)}_{m}
=&\,
\left\{
-4m(2m-3+N)
+\left[3N-6-4k(k+1)\right]
\right\}
b^{(2k+2)}_{m}
\\
&\,
+2m(2m-1)
b^{(2k+2)}_{m+1}
\\
&\,
+(2m-3+N)(2m-2+N)
b^{(2k+2)}_{m-1}
\\
&\,
-4\omega
\sum_{n=0}^{N/2-1}
n
\left[\mathcal{M}^{\ }_{\mathrm{e};N,2}\right]^{k}_{n}
a^{(2n+2)}_{m}
\end{split}
\end{equation}
with $k=0,1,\cdots,(N/2)-1$
and $m=2,3,\cdots$.
Here, the initial conditions are
\begin{equation}
\begin{split}
4  \omega b^{(2k+1)}_1
=&\,
+ \sqrt{2} N (N+1)
- \left[ 3N +6 + 4k(k+1) \right] b^{(2k+1)}_1
+ 6 b_2^{(2k+1)}
\\
&\,
- 4   \omega
\sum_{n=0}^{N/2-1}
n
\left[\mathcal{M}^{\ }_{\mathrm{e};N,1}\right]^{k}_{n}
a_1^{(2n+1)}
\end{split}
\end{equation}
with $k=0,1,\cdots,(N/2)-1$ and
\begin{equation}
\begin{split}
4  \omega  b_1^{(2k+2)}
=&\,
- \sqrt{2}N(N-1)
- \left[N+2+4k(k+1)\right]
b^{(2k+2)}_1
+ 2 b_2^{(2k+2)}
\\
&\,
-
4   \omega
\sum_{n=0}^{N/2-1}
n
\left[\mathcal{M}^{\ }_{\mathrm{e};N,2}\right]^{k}_{n}
a_1^{(2n+2)}
\end{split}
\end{equation}
\end{subequations}
with $k=0,1,2,\cdots,(N/2)-1$.
We aim at a solution of Eq.~\re{eq: rc rel for N even}
for asymptotically small frequency~$\omega$.
As in Sec.~\ref{subsec: Density of states for two channels N=2},
our strategy is to solve Eq.~\re{eq: rc rel for N even}
in the two limits $N\ll m$ and $m\ll\omega^{-1}$.
In the former limit, $m \gg N$, we treat $m$ as a continuous variable.
Then, Eq.~\re{eq: rc rel for N even}
decouples, provided we assume that $a^{(n)}_m$ decays
rapidly for $m \gg N$.
In the other limit, $m \ll \omega^{-1}$,
we can neglect $\omega$ in Eq.~\re{eq: rc rel for N even},
which again decouples
Eq.~\re{eq: rc rel for N even}.
In this way,
it is possible to find approximate solutions in the two regions
$N\ll m$ and $m\ll\omega^{-1}$
that are uniquely fixed up to
some multiplicative coefficients and initial values,
respectively.
For asymptotically small $\omega$,
the overlap between these two regions
$N\ll m\ll\omega^{-1}$ is large.
We can then match the two approximate
solutions in the overlapping region.
This gives a unique and approximate
solution for the ground state wave function,
which in turn determines the density
of states~\re{eq: dos in ground state} in the long wire limit.

\subsubsection{
Solution when $m\gg N$
              }

First, we treat the limit $m \gg N$.
If we assume that $a^{(n)}_m$ decays rapidly for large $m$,
we can drop the last line on the right-hand side
of Eqs.~\re{eq: rec rel decop even N 1}
and~\re{eq: rec rel decop even N 2}, respectively.
Consequently, in terms of the finite differences
\begin{equation}
\Delta b^{(i)}_m :=
b^{(i)}_{m+1} - b^{(i)}_m,
\qquad
\qquad
\Delta^2 b^{(i)}_m:=
b^{(i)}_{m+2} - 2 b^{(i)}_{m+1} + b^{(i)}_m ,
\end{equation}
with $i=1,\cdots,N$,
the recursion relation~(\ref{eq: rc rel for N even})
reads
\begin{subequations}
\label{eq: rc rel for N even large omega limit}
\begin{equation}
\begin{split}
(m+1) 4 \omega  \left( \Delta b^{(2k+1)}_m + b^{(2k+1)}_m \right)
=&\,
-
2 (2m+2)(N-2) \Delta b^{(2k+1)}_m
\\
&\,
+
(2m+2)(2m+3) \Delta^2 b_m^{(2k+1)}
\\
&\,
+
(N-1)(N-2) b^{(2k+1)}_m
\\
&\,
+
\left[ N-2-4k(k+1) \right]
\left( \Delta b^{(2k+1)}_m + b^{(2k+1)}_m \right)
\end{split}
\label{eq: N even m>>N rec a}
\end{equation}
with $k=0,1,\cdots,(N/2)-1$ and
\begin{equation}
\begin{split}
(m+1)  4 \omega \left( \Delta b^{(2k+2)}_m + b^{(2k+2)}_m \right)
=&\,
-2 (2m+2)(N-2) \Delta b_m^{(2k+2)}
\\
&\,
+ (2m+2)(2m+1) \Delta^2 b^{(2k+2)}_m
\\
&\,
+ (N-2)(N-3) b^{(2k+2)}_m
\\
&\,
+ \left[3N-6-4k(k+1)\right]
\left( \Delta b^{(2k+2)}_m + b^{(2k+2)}_m \right)
\end{split}
\label{eq: N even m>>N rec b}
\end{equation}
\end{subequations}
with $k=0,1,\cdots,(N/2)-1$.
Observe that
Eqs.~(\ref{eq: N even m>>N rec a})
and~(\ref{eq: N even m>>N rec b})
only differ from
Eqs.~(\ref{eq: N odd m>>N rec a})
and~(\ref{eq: N odd m>>N rec b})
through the coefficients
$\left[N-2-4k(k+1)\right]$
and
$\left[3N-6-4k(k-1)\right]$,
respectively.
In the limit  $m \gg N$,
we can neglect terms of order $N$
compared to $m$ and replace finite differences by derivatives.
In place of
Eq.~(\ref{eq: rc rel for N even large omega limit})
and if we assume that
$b^{(i)}_{m}\to b^{(i)}_{\vphantom{m}}(m)$
with $i=1,2,\cdots,N$ is slowly varying,
we get
\begin{subequations}
\label{eq: diff eq for lage m N even}
\begin{equation}
\begin{split}
4 m  \omega   b^{(2k+1)}
=&\,
-4 m (N-2)  \frac{d b^{(2k+1)}}{d m}
 + 4m^2 \frac{d^{2}b^{(2k+1)}}{d m^{2}}
+\left[ (N-1)^2 - (2k+1)^2  \right] b^{(2k+1)}
\end{split}
\label{eq: diff eq for lage m N even a}
\end{equation}
and
\begin{equation}
4m  \omega   b^{(2k+2)}
=
-4m(N-2)
\frac{d b^{(2k+2)}}{d m}
+4m^2\frac{d^{2} b^{(2k+2)}}{d m^{2}}
+\left[(N-1)^2-(2k+1)^2\right]
b^{(2k+2)}
\end{equation}
\end{subequations}
with $k=0,1,\cdots,(N/2)-1$.
By use of the substitution
\begin{subequations}
\label{eq: def x as fct m and omega if large m}
\begin{equation}
x^2 := 4 \omega m
\label{eq: def x as fct m and omega}
\end{equation}
whereby
\begin{equation}
\frac{x^{2}}{4\omega}\gg N,
\label{eq: large m condition on x}
\end{equation}
\end{subequations}
we find that the solutions
to Eq.~\re{eq: diff eq for lage m N even}
are given by the linear combinations
\begin{subequations}
\label{eq: sol large m N even}
\begin{equation}
b^{(2k+1)}(x)=
c^{(2k+1)}_{\textrm{e},1}
\left( \frac{x}{2} \right)^{N-1}
K^{\ }_{2k+1}(x)
+
c^{(2k+1)}_{\textrm{e},2}
\left(\frac{x}{2}\right)^{N-1}
I^{\ }_{2k+1}(x)
\end{equation}
and
\begin{equation}
b^{(2k+2)}(x)
=
c^{(2k+2)}_{\textrm{e},1}
\left( \frac{x}{2} \right)^{N-1}
K^{\ }_{2k+1}(x)
+
c^{(2k+2)}_{\textrm{e},2}
\left( \frac{x}{2} \right)^{N-1}
I^{\ }_{2k+1}(x)
\end{equation}
\end{subequations}
with $k=0,1,\cdots,(N/2)-1$,
of modified Bessel functions $K^{\ }_{2k+1}$ and $I^{\ }_{2k+1}$.
We shall demand that $b^{(i)}_{m}$
with $i=1,\cdots,N$ decay to zero as $m\to\infty$,
i.e., we must set
\begin{equation}
c^{(i)}_{\textrm{e},2}=0
\label{eq: c(i) e,2=0}
\end{equation}
with $i=1,\cdots,N$
in Eq.~\re{eq: sol large m N even} .
The remaining $N$
coefficients
$c^{(i)}_{\textrm{e},1}$
with $i=1,\cdots,N$
of the modified Bessel functions
$K^{\ }_{1},K^{\ }_{3},K^{\ }_{5}, \cdots,K^{\ }_{N-1}$
are fixed by matching
solutions \re{eq: sol large m N even}
to the solutions in the $m \ll \omega^{-1}$  region.
Thereto, we need to extract the terms that are of order
$x^{N+2k}$
and
$x^{N+2k-2}$
from the expansion of Eq.~\re{eq: sol large m N even} when $x$ small
(see chapter 9.6 in Ref.~\cite{Abramowitz65}).
This gives
\begin{subequations}
\label{eq: Laurent expansion N even}
\begin{equation}
\begin{split}
b^{(2k+1)} (x)
\sim &\,
c^{(2k+1)}_{\textrm{e},1}
\left[
 \frac{ (x/2)^{N+2k} }{2 (2k+1)!}
\left(
\gamma+\ln\frac{x^2}{4}-\Psi(2k+2)
\right)
+
\frac{(x/2)^{N+2k-2}}{2(2k)!}
\right]
\\
=&\,
c^{(2k+1)}_{\textrm{e},1}
\left[
\frac{(m\omega)^{\frac{N}{2}+k}}{2(2k+1)!}
\ln\left(C^{(\textrm{e})}_km\omega\right)
+
\frac{(m\omega)^{\frac{N}{2}+k-1}}{2(2k)!}
\right]
\end{split}
\label{eq: Laurent expansion N even a}
\end{equation}
and
\begin{equation}
\begin{split}
b^{(2k+2)}(x)
\sim&\,
c^{(2k+2)}_{\textrm{e},1}
\left[
\frac{(x/2)^{N+2k}}{2(2k+1)!}
\left(
\gamma+\ln\frac{x^2}{4}-\Psi(2k+2)
\right)
+
\frac{(x/2)^{N+2k-2}}{2(2k)!}
\right]
\\
=&\,
c^{(2k+2)}_{\textrm{e},1}
\left[
\frac{(m \omega)^{\frac{N}{2}+k}}{2(2k+1)!}
\ln\left(C^{(\textrm{e})}_km\omega\right)
+
\frac{(m\omega)^{\frac{N}{2}+k-1}}{2(2k)!}
\right]
\end{split}
\label{eq: Laurent expansion N even b}
\end{equation}
with $k=0,1,\cdots,(N/2)-1$ when $x\ll1$, and where
\begin{equation}
C^{(\textrm{e})}_k
:=
\exp \left(\gamma-\Psi(2k+2)\right).
\end{equation}
\end{subequations}

\subsubsection{
Solution when $1\leq m\ll\omega^{-1}$
              }

Second, we treat the limit $1\leq m\ll\omega^{-1}$,
in which case we can neglect the $\omega$ terms
in Eq.~\re{eq: rc rel for N even}.
In doing so, Eq.~\re{eq: rc rel for N even}
becomes
\begin{subequations}
\label{eq: rec rel for small m}
\begin{equation}
\begin{split}
0=&\,
\left\{-4m(2m-1+N)+\left[N-2-4k(k+1)\right]\right\}
b^{(2k+1)}_m
+ 2m(2m+1) b^{(2k+1)}_{m+1}
\\
&\,
+ (2m-2+N)(2m-1+N) b^{(2k+1)}_{m-1}
\end{split}
\label{eq: rec rel for small m 1}
\end{equation}
and
\begin{equation}
\begin{split}
0=&\,
\left\{-4m(2m-3+N)
+\left[3N-6-4k(k+1)\right]
\right\}
b^{(2k+2)}_{m}
+2m(2m-1)
b^{(2k+2)}_{m+1}
\\
&\,
+(2m-3+N)(2m-2+N) b^{(2k+2)}_{m-1}
\end{split}
\label{eq: rec rel for small m 2}
\end{equation}
with $k=0,1,\cdots,(N/2)-1$
and $m=2,3,\cdots,$
together with the initial conditions
\begin{equation}
\label{eq: initial cond N even small m 1}
0=
+ \sqrt{2} N(N+1)
- \left[ 3N +6 + 4k(k+1) \right] b^{(2k+1)}_1
+ 6 b_2^{(2k+1)}
\end{equation}
and
\begin{equation}
\label{eq: initial cond N even small m 2}
0=
- \sqrt{2}N(N-1)
- \left[N+2+4k(k+1)\right]
b^{(2k+2)}_1
+ 2
b_2^{(2k+2)}.
\end{equation}
\end{subequations}
The solution to the recursion relation
\re{eq: rec rel for small m}
can be expressed in terms of
generalized hypergeometric functions
\begin{subequations}
\label{eq: sol N even w=0}
\begin{equation}
\label{eq: sol N even w=0a}
\begin{split}
b_m^{(2k+1)}
=&\,
\mathcal{C}^{(\mathrm{e},1)}_k
\frac{ \left( m+k -1 + \frac{N}{2} \right)! }
     { \left( m-1 \right)! }
{}_3 F_2
\left[
\begin{array}{c}
 -k, \frac{1}{2} -k , 1-m \\
\frac{3}{2}, 1-k-m-\frac{N}{2}
\end{array} ; 1
\right]
\\
&\,
+
\mathcal{C}^{(\mathrm{e},2)}_k
\frac{ \left( m+k - \frac{1}{2} + \frac{N}{2} \right)! }
     { \left( m - \frac{1}{2} \right)!
  }
{}_3 F_2
\left[
\begin{array}{c}
 -k , -\frac{1}{2} -k ,  \frac{1}{2} -m \\
\frac{1}{2} , \frac{1}{2} -k -m- \frac{N}{2}
\end{array}; 1
\right]
\end{split}
\end{equation}
and
\begin{equation}
\begin{split}
b_m^{(2k+2)}
=&\,
\mathcal{C}^{(\mathrm{e},3)}_k
\frac{ \left( m+k-1  + \frac{N}{2} \right)!}
     { (m-1)! }
 {}_3 F_2
\left[
\begin{array}{c}
 -k,  -\frac{1}{2} -k, 1-m \\
 \frac{1}{2} , 1-k-m- \frac{N}{2}
\end{array}; 1
\right]
\\
&\,
+
\mathcal{C}^{(\mathrm{e},4)}_k
\frac{ \left(  m+k - \frac{3}{2} + \frac{N}{2} \right)!}
     { \left( m - \frac{3}{2} \right)! }
{}_3 F_2
\left[
\begin{array}{c}
 -k, \frac{1}{2} -k, \frac{3}{2} -m \\
\frac{3}{2}, \frac{3}{2} -k -m- \frac{N}{2}
\end{array} ; 1
\right]
\end{split}
\label{eq: sol N even w=0b}
\end{equation}
\end{subequations}
with $k=0,1,\cdots,(N/2)-1$
and $m=2,3,\cdots$.
The prefactors $\mathcal{C}^{(\mathrm{e},i)}_k$
$i=1,2,3,4$
to the hypergeometric function ${}_3 F_2$
are determined by the
initial conditions \re{eq: initial cond N even small m 1}
and \re{eq: initial cond N even small m 2}
through
\begin{subequations}
\label{eq: constants in sol N even w=0}
\begin{equation}
\begin{split}
\mathcal{C}^{(\mathrm{e},1)}_k
=&\,
\frac{1}{\left( k + \frac{N}{2} \right)! }
\left\{
b_1^{(2k+1)}
-
\mathcal{C}^{(\mathrm{e},2)}_k
\frac{2}{\sqrt{\pi}}
\left( k + \frac{N+1}{2}  \right)! \,
{}_3 F_2 \left[ \begin{array}{c}
-k, - \frac{1}{2} -k , -\frac{1}{2} \\
\frac{1}{2}, -\frac{1}{2} -k - \frac{N}{2}
\end{array}; 1
\right]
\right\},
\\
\mathcal{C}^{(\mathrm{e},2)}_k
=&\,
\sqrt{\pi} 2^{k+\frac{1}{2}}
\frac{( N-2k-2)!!}
{(N-2)!! \left(\frac{N-1}{2} \right)!}
,
\end{split}
\label{eq: constants in sol N even w=0a}
\end{equation}
and
\begin{equation}
\begin{split}
\mathcal{C}^{(\mathrm{e},3)}_k
=&\,
\frac{1}{ \left( k + \frac{N}{2} \right)!}
\left\{
b^{(2k+2)}_1
-
\mathcal{C}^{(\mathrm{e},4)}_k
\frac{1}{\sqrt{\pi}}
\left( k + \frac{N-1}{2} \right)! \,
{}_3 F_2 \left[ \begin{array}{c}
  -k , \frac{1}{2}-k,  \frac{1}{2}\\
\frac{3}{2}, \frac{1}{2} -k - \frac{N}{2}
\end{array}; 1 \right]
\right\},
\\
\mathcal{C}^{(\mathrm{e},4)}_k
=&\,
\sqrt{\pi}2^{k+\frac{3}{2}}
\frac{ (N-2k-2)!!}{ (N-2)!! \left( \frac{ N -3 }{2} \right)! } ,
\end{split}
\label{eq: constants in sol N even w=0b}
\end{equation}
\end{subequations}
with $k=0,1,\cdots,(N/2)-1$.
It is interesting to note that
the explicit $k$ dependences
on the right-hand sides of
Eqs.~\re{eq: sol N even w=0a}
and \re{eq: sol N even w=0b}
are related to those on the right-hand sides of
Eqs.~\re{solution N ODDb}
and \re{solution N ODDc}
by letting $k \to k-(1/2)$
and $k \to k+(1/2)$, respectively.

In order to match this solution
when $1\leq m\ll 1/\omega$
to the solution when $N\ll m$,
we need the leading and subleading terms of the large $m$
behavior of
$b^{(2k+1)}_m$ and $b^{(2k+2)}_m$.
For $b^{(1)}_m$ and $b^{(2)}_m$,
we have
\begin{subequations}
\label{eq: asym soll m small N even}
\begin{eqnarray}
b^{(1)}_m
&\simeq&
\left(
\mathcal{C}^{(\mathrm{e},1)}_0
+
\mathcal{C}^{(\mathrm{e},2)}_0
\right)
m^{\frac{N}{2}}
+
\frac{N}{8}
\left(
(N-2) \mathcal{C}^{(\mathrm{e},1)}_0
+
N \mathcal{C}^{(\mathrm{e},2)}_0
\right)
m^{\frac{N}{2}-1} ,
\label{eq: asym soll m small N even a}
\\
b^{(2)}_m
&\simeq&
\left(
\mathcal{C}^{(\mathrm{e},3)}_0
+
\mathcal{C}^{(\mathrm{e},4)}_0
\right)
m^{\frac{N}{2}}
+
\frac{N}{8}
\left(
(N-2) \mathcal{C}^{(\mathrm{e},3)}_0
+
(N-4)  \mathcal{C}^{(\mathrm{e},4)}_0
\right)
m^{\frac{N}{2}-1} ,
\label{eq: asym soll m small N even b}
\end{eqnarray}
to leading and subleading order in $m$,
respectively.
In general
we have for the odd-numbered coefficients
\begin{equation}
\begin{split}
b^{(2k+1)}_m
\simeq &\,
4^k
\left(
\frac{ \mathcal{C}^{(\mathrm{e},1)}_k }{2k+1}
+
\mathcal{C}^{(\mathrm{e},2)}_k
\right)
m^{\frac{N}{2}+k}
\\
&\,
+
4^{k-\frac{3}{2}}(N+2k)(N-1)
\left(
\frac{ \mathcal{C}^{(\mathrm{e},1)}_k }{2k+1}
+
\mathcal{C}^{(\mathrm{e},2)}_k
\right)
m^{\frac{N}{2}+k-1}
\end{split}
\label{eq: asym soll m arbitrary N odd c}
\end{equation}
and for the even-numbered coefficients
\begin{equation}
\begin{split}
b^{(2k+2)}_m
\simeq &\,
4^k
\left(
\mathcal{C}^{(\mathrm{e},3)}_k
+
\frac{ \mathcal{C}^{(\mathrm{e},4)}_k}{2k+1}
\right)
m^{\frac{N}{2} +k }
\\
&\,
+
4^{k-\frac{3}{2}}(N+2k)(N-3)
\left(
\mathcal{C}^{(\mathrm{e},3)}_k
+
\frac{ \mathcal{C}^{(\mathrm{e},4)}_k}{2k+1}
\right)
m^{\frac{N}{2} +k -1}
\end{split}
\label{eq: asym soll m arbitrary N even d}
\end{equation}
\end{subequations}
with $k=1,\cdots,(N/2)-1$,
to leading and subleading order in $m$.
We emphasize that these two formula are only valid for $k \geq 1$.
Observe that the large $m$ subleading behavior
of $b^{(1)}_m$ and  $b^{(2)}_m$
differ from those of
$b^{(2k+1)}_m$ and  $b^{(2k+2)}_m$ with $k\geq1$, respectively.
We note again that
the explicit $k$ dependences
on the right-hand sides of Eqs.%
~\re{eq: asym soll m arbitrary N odd c}
and
\re{eq: asym soll m arbitrary N even d}
are related to those on the right-hand sides of Eqs.%
~\re{asymptotics N ODDb}
and
\re{asymptotics N ODDc}
by letting
$k\to k-(1/2)$
and
$k\to k+(1/2)$,
respectively.

\subsubsection{
Matching solutions when $N \ll m\ll\omega^{-1}$
              }

Having solved the recursion relations%
~\re{eq: rc rel for N even} in the
two limits $N \ll m$ and $1\ll m\ll\omega^{-1}$,
we are going to match
the $m=x^{2}/(4\omega)$ dependences of the solutions
\re{eq: Laurent expansion N even}
and
\re{eq: asym soll m small N even}
in the overlapping
region $N \ll m \ll  \omega^{-1}$,
where both solutions are valid.
We start with the coefficients
$b^{(1)}_m$ and $b^{(2)}_m$.
Matching Eq.~\re{eq: Laurent expansion N even a}
when $k=0$
with Eq.~\re{eq: asym soll m small N even a}
gives two equations in the two unknowns
$b^{(1)}_1$
and
$c^{(1)}_{\textrm{e},1}$
(whereby we neglect
$\ln m$
and
$\ln C^{(\textrm{e})}_k$
compared to
$\ln \omega$)
\begin{subequations}
\begin{equation}
\begin{split}
c^{(1)}_{\textrm{e},1}
\frac{1}{2} \omega^{\frac{N}{2}}
\ln \omega
=& \,
\mathcal{C}^{(\textrm{e},1)}_0
+
\mathcal{C}^{(\textrm{e},2)}_0,
\\
c^{(1)}_{\textrm{e},1}
\frac{1}{2}
\omega^{\frac{N}{2}-1}
=& \,
\frac{N}{8}
\left[
(N-2) \mathcal{C}^{(\textrm{e},1)}_0
+
N \mathcal{C}^{(\textrm{e,2})}_0
\right] .
\end{split}
\end{equation}
Matching Eq.~\re{eq: Laurent expansion N even b}
when $k=0$
with Eq.~\re{eq: asym soll m small N even b}
gives two equations in the two unknowns
$b^{(2)}_1$
and
$c^{(2)}_{\textrm{e},1}$,
\begin{equation}
\begin{split}
c^{(2)}_{\textrm{e},1}
\frac{1}{2}
\omega^{\frac{N}{2} } \ln \omega
=& \,
\mathcal{C}^{(\textrm{e},3)}_0
+
\mathcal{C}^{(\textrm{e},4)}_0,
\\
c^{(2)}_{\textrm{e},1}
\frac{1}{2} \omega^{\frac{N}{2}-1}
=&\,
\frac{N}{8}
\left[
(N-2)\mathcal{C}^{(\textrm{e},3)}_0
+
(N-4)\mathcal{C}^{(\textrm{e},4)}_0
\right].
\end{split}
\end{equation}
\end{subequations}
Solving for
$b^{(1)}_1$,
$b^{(2)}_1$,
$c^{(1)}_{\textrm{e},1}$,
and
$c^{(2)}_{\textrm{e},1}$,
we determine their
$\omega$ dependences,
\begin{subequations}
\label{eq: leading omega dependence of initial conditions k leq 2}
\begin{equation}
\label{eq: leading omega dependence of initial conditions k leq 2 a}
\begin{split}
c^{(1)}_{\textrm{e},1}
=& \,
\frac{
+ 4 \mathcal{C}^{(\textrm{e},2)}_0 N \omega^{1- \frac{N}{2} }
}
{ 8 - N(N-2) \omega \ln \omega },
\\
c^{(2)}_{\textrm{e},1}
=& \,
\frac{
-4 \mathcal{C}^{(\textrm{e},4)}_0 N \omega^{1-\frac{N}{2} }
     }
     {
8 - N (N-2) \omega \ln \omega
     },
\end{split}
\end{equation}
and
\begin{equation}
\label{eq: leading omega dependence of initial conditions k leq 2 b}
\begin{split}
b^{(1)}_1
=& \,
\sqrt{2} (N+1)
-
\frac{
\mathcal{C}^{(\textrm{e},2)}_0
\left( \frac{N}{2} \right)!
\left[ 8 - N^2 \omega \ln \omega \right]
     }
     {
8 - N (N-2) \omega \ln \omega
     },
\\
b^{(2)}_1
=& \,
\sqrt{2} (N-1)
-
\frac{
\mathcal{C}^{(\textrm{e},4)}_0
\left( \frac{N}{2} \right)!
\left[ 8 - N(N-4) \omega \ln \omega \right]
}
{8 - N (N-2) \omega \ln \omega} ,
\end{split}
\end{equation}
\end{subequations}
respectively.
For $k \geq 1$,
matching Eq.~\re{eq: Laurent expansion N even a}
with Eq.~(\ref{eq: asym soll m arbitrary N odd c})
gives two equations for the two unknowns
$b^{(2k+1)}_1$
and
$c^{(2k+1)}_{\textrm{e},1}$
(whereby we neglect
$\ln m$
and
$\ln C^{(\textrm{e})}_k$
compared to
$\ln \omega$)
\begin{subequations}
\label{eq: matching conditions N even}
\begin{equation}
\begin{split}
c^{(2k+1)}_{\textrm{e},1}
\frac{ \omega^{\frac{N}{2}+k}}
{2(2k+1)!}\ln\omega
=&\,
4^k
\left(
\frac{\mathcal{C}^{(\textrm{e},1)}_k }{2k+1}
+
\mathcal{C}^{(\textrm{e},2)}_k
\right),
\\
c^{(2k+1)}_{\textrm{e},1}
\frac{ \omega^{\frac{N}{2}+k-1}}
{2(2k)!}
=&\,
4^{k-\frac{3}{2}}
(N+2k)(N-1)
\left(
\frac{\mathcal{C}^{(\textrm{e},1)}_k }{2k+1}
+
\mathcal{C}^{(\textrm{e},2)}_k
\right),
\end{split}
\end{equation}
while matching Eq.~\re{eq: Laurent expansion N even b}
with Eq.~(\ref{eq: asym soll m arbitrary N even d})
gives two equations for the two unknowns
$b^{(2k+2)}_1$
and
$c^{(2k+2)}_{\textrm{e},1}$
\begin{equation}
\begin{split}
c^{(2k+2)}_{\textrm{e},1}
\frac{ \omega^{\frac{N}{2}+k}}{2(2k+1)!}
\ln\omega
=&\,
4^k
\left(
\mathcal{C}^{(\textrm{e},3)}_k
+
\frac{\mathcal{C}^{(\textrm{e},4)}_k }{2k+1}
\right),
\\
c^{(2k+2)}_{\textrm{e},1}
\frac{\omega^{\frac{N}{2}+k-1}}{2(2k)!}
=&\,
4^{k-\frac{3}{2}}
(N+2k)(N-3)
\left(\mathcal{C}^{(\textrm{e},3)}_k
+
\frac{\mathcal{C}^{(\textrm{e},4)}_k }{2k+1}
\right),
\end{split}
\end{equation}
\end{subequations}
with $k=1,2,\cdots,(N/2)-1$.
Solving for
$b^{(i)}_1$
and
$c^{(i)}_{\textrm{e},1}$,
we find
\begin{subequations}
\label{eq: leading omega dependence of initial conditions k geq 2}
\begin{equation}
\begin{split}
c^{(2k+1)}_{\textrm{e},1}
=&\,0,
\qquad
c^{(2k+2)}_{\textrm{e},1}
=\,0,
\end{split}
\label{eq: leading omega dependence of initial conditions k geq 2 a}
\end{equation}
\begin{equation}
\begin{split}
b^{(2k+1)}_1
=&\,
\mathcal{C}^{(\textrm{e},2)}_k\!
\left\{\!
-(2k+1)\left(\frac{N}{2}+k\right)!
\!+\!
\frac{2}{\sqrt{\pi}}\left(\!\frac{N+1}{2}\!+\!k\!\right)!
{}_3 F_2
\left[
\begin{array}{c}
-k ,  - \frac{1}{2} -k,  -\frac{1}{2}\\
\frac{1}{2},   - \frac{1}{2} -k - \frac{N}{2}
\end{array}; 1
\right]
\right\},
\label{eq: leading omega dependence of initial conditions k geq 2 c}
\end{split}
\end{equation}
and
\begin{equation}
\begin{split}
b^{(2k+2)}_1
=& \,
\mathcal{C}^{(\textrm{e},4)}_k
\left\{
\frac{-1}{2k+1}
\left(\frac{N}{2}+k\right)!
+
\frac{1}{\sqrt{\pi}}
\left(
\frac{N-1}{2}+k
\right)!
{}_3 F_2
\left[
\begin{array}{c}
-k ,\frac{1}{2}-k,\frac{1}{2}\\
\frac{3}{2},\frac{1}{2}-k-\frac{N}{2}
\end{array};1
\right]
\right\} ,
\label{eq: leading omega dependence of initial conditions k geq 2 d}
\end{split}
\end{equation}
\end{subequations}
for $k=1,2,\cdots,(N/2)-1$.
We note that  the explicit $k$ dependences
on the right-hand sides of
Eq.~\re{eq: leading omega dependence of initial conditions k geq 2 c}
and
Eq.~\re{eq: leading omega dependence of initial conditions k geq 2 d}
are related to those on the right-hand sides of
Eq.~\re{eq: N odd 2k+1 b}
and
Eq.~\re{eq: N odd 2k b}
by letting
$k\to k-(1/2)$
and
$k\to k+(1/2)$, respectively.
Observe that
the coefficients%
~(\ref{eq: leading omega dependence of initial conditions k geq 2 a})
of the modified Bessel functions
$K^{\ }_{1},K^{\ }_{3},\cdots,K^{\ }_{N-1}$
are trivially independent of $\omega$,
while the initial values%
~\re{eq: leading omega dependence of initial conditions k geq 2 c}
and~\re{eq: leading omega dependence of initial conditions k geq 2 d}
of the $\omega=0$ recursion relation
are independent of $\omega$
in view of
Eqs.%
~(\ref{eq: constants in sol N even w=0a})
and~(\ref{eq: constants in sol N even w=0b}),
respectively.

\subsubsection{
$\omega$ dependence of
the recursion relation~(\ref{eq: rc rel for N even})
              }

Equipped with the solutions
~(\ref{eq: leading omega dependence of initial conditions k leq 2})
and~(\ref{eq: leading omega dependence of initial conditions k geq 2})
for the $\omega$ dependence of
$b^{(i)}_{1}$
and
$c^{(i)}_{\textrm{e},1}$
with $i=1,2,\cdots, N$,
we can derive the
$\omega$ dependence
of the coefficients
$b^{(i)}_m$
with $m=2,3,\cdots$
in the limit $1\leq m\ll\omega^{-1}$
from Eq.~(\ref{eq: sol N even w=0}).
Insertion of Eq.%
\re{eq: leading omega dependence of initial conditions k leq 2 b}
into the prefactors%
~(\ref{eq: constants in sol N even w=0})
appearing in Eq.~(\ref{eq: sol N even w=0})
when $k=0$
gives
\begin{subequations}
\label{w dependence of b1 and b2 N even}
\begin{eqnarray}
b^{(1)}_m
&=&
\mathcal{C}^{(e,2)}_0
\frac{ \left( m - \frac{1}{2} + \frac{N}{2} \right)! }
     { \left( m - \frac{1}{2} \right)! }
-
\mathcal{C}^{(e,2)}_0
\frac{ \left( m -1 + \frac{N}{2} \right)! }
     { 4 (m-1)! }
\left(
4 - N \omega \ln \omega
\right),
\label{bmEins final}
\\
b^{(2)}_m
&=&
\mathcal{C}^{(e,4)}_0
\frac{ \left( m - \frac{3}{2} + \frac{N}{2} \right)! }
     { \left( m - \frac{3}{2} \right)! }
-
\mathcal{C}^{(e,4)}_0
\frac{ \left( m -1+ \frac{N}{2} \right)!}
     {4 \left( m-1 \right)! }
\left( 4 + N \omega \ln \omega \right) ,
\label{bmZwei final}
\end{eqnarray}
where we have disregarded
terms of order $\omega^2$ and higher.
Insertion of Eq.%
~(\ref{eq: leading omega dependence of initial conditions k geq 2 c})
into the prefactor%
~(\ref{eq: constants in sol N even w=0a})
appearing in Eq.~(\ref{eq: sol N even w=0a})
gives
\label{bm final}
\begin{equation}
\label{bmOdd final}
\begin{split}
b^{(2k+1)}_m
=&\,
\mathcal{C}^{(\textrm{e},2)}_k
\frac{ \left( m+k- \frac{1}{2} + \frac{N}{2} \right)! }
     { \left( m - \frac{1}{2} \right)! }
{}_3 F_2 \left[ \begin{array}{c}
 -k, -\frac{1}{2} -k,  \frac{1}{2} -m \\
\frac{1}{2} , \frac{1}{2} -k-m-\frac{N}{2}
\end{array}; 1
\right]
\\
&\,
-
\mathcal{C}^{(\textrm{e},2)}_k
(2k+1)
\frac{ \left( m+k-1 + \frac{N}{2} \right)! }
     { (m-1)! }
{}_3 F_2
\left[
\begin{array}{c}
 -k,  \frac{1}{2} -k,  1-m \\
\frac{3}{2}, 1 -k -m - \frac{N}{2}
\end{array}; 1
\right]
\end{split}
\end{equation}
with $k=1,2,\cdots,(N/2)-1$.
Insertion of Eq.%
~(\ref{eq: leading omega dependence of initial conditions k geq 2 d})
into the prefactor%
~(\ref{eq: constants in sol N even w=0b})
appearing in Eq.~(\ref{eq: sol N even w=0b})
gives
\begin{equation}
\begin{split}
\label{bmEven final}
b^{(2k+2)}_m
=& \,
\mathcal{C}^{(\textrm{e},4)}_k
\frac{ \left( m+k -\frac{3}{2} + \frac{N}{2} \right)! }
     { \left( m - \frac{3}{2} \right)! }
{}_3 F_2
\left[
\begin{array}{c}
  -k,  \frac{1}{2} -k, \frac{3}{2}-m \\
\frac{3}{2}, \frac{3}{2} -k -m - \frac{N}{2}
\end{array}; 1
\right]
\\
&\,
-
\mathcal{C}^{(\textrm{e},4)}_k
\frac{1}{2k+1}
\frac{ \left( m +k -1 + \frac{N}{2} \right)! }
     {  (m-1)!}
{}_3 F_2
\left[
\begin{array}{c}
 -k, - \frac{1}{2} -k,  1-m \\
\frac{1}{2}, 1 -k -m - \frac{N}{2}
\end{array}; 1
\right]
\end{split}
\end{equation}
\end{subequations}
with $k=1,2,\cdots,(N/2)-1$.
We emphasize again that
the explicit $k$ dependences on the right-hand sides
of Eq.~\re{bmOdd final} and Eq.~\re{bmEven final} are related to those
on the right-hand sides of Eq.~\re{eq: rec rel sol N odd small m b}
and Eq.~\re{eq: rec rel sol N odd small m c}
by letting $k \to k -(1/2)$ and $k \to k + (1/2)$, respectively.

In conclusion, the $\omega$ dependence of
the solution to the recursion relation%
~(\ref{eq: rc rel for N even})
follows from combining
Eqs.~\re{eq: sol large m N even},
(\ref{eq: c(i) e,2=0}),
(\ref{eq: leading omega dependence of initial conditions k leq 2}),
(\ref{eq: leading omega dependence of initial conditions k geq 2 a})
when $m\gg N$ together with
Eq.~(\ref{bm final})
when $m\omega\ll1$.
This gives
\begin{equation}
\begin{split}
b^{(1)}_m
=&\,
\left\{
\begin{array}{l l}
+4\mathcal{C}^{( \textrm{e},2)}_0
\frac{
N\sqrt{\omega}
     }
     {
8-N(N-2)\omega\ln\omega
     }
m^{\frac{N-1}{2}}
K^{\ }_1
\left(
2\sqrt{m\omega}
\right) ,
&
\quad
N\ll m,
\\
\mathcal{C}^{(e,2)}_0
\left(
\frac{\left(m-\frac{1}{2}+\frac{N}{2}\right)!}
     {\left(m-\frac{1}{2}\right)!}
-
\frac{\left(m-1+\frac{N}{2}\right)!}
     {4(m-1)!}
\left(
4-N\omega\ln\omega
\right)
\right),
&
\quad
m\ll\omega^{-1},
\end{array}
\right.
\\
b^{(2)}_m
=&\,
\left\{
\begin{array}{l l}
-  4 \mathcal{C}^{(e,4)}_0 \frac{
 N  \sqrt{\omega}
     }
     {
8 - N(N-2) \omega \ln \omega
     }
m^{\frac{N-1}{2}}
K^{\ }_1
\left(
2 \sqrt{ m \omega }
\right) ,
&
\quad
N\ll m,
\\
\mathcal{C}^{(e,4)}_0
\left(
\frac{ \left( m - \frac{3}{2} + \frac{N}{2} \right)! }
     { \left( m - \frac{3}{2} \right)! }
-
\frac{ \left( m -1+ \frac{N}{2} \right)!}
     {4 \left( m-1 \right)! }
\left( 4 + N \omega \ln \omega \right)
\right),
&
\quad
m\ll\omega^{-1},
\end{array}
\right.
\\
b^{(2k+1)}_m
=&\,
\left\{
\begin{array}{l l}
0,
&
\quad
N\ll m,
\\
\re{bmOdd final},
&
\quad
m\ll\omega^{-1},
\end{array}
\right.
\\
b^{(2k+2)}_m
=&\,
\left\{
\begin{array}{l l}
0,
&
\quad
N\ll m,
\\
\re{bmEven final},
&
\quad
m\ll\omega^{-1},
\end{array}
\right.
\end{split}
\label{eq: limiting solutions to rec rel N even}
\end{equation}
with $k=1,2,\cdots,(N/2)-1$.
Equation~(\ref{eq: limiting solutions to rec rel N even})
should be compared to Eq.~(\ref{eq: asymptotic solutions n=2}).
Observe that
Eqs.~(\ref{bmOdd final})
and~(\ref{bmEven final})
are independent of $\omega$.

\subsubsection{
Leading energy dependence of the density of states
              }
We are ready to extract the leading behavior
of the average density of
states $\nu (\varepsilon)$,
Eq.~(\ref{eq: dos in ground state}),
for asymptotically small energies $\varepsilon$.
The density of states in the long wire limit Eq.~\re{eq: QM DOS} is given by
\begin{equation}
\nu ( \varepsilon )
=
\lim_{ \omega \to - i \varepsilon  }
\pi^{-1} \mathrm{Re}\
{\vphantom{A}}^{\ }_{\mathrm{L}}
\left\langle \varphi^{\ }_{0} \right|
\left(\widebar{B} - B\right)
\left| \varphi^{\ }_{0} \right\rangle^{\ }_{\mathrm{R}},
\end{equation}
with  the left and right
ground states%
~(\ref{eq: expansion of eigenstatesa})
and~(\ref{eq: expansion of eigenstatesb}),
respectively.
The normalizations%
~(\ref{eq: norm of basis states})
and%
~(\ref{eq: normalization a(0)}),
the biorthogonal relations%
~(\ref{eq: zero overlap}),
and the identities
\begin{equation}
\begin{split}
&
\left( \widebar{B} - B \right) \left| 0 \right\rangle
=
N \left| 0 \right\rangle,
\\
&
\left( \widebar{B} - B \right) \left| m \right\rangle^{(2n+1)}_{\mathrm{R}}
=
( N - 4n -1) \left| m \right\rangle^{(2n+1)}_{\mathrm{R}}
+
\left| m \right\rangle^{(2n+1)}_{\mathrm{L}} ,
\\
&
\left( \widebar{B} - B \right) \left| m \right\rangle^{(2n+2)}_{\mathrm{R}}
=
(N-4n -3)  \left| m \right\rangle^{(2n+2)}_{\mathrm{R}}
-  \left| m \right\rangle^{(2n+2)}_{\mathrm{L}},
\end{split}
\label{eq: GS bar B - B expansion}
\end{equation}
deliver the expectation value
\begin{equation}
\label{sum over coeff N EVEN}
\begin{split}
{\vphantom{A}}^{\ }_{\mathrm{L}}
\left\langle \varphi^{\ }_{0} \right|
\left(\widebar{B} - B\right)
\left| \varphi^{\ }_{0} \right\rangle^{\ }_{\mathrm{R}}
=&\,
N
+
\sum_{m=1}^{\infty}
\sum_{n=0}^{N/2-1 }
\mathcal{N}^{(1)}_{m,n}
\left( a^{(2n+1)}_{m}\right )^2
-
\sum_{m=1}^{\infty}
\sum_{n=0}^{N/2-1 }
\mathcal{N}^{(2)}_{m,n}
\left( a^{(2n+2)}_{m}\right )^2
\\
=&\,
N
+
\sum\limits_{m=1}^{\infty}
\sum\limits_{n=0}^{N/2-1}
\mathcal{N}^{(1)}_{m,n}
\left(
\sum\limits_{k=0}^{N/2-1}
\left[\mathcal{M}^{-1}_{\mathrm{e};N,1}\right]^{n}_{k}
b^{(2k+1)}_{m}
\right)^2
\\
&\,
-
\sum\limits_{m=1}^{\infty}
\sum\limits_{n=0}^{N/2-1}
\mathcal{N}^{(2)}_{m,n}
\left(
\sum\limits_{k=0}^{N/2-1}
\left[\mathcal{M}^{-1}_{\mathrm{e};N,2}\right]^{n}_{k}
b^{(2k+2)}_{m}
\right)^2.
\end{split}
\end{equation}

The matrices
$\mathcal{M}^{\ }_{\mathrm{e};N,1}$
and
$\mathcal{M}^{\ }_{\mathrm{e};N,2}$
and their inverses are given in Eq.~\re{eq: transf N even}
while
the normalization factors $\mathcal{N}^{(1)}_{m,n}$
and
$\mathcal{N}^{(2)}_{m,n}$
are given by Eq.~(\ref{eq: norm of basis states}).
In order to evaluate Eq.~\re{sum over coeff N EVEN}
it is necessary to break the sum over $m$
into two parts separated by the integer $1 \ll m^{\ }_{0} \sim \omega^{-1}$.
This gives
\begin{subequations}
\label{fineal exp value N gerade}
\begin{equation}
{\vphantom{A}}^{\ }_{\mathrm{L}}
\left\langle \varphi^{\ }_{0} \right|
\left(\widebar{B} - B\right)
\left| \varphi^{\ }_{0} \right\rangle^{\ }_{\mathrm{R}}
=
N
+
S_{1}
+
S_{2}
\end{equation}
with
\begin{eqnarray}
\label{sum S1 N even}
S^{\ }_{1}
&:=&
\sum_{m=1}^{m^{\ }_{0}}
\sum_{n=0}^{N/2-1 }
\mathcal{N}^{(1)}_{m,n}
\left( a^{(2n+1)}_{m}\right)^2
-
\sum_{m=1}^{m^{\ }_{0}}
\sum_{n=0}^{N/2-1 }
\mathcal{N}^{(2)}_{m,n}
\left( a^{(2n+2)}_{m}\right)^2,
\\
\label{sum S2 N even}
S^{\ }_{2}
&:=&
\sum\limits_{m=m^{\ }_{0}+1}^{\infty}
\sum\limits_{n=0}^{N/2-1}
\left[
\mathcal{N}^{(1)}_{m,n}
\left(
\left[\mathcal{M}^{-1}_{\mathrm{e};N,1}\right]^{n}_{0}
 b^{(1)}_{m}
 \right)^2
-
\mathcal{N}^{(2)}_{m,n}
\left(
\left[\mathcal{M}^{-1}_{\mathrm{e};N,2}\right]^{n}_{0}
  b^{(2)}_{m}
  \right)^2
\right].
\end{eqnarray}
\end{subequations}
Here, we made use of the fact that in the large $m$ limit
the coefficients $b^{(2k+1)}_m$ and $b^{(2k+2)}_m$ all vanish
except for $b^{(1)}_m$ and $b^{(2)}_m$.
We are going to
compute the sum $S^{\ }_{1}$ for $m \leq m^{\ }_{0}$ and
the sum $S^{\ }_{2}$ for $m > m^{\ }_{0}$ separately.

To compute the sum $m \leq m^{\ }_{0}$, $S^{\ }_{1}$,
we first need to derive the leading $\omega$ dependence of the
coefficients $a^{(2n+1)}_m$ and $a^{(2n+2)}_m$.
Inserting Eq.~\re{bmEins final}
and \re{bmOdd final}
into Eq.~\re{eq: transf N even a bis} yields
\begin{equation}
\begin{split}
a^{(2n+1)}_m
=& \,
+ \frac{ (-1)^n \sqrt{2}    \left(  \frac{N}{2} + m -1 \right)! }
{ (m+n)  (m-1)!  \left( \frac{N}{2} - n -1 \right)!  n! }
\left[
1 + ( m + n) \omega \ln \omega
\right] ,
\end{split}
\end{equation}
where $n=0,1,\cdots, (N/2)-1$.
Inserting Eq.~\re{bmZwei final} and \re{bmEven final}
into Eq.~\re{eq: transf N even a bis} gives
\begin{equation}
\begin{split}
&
a^{(2n+2)}_m
=
- \frac{ (-1)^n \sqrt{2} \left( \frac{N}{2} + m -1 \right)! }
{ (m+n) (m-1)! \left( \frac{N}{2} -n -1 \right)!  n! }
\left[
1 +(m+n) \omega \ln \omega
\right] ,
\end{split}
\end{equation}
where $n=0,1,\cdots, (N/2)-1$.
With this, the normalization factors%
~\re{eq: norm of basis states},
and the choice $m^{\ }_{0} = \omega^{-1}$,
we find
\begin{equation}
\label{eq: small m sum}
\begin{split}
S^{\ }_{1}
=&\,
\sum_{m=1}^{1/\omega}
\sum_{n=0}^{N/2-1 }
\left(
\mathcal{N}^{(1)}_{m,n}
\left(a^{(2n+1)}_{m}\right)^2
-
\mathcal{N}^{(2)}_{m,n}
\left(  a^{(2n+2)}_{m}\right)^2
\right)
\\
\approx &\,
\frac{2}{\pi}
\sum\limits_{m=1}^{1/ \omega }
\frac{ \left( \frac{N}{2} + m -1 \right)!}{(m-1)!}
\sum\limits_{n=0}^{N/2-1}
\frac{ \Big( 1 + (m+n) \omega \ln \omega \Big)^2 }
{ (m+n) \left( \frac{N}{2}-n-1 \right)! n! }
\\
&\, \;
\times
\left(
\frac{
\left( m -\frac{1}{2} \right)!
\left( \frac{N}{2} -n -\frac{1}{2} \right)!
\left( n - \frac{1}{2} \right)! }
     {
\left( \frac{N}{2} +m -\frac{1}{2} \right)!
     }
-
\frac{
\left( m -\frac{3}{2} \right)!
\left( \frac{N}{2} -n -\frac{3}{2} \right)!
\left( n + \frac{1}{2} \right)!
     }
     {
\left( \frac{N}{2} +m -\frac{3}{2} \right)!
     }
\right),
\end{split}
\end{equation}
which is to be compared with the telescopic expansion%
~(\ref{eq: telescopic expansion N=2}).
To compute the leading behavior of the density of states it is
sufficient to retain only the lowest order in $\omega$.
This gives
\begin{equation}
S^{\ }_{1}
\approx
-
N
-
\textrm{const} \times
\omega \ln^2 \omega,
\label{eq: final S1}
\end{equation}
to leading and subleading orders in $\omega$.

For the sum over $m>m^{\ }_{0}$, $S^{\ }_{2}$,
we obtain by inserting the definition of
$\mathcal{M}^{-1}_{\mathrm{e};N,1}$,
$\mathcal{M}^{-1}_{\mathrm{e};N,2}$,
$\mathcal{N}^{(1)}_{m,n}$, and
$\mathcal{N}^{(2)}_{m,n}$,
as well as the solutions for $b^{(1)}_{m}$ and
$b^{(2)}_{m}$ from Eq.%
~\re{eq: limiting solutions to rec rel N even}
\begin{subequations}
\label{S2 evaluated}
\begin{equation}
\begin{split}
S^{\ }_{2}
=& \,
\sum_{m=m^{\ }_{0}+1}^{\infty}
\mathcal{S}^{(\textrm{e})}_{N} ( m)
\frac{ (4N)^2 \omega m^{N-1} K^2_1 ( 2 \sqrt{m \omega} ) }
{ \left[ 8 - N(N-2) \omega \ln \omega \right]^2 }
\end{split}
\end{equation}
with the combinatorial factor
\begin{equation}
\begin{split}
\mathcal{S}^{(\textrm{e})}_{N} ( m)
:=& \,
\sum_{n=0}^{N/2-1}
\left[
\mathcal{N}^{(1)}_{m,n}
\left(
\mathcal{C}^{(e,2)}_0
\left[\mathcal{M}^{-1}_{e,N,1}\right]_0^n
\right)^2
-
\mathcal{N}^{(2)}_{m,n}
\left(
\mathcal{C}^{(e,4)}_0
\left[\mathcal{M}^{-1}_{e,N,2}\right]_0^n
\right)^2
\right].
\end{split}
\end{equation}
\end{subequations}
After converting the sum in $S^{\ }_{2}$
into an integral,
we find
\begin{equation}
S^{\ }_{2}
=
\int\limits_{m^{\ }_{0}+1}^{\infty} dm \,
\mathcal{S}^{(\textrm{e})}_{N} ( m)
\frac{ (4N)^2 \omega m^{N-1} K^2_1 ( 2 \sqrt{m \omega} ) }
{ \left[ 8 - N(N-2) \omega \ln \omega \right]^2 } .
\end{equation}
Using the substitution $x^2=4 m \omega$ the integral transforms into
\begin{equation}
\begin{split}
S^{\ }_{2}
=& \,
\frac{ (4N)^2   \omega^{1-N}  }
{ \left[ 8 - N(N-2) \omega \ln \omega \right]^2 }
\int\limits_{x_0}^{\infty}
d x \,
\mathcal{S}^{(\textrm{e})}_{N} \left( \frac{x^2}{4 \omega}
\right) \left( \frac{x}{2} \right)^{2N-1}   K^2_1 ( x ) ,
\end{split}
\end{equation}
with $x_0 = 2 \sqrt{( m^{\ }_{0} +1) \omega} \sim \mathcal{O} (1)$,
since we chose $m^{\ }_{0} = \omega^{-1}$.
Let us now expand
$\mathcal{S}^{(\textrm{e})}_{N}$
in small $\omega$ (i.e., large argument).
This gives
\begin{subequations}
\begin{equation}
\mathcal{S}^{(\textrm{e})}_{N}
\left(\frac{x^2}{4\omega}\right)
\approx
\mathcal{S}^{(\textrm{e},1)}_{N}\,
\times
\left( \frac{2 \omega}{x^2} \right)^{N-1}
+
\mathcal{O}
\left[
\left( \frac{ 2 \omega}{x^2} \right)^N
\right],
\end{equation}
where the prefactor to the $\omega^{N-1}$ term,
\begin{equation}
\begin{split}
\mathcal{S}^{(\textrm{e},1)}_{N}
:=& \,
\frac{ 8 }
     { \left( \frac{N}{2} \right)!^2 }
\sum_{n=0}^{N/2-1}
\Bigg\{
\frac{
\left[(N-1)! \right]^2
\left(
{}_2 F_1
\left[
-\frac{N}{2}, -n , \frac{1}{2} - \frac{N}{2} , 1
\right]
\right)^2
     }
     {
  \left( N -2n -1 \right)! (2n)!
     }
\\
&\,
\qquad \qquad \qquad
-
\frac{
\left[ (N-2)! \right]^2
\left(
{}_2 F_1
\left[
- \frac{N}{2}, -n , \frac{3}{2} - \frac{N}{2} ,1
\right]
\right)^2
     }
     {
  \left(N  -2n -2 \right)! ( 2n +1)!
     }
\Bigg\}
\\
=&\,0,
\end{split}
\end{equation}
\end{subequations}
is in fact vanishing.
Here,
we have introduced the Gauss hypergeometric function ${}_2 F_1$.
Hence, we conclude that the leading term of
$S^{\ }_{2}$ is (at most) of order $\omega$,
and can therefore be neglected compared to $S^{\ }_{1}$,
whose subleading $\omega$ dependence
is of order $\omega \ln \omega$.
The above reasoning also justifies the neglect of the
$m>m^{\ }_{0}$
sum in Eq.%
~(\ref{eq: dropping S2 N=2}).

Combining Eq.~\re{fineal exp value N gerade}
and \re{eq: final S1},
we find that the density of states, for asymptotically small energies,
is given by
\begin{equation}
\begin{split}
\nu(\varepsilon)
=&\,
\lim_{ \omega \to -i \varepsilon  }
\pi^{-1} \mathrm{Re}
\left( N + S^{\ }_{1} + S^{\ }_{2} \right)
\\
\propto&\,
\lim_{ \omega \to -i \varepsilon  }
\pi^{-1} \mathrm{Re}
\left( - \omega \ln^2 \omega \right)
\\
\propto&\
\varepsilon \left|  \ln  \varepsilon \right| .
\end{split}
\end{equation}
Recalling that $\varepsilon$ is measured in units
of the disorder strength, we have recovered
Eq.~(\ref{eq: main result paper})
when the number of channels $N$ is even.

\section{
Discussion
        }
\label{sec: Discussion}

The asymptotic limit $0<\varepsilon\ll1$ for the density of states%
~(\ref{eq: main result paper})
of a disordered quasi-one-dimensional quantum wire
in the chiral-unitary symmetry class
was first derived
in Refs.\ \cite{Brouwer00} and \cite{Titov01}
using the DMPK approach.
This approach is based on finding the stationary
solution of a forced diffusive process obeyed
by the Lyapunov exponents of a transfer matrix,
whereby the length $L$ of the quasi-one-dimensional
quantum wire plays the role of time
and the forcing term is proportional to
the frequency $\omega$ related to the energy
$\varepsilon$ by the analytical continuation
$\omega\to-\mathrm{i}\varepsilon$.
The DMPK approach is geometric, for
the Lyapunov exponents are none
but the radial coordinates of a smooth manifold determined by
the symmetry class to which the disorder belongs.
This symmetry requirement,
the quasi-one-dimensionality,
and the implicit weakness of the disorder
determines in a unique way the forced diffusive process
in the thick quantum wire limit $N\to\infty$
\cite{Brouwer00,Brouwer00b}.
In this paper, we have derived the density of states%
~(\ref{eq: main result paper})
using the alternative superspin approach.

The superspin approach relies on
interpreting the disorder-average retarded Green function
as a ``quantum thermal average'',
whereby the role of temperature is played by the length $L$
of the disordered quantum wire and,
for a disordered quasi-one-dimensional quantum wire,
the quantum partition function is given by Eqs.%
~(\ref{eq: final hamiltonian n chains})
and~(\ref{eq: partition fun 2}).
The quantum evolution,
although unitary according to
Eq.~(\ref{eq: proof unitarity quantum evolution}),
is here generated by the supersymmetric
and \textit{non-Hermitian} Hamiltonian%
~(\ref{eq: final hamiltonian n chains}).
In the chiral-unitary symmetry class,
the supersymmetric and non-Hermitian Hamiltonian%
~(\ref{eq: final hamiltonian n chains}) simplifies to the
supersymmetric and non-Hermitian Hamiltonian~(\ref{eq: final
hamiltonian n chains chiral-unitary}) so that it can be
interpreted as a $gl(1|1)$ quantum superspin Hamiltonian, for
it reduces to a bilinear form in the generators of the
superalgebra $gl(2|2)$ and, furthermore, it commutes with a
$gl(1|1)$ sub-superalgebra of $gl(2|2)$. In the thermodynamic
limit $L\to\infty$, the quantum statistical average is solely
determined by the non-degenerate right and left ground states
of Hamiltonian~(\ref{eq: final hamiltonian n chains
chiral-unitary}), whose existence is guaranteed by
supersymmetry. Thus, the computation of the density of states
in the thermodynamic limit amounts
to the construction of the right and left ground states%
~(\ref{eq: expansion of eigenstates})
for the supersymmetric and non-Hermitian Hamiltonian%
~(\ref{eq: final hamiltonian n chains chiral-unitary}).

Solving for these right and left ground states is
difficult because the right and left Hilbert spaces
on which the quantum superspin Hamiltonian is defined are
infinite dimensional. For comparison, the irreducible
Hilbert space of a single $SU(2)$-spin  is $(2s+1)$-dimensional.
For any given number $N$ of channels
in the disordered quasi-one-dimensional quantum wire,
the right and left ground states are constructed from
two limiting solutions to the recursion relation%
~(\ref{eq: recursion relations gen N})
for their expansion coefficients $a^{(n)}_{m}$.
Here, $n=1,\cdots,N$ is the channel index and
$m=1,2,\cdots$ is the basis index in the right Hilbert space.
More precisely, the recursion relation%
~(\ref{eq: recursion relations gen N})
was solved independently in the large $m\gg N$
and in the small $m\omega\ll 1$ limits, respectively,
with the help of a change of basis in the right Hilbert space,
i.e., by trading the expansion coefficients
$a^{(n)}_{m}$
for the expansion coefficients
$b^{(k)}_{m}$, $k=1,\cdots,N$.
In the large $m\gg N$ limit, the recursion relation%
~(\ref{eq: recursion relations gen N})
reduces to $N$ modified Bessel equations
whose leading solution for small $\omega$ is governed by
(i) the modified Bessel function $K^{\ }_{0}$ when $N$ is odd
or (ii) the modified Bessel function $K^{\ }_{1}$ when $N$ is even.
Evidently, this parity effect also characterizes the
solution to the recursion relation in the limit $m\omega\ll1$
provided $N\ll m\ll 1/\omega$, after both solutions have been
matched in their overlapping regime  $N\ll m\ll 1/\omega$ of validity.
In turn, the same parity effect has a counterpart for the density of states
for asymptotically small values of $\varepsilon$, since,
upon the analytical continuation $\omega\to-\mathrm{i}\varepsilon$,
(i) the coefficients $b^{(1)}_{m}$ dominate
over the coefficients  $b^{(k)}_{m}$ with $k=2,\cdots,N$
in the density of states
when $N$ is odd,
while (ii) the coefficients $b^{(1)}_{m}$ and $b^{(2)}_{m}$
dominate over the coefficients  $b^{(k)}_{m}$ with $k=3,\cdots,N$
in the density of states
when $N$ is even.

In the superspin approach, the expansion coefficients
$b^{(1)}_{m}$ when $N$ is odd and
$b^{(1)}_{m}$ and $b^{(2)}_{m}$ when $N$ is even
play the role, in the DMPK approach,
of the smallest Lyapunov exponent when $N$ is odd
and the first two  smallest Lyapunov exponents when $N$ is even,
respectively. The only effect from the number $N$
of channels is to determine
the size of the asymptotic regime $\varepsilon\ll1$ for which
Eq.~(\ref{eq: main result paper})
holds through the choice
$g^{\ }_{2}\propto v^{\ }_{\mathrm{F}}/(N^{2}\ell)$,
where $v^{\ }_{\mathrm{F}}$ is the Fermi velocity in the clean limit
and $\ell$ is the mean free path in the Born approximation,
for the variance in Eq.~(\ref{eq: gaussian disorder N chains}).

\medskip

\textbf{Acknowledgments}

We thank A.\ W.\ W.\ Ludwig for helpful discussions.
This research was supported in
part by the National Science Foundation under
Grant No. PHY05-51164.

\appendix

\section{
Symmetries of the Hamiltonian
        }
\label{app sec: Symmetries of the Hamiltonian}

In this appendix we will show that $H_{2}$ [as defined in
Eq.~\re{eq: final hamiltonian n chains chiral-unitary}]
commutes with the eight generators of the Lie superalgebra 
$gl(1|1)\oplus gl(1|1)$, while $H^{\ }_\omega$
commutes with the four generators of the diagonal sub-algebra
$gl(1|1)\subset gl(1|1)\oplus gl(1|1)$. It thus follows
that $H$ is invariant under all transformations of
$
(\mathfrak{F}^{\ }_{\mathrm{L}}, 
 \mathfrak{F}^{\ }_{\mathrm{R}})
$ 
induced by the Lie supergroup
$GL(1|1)$ generated by the Lie superalgebra
$gl(1|1)$. 
We refer to Ref.~\cite{DeWitt92} for an introduction on supermanifolds, 
to Ref.~\cite{Freund86} for an introduction on Lie superalgebra, 
and to Ref.~\cite{Frappat00} for a dictionary of Lie superalgebras.

The first step in the proof consists in rewriting $H^{\ }_{2}$
as a quadratic form in the generators of the
Lie superalgebra $gl(2|2)$. 
To this end, define the 16 operators $E^{\ }_{ab}$ with $a,b=1,\cdots,4$ by:

\noindent\textit{bosonic operators in the nonbar sector}
\begin{equation}
\begin{array}{lcl}
E^{\ }_{11}
\equiv -B^{}_{}
:=
-\sum\limits_{i=1}^{N}
f^{\dag}_{i}
f^{\ }_{i}
+\frac{1}{2}N,
&
\hphantom{AAAA}
&
E^{\ }_{33}
\equiv
-Q^{}_{}
:=
-\sum\limits_{i=1}^{N}
b^{\dag}_{i}
b^{\   }_{i}
-
\frac{1}{2}N,
\end{array}
\label{eq: SUSY generators a}
\end{equation}

\noindent\textit{fermionic operators in the nonbar sector}
\begin{equation}
\begin{array}{lcl}
E^{\ }_{13}
\equiv
+W^{}_{+}:=
+\sum\limits_{i=1}^{N}
b^{\dag}_{i}
f^{\ }_{i},
&
\hphantom{AAAAAAA}
&
E^{\ }_{31}
\equiv
-W^{}_{-}:=
-\sum\limits_{i=1}^{N}
f^{\dag}_{i}
b^{\ }_{i},
\end{array}
\label{eq: SUSY generators b}
\end{equation}

\noindent\textit{bosonic operators in the bar sector}
\begin{equation}
\begin{array}{lcl}
E^{\ }_{22}
\equiv
-\widebar{B}^{}_{}:=
-\sum\limits_{i=1}^{N}
\bar{f}^{\ }_{i}
\bar{f}^{\dag}_{i}
+
\frac{1}{2}N,
&
\hphantom{AAAA}
&
E^{\ }_{44}
\equiv
-\widebar{Q}^{}_{}
:=
+\sum\limits_{i=1}^{N}
\bar{b}^{\   }_{i}
\bar{b}^{\dag}_{i}
-
\frac{1}{2}N,
\end{array}
\label{eq: SUSY generators c}
\end{equation}

\noindent\textit{fermionic operators in the bar sector}
\begin{equation}
\begin{array}{lcl}
E^{\ }_{24}\equiv
+\widebar{W}^{}_{+}
:=
-\sum\limits_{i=1}^{N}
\bar{b}^{\   }_{i}
\bar{f}^{\dag}_{i},
&
\hphantom{AAAAAAA}
&
E^{\ }_{42} \equiv
-\widebar{W}^{}_{-}:=
-\sum\limits_{i=1}^{N}
\bar{f}^{\   }_{i}
\bar{b}^{\dag}_{i},
\end{array}
\label{eq: SUSY generators d}
\end{equation}

\noindent\textit{bosonic operators in the mixed bar and nonbar sector}
\begin{equation}
\begin{array}{lcl}
E^{\ }_{12}\equiv
-A^{}_{+}:=
-\sum\limits_{i=1}^{N}
     f ^{\ }_{i}
\bar{f}^{\ }_{i},
&
\hphantom{AAAAAAA:}
&
E^{\ }_{21}\equiv
+A^{}_{- }:=
+\sum\limits_{i=1}^{N}
     f ^{\dag}_{i}
\bar{f}^{\dag}_{i},
\\
E^{\ }_{34}\equiv
+D^{}_{+}:=
-\sum\limits_{i=1}^{N}
     b ^{\ }_{i}
\bar{b}^{\ }_{i},
&
\hphantom{AAAA}
&
E^{\ }_{43}\equiv
+D^{}_{-}:=
+\sum\limits_{i=1}^{N}
     b ^{\dag}_{i}
\bar{b}^{\dag}_{i},
\end{array}
\label{eq: SUSY generators e}
\end{equation}

\noindent\textit{fermionic operators in the mixed bar and nonbar sector}
\begin{equation}
\begin{array}{lcl}
E^{\ }_{14}\equiv
-S^{}_{+}:=
+\sum\limits_{i=1}^{N}
     f ^{\ }_{i}
\bar{b}^{\ }_{i},
&
\hphantom{AAAAAAAA}
&
E^{\ }_{41}\equiv
+S^{}_{- }:=
+\sum\limits_{i=1}^{N}
     f ^{\dag}_{i}
\bar{b}^{\dag}_{i},
\\
E^{\ }_{23}\equiv
-\widebar{S}^{}_{+}:=
-\sum\limits_{i=1}^{N}
     b ^{\dag}_{i}
\bar{f}^{\dag}_{i},
&
\hphantom{AAAA}
&
E^{\ }_{32}\equiv
+\widebar{S}^{}_{-}:=
+\sum\limits_{i=1}^{N}
     b ^{\ }_{i}
\bar{f}^{\ }_{i}.
\end{array}
\end{equation}
The 16 operators $E^{\ }_{ab}$ with $a,b=1,\cdots,4$
are either bosonic or fermionic.
Hence, any  $E^{\ }_{ab}$ with $a,b=1,\cdots,4$
can be assigned the degree (or grade)
$0$ or $1$ if they are bosonic or fermionic, respectively,
a mapping that we shall denote by
$\mathrm{deg}(E^{\ }_{ab})$.
Our rational for the definition of the 16 operators
$E^{\ }_{ab}$ with $a,b=1,\cdots,4$
is that if we define
\begin{equation}
[a]:=
\left\{
\begin{array}{ll}
0,
&
\hbox{ if $a=1,2$,}
\\
1,
&
\hbox{ if $a=3,4$,}
\end{array}
\right.
\end{equation}
then it follows that
\begin{equation}
\mathrm{deg}(E^{\ }_{ab})=
\big([a]+[b]\big)\, \mathrm{mod}\, 2.
\end{equation}
We can now define the 256 supercommutators
\begin{equation}
\left[
E^{\ }_{ab},
E^{\ }_{cd}
\right]:=
E^{\ }_{ab}
E^{\ }_{cd}
-
(-1)^{\mathrm{deg}(E^{\ }_{ab})\,\mathrm{deg}(E^{\ }_{cd})}
E^{\ }_{cd}
E^{\ }_{ab},
\qquad
a,b,c,d=1,\cdots,4,
\end{equation}
and verify that they define the Lie superalgebra 
of $gl(2|2)$ in the \emph{standard basis}~\cite{Zhang05}, i.e.,
\begin{equation}
\left[E^{\ }_{ab},E^{\ }_{cd}\right]=
\delta^{\ }_{bc}
E^{\ }_{ad}
-
(-1)^{\left([a]+[b]\right)\left([c]+[d]\right)}
\delta^{\ }_{ad}
E^{\ }_{cb},
\qquad
a,b,c,d=1,\cdots,4.
\end{equation}
The first step of the proof is then completed after verifying that
$H_2$, as given by Eqs.\
\re{eq: final hamiltonian n chains chiral-unitary} and \re{def of Jij},
can be rewritten as
\begin{equation}
\begin{split}
&
H^{\ }_{2}
=
g^{\ }_{2} \bigg[
\left( A_+ \right)^2
+ 2 B \widebar{B}
- 2 S_- \widebar{S}_+
-  2 W_- \widebar{W}_+
+  \left( A_- \right)^2
\\
&
\hphantom{H^{\ }_{\mathrm{V}}=g^{\ }_{2}\Bigg]}
+ 2 \widebar{W}_- W_+
+ 2 \widebar{S}_-  S_+
- \left( D_+ \right)^2
- 2 Q \widebar{Q}
- \left( D_- \right)^2
\bigg],
\end{split}
\end{equation}
or alternatively
\begin{equation}
\begin{split}
&
H^{\ }_{2}
=
g^{\ }_{2}
\bigg[
\left( E^{\ }_{12} \right)^2
+ 2 E^{\ }_{11} E^{\ }_{22}
+ 2 E^{\ }_{41} E^{\ }_{23}
+ 2 E^{\ }_{31} E^{\ }_{24}
+ \left( E^{\ }_{21} \right)^2
\\
&
\hphantom{H^{\ }_{\mathrm{V}}=g^{\ }_{2}\Bigg]}
+ 2 E^{\ }_{42} E^{\ }_{13}
+ 2   E^{\ }_{32} E^{\ }_{14}
- \left( E^{\ }_{34} \right)^2
- 2 E^{\ }_{33} E^{\ }_{44}
- \left( E^{\ }_{43} \right)^2
\bigg].
\end{split}
\end{equation}
Observe that
\begin{equation}
H^{\ }_{\omega}=
\omega
\left(
B-\bar{B}
+
Q-\bar{Q}
\right)
=
\omega
\left(
E^{\ }_{22}
-
E^{\ }_{11}
+
E^{\ }_{44}
-
E^{\ }_{33}
\right).
\end{equation}

In the second step, one verifies that
$H^{\ }_{2}$
commutes with the 8 operators
\begin{equation}
\begin{split}
&
\mathcal{G}^{\ }_{1}:=
\frac{1}{2}
\left(
-E^{\ }_{33} - E^{\ }_{44}
+ E^{\ }_{34} + E^{\ }_{43}
\right),
\quad
\mathcal{H}^{\ }_{1}:=
\frac{1}{2}
\left(
- E^{\ }_{11} - E^{\ }_{22}
+ E^{\ }_{12}+E^{\ }_{21}
\right),
\\
&
\mathcal{E}^{+}_{1}:=
\frac{1}{2}
\left(
+E^{\ }_{13}+E^{\ }_{24}
-E^{\ }_{14}-E^{\ }_{23}
\right),
\quad
\mathcal{E}^{-}_{1}:=
\frac{1}{2}
\left(
-E^{\ }_{31}-E^{\ }_{42}
+E^{\ }_{41}+E^{\ }_{32}
\right),
\\
&
\mathcal{G}^{\ }_{2}:=
\frac{1}{2}
\left(
-E^{\ }_{33} - E^{\ }_{44}
- E^{\ }_{34} - E^{\ }_{43}
\right),
\quad
\mathcal{H}^{\ }_{2}:=
\frac{1}{2}
\left(
- E^{\ }_{11} - E^{\ }_{22}
- E^{\ }_{12} - E^{\ }_{21}
\right),
\\
&
\mathcal{E}^{+}_{2}:=
\frac{1}{2}
\left(
+E^{\ }_{13}+E^{\ }_{24}
+E^{\ }_{14}+E^{\ }_{23}
\right),
\quad
\mathcal{E}^{-}_{2}:=
\frac{1}{2}
\left(
-E^{\ }_{31}-E^{\ }_{42}
-E^{\ }_{41}-E^{\ }_{32}
\right),
\end{split}
\end{equation}
that generate the Lie superalgebra
$gl(1|1)\oplus gl(1|1)$
with the only nonvanishing supercommutators%
~\cite{Goetz07,Frappat98}
\begin{equation}
\begin{split}
&
\left[\mathcal{G}^{\ }_{1},\mathcal{E}^{\pm}_{1}\right]=
\pm \mathcal{E}^{\pm}_{1},
\quad
\left[\mathcal{H}^{\ }_{1},\mathcal{E}^{\pm}_{1}\right]=
\mp \mathcal{E}^{\pm}_{1},
\quad
\left\{\mathcal{E}^{+}_{1},\mathcal{E}^{-}_{1}\right\}=
\mathcal{G}^{\ }_{1}+\mathcal{H}^{\ }_{1},
\\
&
\left[\mathcal{G}^{\ }_{2},\mathcal{E}^{\pm}_{2}\right]=
\pm\mathcal{E}^{\pm}_{2},
\quad
\left[\mathcal{H}^{\ }_{2},\mathcal{E}^{\pm}_{2}\right]=
\mp\mathcal{E}^{\pm}_{2},
\quad
\left\{\mathcal{E}^{+}_{2},\mathcal{E}^{-}_{2}\right\}=
\mathcal{G}^{\ }_{2}+\mathcal{H}^{\ }_{2}.
\end{split}
\end{equation}

At last, one verifies that the full Hamiltonian
$H=H^{\ }_{\omega}+H^{\ }_{2}$
commutes with the 4 operators
\begin{equation}
\begin{split}
&
\mathcal{B}:=
\mathcal{H}_{1}+ \mathcal{H}_{2}=
\sum_{i=1}^{N}
\left(
f^{\dag}_{i}
f^{\   }_{i}
-
\bar f^{\dag}_{i}
\bar f^{\   }_{i}
\right),
\\
&
\mathcal{Q}:=
\mathcal{G}_{1}+\mathcal{G}_{2}=
\sum_{i=1}^{N}
\left(
b^{\dag}_{i}
b^{\   }_{i}
+
\bar b^{\dag}_{i}
\bar b^{\   }_{i}
\right),
\\
&
\mathcal{W}^{\ }_{+}:=
\mathcal{E}^{+}_{1}+\mathcal{E}^{+}_{2}=
\sum_{i=1}^{N}
\left(
b^{\dag}_{i}
f^{\   }_{i}
+
\bar f^{\dag}_{i}
\bar b^{\   }_{i}
\right),
\\
&
\mathcal{W}^{\ }_{-}:=
\mathcal{E}^{-}_{1}+\mathcal{E}^{-}_{2}=
\sum_{i=1}^{N}
\left(
f^{\dag}_{i}
b^{\   }_{i}
+
\bar b^{\dag}_{i}
\bar f^{\   }_{i}
\right),
\end{split}
\label{eq: gl(1|1)-generators for AIII}
\end{equation}
that generate the Lie superalgebra $gl(1|1)$
whose only nonvanishing supercommutators are
\begin{equation}
\begin{split}
\left[
\mathcal{Q}
-
\mathcal{B},
\mathcal{W}^{\ }_{\pm}
\right]=
\pm
2\mathcal{W}^{\ }_{\pm},
\qquad
\left\{
\mathcal{W}^{\ }_{+},
\mathcal{W}^{\ }_{-}
\right\}=
\mathcal{Q}
+
\mathcal{B}.
\end{split}
\end{equation}
Thus, $H^{\ }_{\omega}$
breaks the symmetry of $H^{\ }_{2}$
generated by $gl(1|1)\oplus gl(1|1)$ down to
the one generated by $gl(1|1)$.

\section{
Useful identities
        }
\label{app sec: Useful identities}

The generalized hypergeometric functions
have the series expansion~\cite{Gradshteyn94}
\begin{subequations}
\begin{equation}
{}_pF_{q}
\left[
\begin{array}{c}
\alpha^{\ }_{1},\cdots,\alpha^{\ }_{p}
\\
\beta^{\ }_{1},\cdots,\beta^{\ }_{p}
\end{array};
z
\right]=
\sum_{n=0}^{\infty}
\frac{
\left(\alpha^{\ }_{1}\right)^{\ }_{n}
\cdots
\left(\alpha^{\ }_{p}\right)^{\ }_{n}
     }
     {
\left(\beta^{\ }_{1}\right)^{\ }_{n}
\cdots
\left(\beta^{\ }_{q}\right)^{\ }_{n}
     }
\frac{z^{n}}{n!}
\end{equation}
where
$\alpha^{\ }_{1},\cdots,\alpha^{\ }_{p}$
and
$\beta^{\ }_{1},\cdots,\beta^{\ }_{q}$
are complex-valued parameters,
$z$ is complex-valued with magnitude less than unity, and
\begin{equation}
\left(\alpha\right)^{\ }_{n}:=
\frac{
\Gamma(\alpha+n)
     }
     {
\Gamma(\alpha)
     },
\end{equation}
\end{subequations}
with $\Gamma(z)$ the gamma function,
denotes the Pochhammer symbol.

We introduce the short-hand notations
\begin{subequations}
\begin{eqnarray}
&&
F_1[k,m,n]
:=
{}_3F_{2}
\left[
\begin{array}{c}
-k, k, -n
\label{prop: HYP GEO 1a}
\\
\frac{1}{2} , -m,
\end{array};1
\right],
\\
&&
F_2[k,m,n]
:=
{}_3F_{2}
\left[
\begin{array}{c}
-k, 2+k, -n
\\
\frac{3}{2} , -m,
\end{array}; 1
\right],
\label{prop: HYP GEO 2a}
\\
&&
F_3[k,m,n]
:=
{}_3F_{2}
\left[
\begin{array}{c}
-k, k+1, -n
\\
\frac{1}{2} , -m,
\end{array}; 1
\right],
\label{prop: HYP GEO 3a}
\\
&&
F_4[k,m,n]
:=
{}_3F_{2}
\left[
\begin{array}{c}
-k, k+1, -n
\\
\frac{3}{2} , -m,
\end{array}; 1
\right],
\label{prop: HYP GEO 4a}
\end{eqnarray}
\end{subequations}
for the family of hypergeometric functions labeled
by the positive integers $k$, $m$, and $n$. They satisfy the identities
\begin{subequations}
\begin{eqnarray}
0&=&
\left[
 4n(2n-2m)
-
(2m - 4 k^2)
\right]
F_1 [k,m,n]
\nonumber\\
&&
+ 2(2n+1)(m-n) F_1 [k,m,n+1]
+ 2n(2m+1-2n) F_1 [k,m,n-1],
\label{prop: HYP GEO 1b}
\\
0&=&
\left\{
4n(2n-2m)
-
[6m + 4 -4 (k+1)^2]
\right\}
F_2[k,m,n]
\nonumber\\
&&
+ 2(2n+3)(m-n) F_2[k,m,n+1]
+ 2n(2m+3-2n) F_2[k,m,n-1],
\label{prop: HYP GEO 2b}
\\
0&=&
\left\{
4n(2n-2m-1)
-
\left[2m - 4 k(k+1)\right]
\right\}
F_3 [k,m,n]
\nonumber\\
&&
+ 2(2n+1)(m-n) F_3 [k,m,n+1]
+ 2n(2m+3-2n) F_3 [k,m,n-1],
\label{prop: HYP GEO 3b}
\\
0&=&
\left\{
4n(2n+1-2m)
-
\left[6m-4k(k+1)\right]
\right\}
F_4[k,m,n]
\nonumber\\
&&
+ (2n+3)(2m-2n) F_4[k,m,n+1]
+ 2n(2m+1-2n) F_4[k,m,n-1],
\label{prop: HYP GEO 4b}
\end{eqnarray}
\end{subequations}
respectively.

Furthermore, we introduce the notation
\begin{subequations}
\begin{eqnarray}
&&
F^{-1}_3 [k,m,n]
:=
\frac{1}{(m-k)! (m+k+1)! }
{}_3F_{2}
\left[
\begin{array}{c}
-k -m-1 , k -m , -n
\\
-m - \frac{1}{2} , -m,
\end{array}; 1
\right],
\\
&&
F^{-1}_4 [k,m,n]
:=
\frac{ (2k+1)^2}
{(m-k)! (m+k+1)!}
{}_3F_{2}
\left[
\begin{array}{c}
-k -m-1 , k -m , -n
\\
-m  , \frac{1}{2} -m,
\end{array}; 1
\right],
\end{eqnarray}
\end{subequations}
with the positive integers
$k$, $m$, and $n$.
They satisfy the identities
\begin{subequations}
\begin{eqnarray} \label{identity for inverse of F3}
&&
\sum_{k=0}^m
F^{-1}_3 [k,m,n^{\phantom{\prime}} ] \,
F^{\vphantom{-1}}_3 [k,m,n^{\prime} ]
=
(-1)^n  4^m
\frac{ n! (m-n)! }{m! (2m+1)!} \,
\delta_{n^{\ },n^{\prime}},
\\
 \label{identity for inverse of F4}
&&
\sum_{k=0}^m
F^{-1}_4 [k,m,n^{\phantom{\prime}} ] \,
F^{\vphantom{-1}}_4 [k,m,n^{\prime} ]
=
(-1)^n  4^m
\frac{ n! (m-n)! }{m! (2m)!} \,
\delta_{n^{\ },n^{\prime}},
\end{eqnarray}
\end{subequations}
for $n$, $n^{\prime} \in \left\{ 0, 1,2,\cdots, m \right\}$ .


\begin{thebibliography}{99}

\bibitem{Anderson58}
P. W. Anderson,
Phys.\ Rev.\ \textbf{109}, 1492 (1958).

\bibitem{Lee85}
P. A. Lee and R. V. Ramakrishnan,
Rev.\ Mod.\ Phys.\ \textbf{57}, 287 (1985).

\bibitem{Kramer93}
B. Kramer and A. MacKinnon,
Rep.\ Prog.\ Phys.\ \textbf{56}, 1469 (1993).

\bibitem{Kramer05}
B.\ Kramer, T.\ Ohtsuki, and S. Kettemann,
Phys.\ Rep.\ \textbf{417}, 211 (2005).

\bibitem{Evers08}
F.\ Evers and A.\ D.\ Mirlin,
Rev.\ Mod.\ Phys.\ \textbf{80}, 1355 (2008).

\bibitem{Frohlich83}
J. Fr\"ohlich and T. Spencer,
Commun.\ Math.\ Phys.\ \textbf{88}, 151 (1983).

\bibitem{Dyson62}
F. J. Dyson,
J.\ Math.\ Phys.\ \textbf{6}, 1199 (1962).

\bibitem{Verbaarschot94}
J. J. M. Verbaarschot,
Phys.\ Rev.\ Lett.\ \textbf{72}, 2531 (1994).

\bibitem{Zirnbauer96}
M. R. Zirnbauer,
J.\ Math.\ Phys.\ \textbf{37}, 4986 (1996). 	

\bibitem{Altland97}
A. Altland and M. R. Zirnbauer,
Phys.\ Rev.\ B \textbf{55}, 1142 (1997).

\bibitem{Caselle04}
M.\ Caselle and W.\ Magnea,
Phys.\ Rep.\ \textbf{394}, 41 (2004).

\bibitem{Heinzner05}
P. Heinzner, A. Huckleberry, and M. R. Zirnbauer,
Commun.\ Math.\ Phys.\ \textbf{257}, 725 (2005).


\bibitem{Helgason78}
S. Helgason,
\textit{Differential Geometry, Lie Groups, and Symmetric Spaces}
(Academic Press, San Diego, 1978).

\bibitem{footnote mobility edge} 
Slightly abusing terminology,
we also call a quantum critical point that separates two
insulating phases (instead of separating a metallic from an
insulating phase) a mobility edge in a problem of Anderson
localization.

\bibitem{Wegner79}
F. J. Wegner,
Z.\ Physik B \textbf{35}, 207 (1979).

\bibitem{Hikami80}
S. Hikami, A. I. Larkin, and Y. Nagaoka,
Prog.\ Theor.\ Phys.\ \textbf{63}, 707 (1980).

\bibitem{Schafer80}
L. Sch\"afer and R. J. Wegner,
Z.\ Physik B \textbf{38}, 113 (1980).

\bibitem{Efetov80}
K.\ B.\ Efetov, A.\ I.\ Larkin, and D.\ E.\ Khemlnitskii,
Zh.\ Eksp.\ Teor.\ Fiz.\ \textbf{79}, 1120 (1980)
[Sov.\ Phys.\ JETP \textbf{52}, 568 (1980)].

\bibitem{Houghton80}
A. Houghton, A. Jevicky, R. D. Kenway, and A. M. M. Pruisken,
Phys.\ Rev.\ Lett.\ \textbf{45}, 394 (1980).

\bibitem{Hikami81}
S. Hikami,
Phys.\ Rev.\ B \textbf{24}, 2671 (1981).

\bibitem{Efetov83}
K. B. Efetov,
Adv.\ Phys.\ \textbf{32}, 53 (1983).

\bibitem{Pruisken84}
H. Levine, S. B. Libby, and A. M. M. Pruisken,
Phys.\ Rev.\ Lett.\ \textbf{51}, 1915 (1983);
A. M. M. Pruisken,
Nucl.\ Phys.\ B \textbf{235}, 277 (1984).

\bibitem{Oppermann90}
R. Oppermann,
Physica (Amsterdam) \textbf{167A}, 301 (1990).

\bibitem{Gade93}
R. Gade,
Nucl.\ Phys.\ B \textbf{398}, 499 (1993);
R. Gade and F. Wegner,
\textit{ibid}. \textbf{360}, 213 (1991).

\bibitem{Zirnbauer92}
M. R. Zirnbauer,
Phys.\ Rev.\ Lett.\ \textbf{69}, 1584 (1992).

\bibitem{Bundschuh98}
R. Bundschuh, C. Cassanello, D. Serban, and M. R. Zirnbauer,
Nucl.\ Phys.\ B \textbf{532}, 689 (1998);
\textit{ibid} Phys.\ Rev.\ B \textbf{59}, 4382 (1999).

\bibitem{Senthil98}
T. Senthil, M. P. A. Fisher, L. Balents, and C. Nayak,
Phys.\ Rev.\ Lett.\ \textbf{81}, 4704 (1998);
T. Senthil and M. P. A. fisher,
Phys.\ Rev.\ B \textbf{60}, 6893 (1999).

\bibitem{Senthil00}
T. Senthil and M. P. A. Fisher,
Phys.\ Rev.\ B \textbf{61}, 9690 (2000).

\bibitem{Zirnbauer00}
M. Bocquet, D. Serban, and M. R. Zirnbauer,
Nucl.\ Phys.\ B \textbf{578}, 628 (2000).

\bibitem{Read00}
N. Read and D. Green,
Phys.\ Rev.\ B \textbf{61}, 10267 (2000).

\bibitem{Guruswamy00}
S. Guruswamy, A. LeClair, and A. W. W. Ludwig,
Nucl.\ Phys.\ B \textbf{583}, 475 (2000).

\bibitem{Fendley01}
P. Fendley,
Phys.\ Rev.\ B \textbf{63}, 104429 (2001).

\bibitem{Altland01}
A. Altland and R. Merkt,
Nucl.\ Phys.\ B \textbf{607}, 511 (2001).

\bibitem{Lamacraft04}
A. Lamacraft, B. D. Simons, and M. R. Zirnbauer,
Phys.\ Rev.\ B \textbf{70}, 075412 (2004).

\bibitem{Ryu06}
S. Ryu, A. Furusaki, A. W. W. Ludwig, and C. Mudry,
Nucl.\ Phys.\ B \textbf{780}, 105 (2007).

\bibitem{Ostrovsky07}
P. M. Ostrovsky, I. V. Gornyi, and A. D. Mirlin,
Phys.\ Rev.\ Lett.\ \textbf{98}, 256801 (2007).

\bibitem{Ryu07}
S. Ryu, C. Mudry, H. Obuse, and A. Furusaki,
Phys.\ Rev.\ Lett.\ \textbf{99}, 116601 (2007).

\bibitem{Chalker88}
J. T. Chalker and P. D. Coddington,
J.\ Phys.\ C \textbf{21}, 2665 (1988).

\bibitem{DKKLee94}
D. K. K. Lee and J. T. Chalker,
Phys.\ Rev.\ Lett.\ \textbf{72}, 1510 (1994);
D. K. K. Lee, J. T. Chalker, and D. Y. K. Ko,
Phys.\ Rev.\ B \textbf{50}, 5272 (1994).

\bibitem{Kim95}
Y.\ B.\ Kim, A.\ Furusaki, and D.\ K. Lee,
Phys.\ Rev.\ B \textbf{52}, 16646 (1995).

\bibitem{Chalker95}
J. T. Chalker and A. Dohmen,
Phys.\ Rev.\ Lett.\ \textbf{75}, 4496 (1995).

\bibitem{Klesse95}
R. Klesse and R. Metzler,
Europhys.~Lett.~\textbf{32}, 229 (1995).

\bibitem{Ho96}
C.-M. Ho and J.T. Chalker,
Phys.~Rev.~B \textbf{54}, 8708 (1996).

\bibitem{Cho97}
S. Cho and M. P. A. Fisher,
Phys.~Rev.~B \textbf{55}, 1025 (1997).

\bibitem{Kagalovsky97}
V. Kagalovsky, B. Horovitz, and Y. Avishai,
Phys.~Rev.~B \textbf{55}, 7761 (1997).

\bibitem{Merkt98}
R. Merkt, M. Janssen, and B. Huckestein,
Phys.\ Rev.\ B \textbf{58}, 4394 (1998).

\bibitem{Kagalovsky99}
V. Kagalovsky, B. Horovitz, Y. Avishai, and J.T. Chalker,
Phys.~Rev.~Lett.~\textbf{82}, 3516 (1999).

\bibitem{Chalker01}
J. T. Chalker, N. Read, V. Kagalovsky,
B. Horovitz, Y. Avishai, and A. W. W. Ludwig,
Phys.~Rev.~B \textbf{65}, 012506 (2001).

\bibitem{Bocquet03}
M.\ Bocquet and J. T. Chalker,
Phys.~Rev.~B \textbf{67}, 054204 (2003).

\bibitem{Obuse07}
H.\ Obuse, A.\ Furusaki, S.\ Ryu, C.\ Mudry,
Phys.\ Rev.\ B \textbf{76}, 075301 (2007);
\textit{ibid},
Phys.~Rev.~B \textbf{78}, 115301 (2008).

\bibitem{Huckestein95}
B. Huckestein,
Rev.\ Mod.\ Phys.\ \textbf{67}, 357 (1995).

\bibitem{Read91} 
N. Read, unpublished, 1991.

\bibitem{DHLee94}
D.-H.\ Lee,
Phys.~Rev.~B \textbf{50}, 10788 (1994).

\bibitem{Zirnbauer94}
M. R. Zirnbauer,
Ann.~Phys.~(Leipzig) \textbf{3}, 513 (1994).

\bibitem{Lee96}
D.-H. Lee and Z. Wang,
Philos.~Mag.~Lett.~ \textbf{73}, 145 (1996).

\bibitem{Kondev97}
J. Kondev and J.B. Marston,
Nucl.~Phys.~B \textbf{497}, 639 (1997).

\bibitem{Zirnbauer97}
M. R. Zirnbauer,
J.\ Math.\ Phys.\ \textbf{38}, 2007 (1997)
and 2197 (1999).

\bibitem{Marston99}
J. B. Marston and S.-W. Tsai,
Phys.\ Rev.\ Lett.\ \textbf{82}, 4906 (1999).

\bibitem{Affleck86}
I. Affleck,
Nucl.\ Phys.\ B \textbf{265}, 409 (1986).

\bibitem{Gruzberg99}
I. A. Gruzberg, A. W. W. Ludwig, and N. Read,
Phys.~Rev.~Lett.~\textbf{82}, 4524 (1999).

\bibitem{Cardy00}
J. Cardy,
Phys.~Rev.~Lett.~\textbf{84}, 3507 (2000).

\bibitem{Beamond02} E. J. Beamond, J. Cardy, and J. T. Chalker,
    Phys. Rev. B~\textbf{65}, 214301 (2002).

\bibitem{Mirlin03} A. D. Mirlin, F. Evers, and A. Mildenberger,
    J. Phys. A~\textbf{36}, 3255 (2003).

\bibitem{Cardy05}
J. Cardy,
Commun.\ Math.\ Phys.\ \textbf{258}, 87 (2005).

\bibitem{Dyson53}
F. J. Dyson,
Phys.\ Rev.\ \textbf{92}, 1331 (1953).

\bibitem{Smith70}
E. R. Smith,
J.\ Phys.\ C \textbf{3}, 1419 (1970).

\bibitem{Theodorou76}
G. Theodorou and M. H. Cohen,
Phys. Rev. B \textbf{13}, 4597 (1976).

\bibitem{Fleishman77}
L.\ Fleishman and D.\ C.\ Licciardello,
J. Phys. C {\bf 10}, L125 (1977).

\bibitem{Ovchinnikov77}
A. A. Ovchinnikov and N. S. \'Erikhman,
Zh.\ Eksp.\ Teor.\ Fiz.\ \textbf{73}, 650 (1977)
[Sov.\ Phys.\ JETP \textbf{46}, 340 (1977)].

\bibitem{Berezinskii73}
V. L. Berezinskii,
Zh. Eksp. Teor. Fiz. \textbf{65}, 1251 (1973)
[Sov.\ Phys.\ JETP \textbf{38}, 620 (1974)].

\bibitem{Gogolin77}
A. A. Gogolin and V. I. Melnikov,
Zh.\ Eksp.\ Teor.\ Fiz.\ \textbf{73}, 706 (1977)
[Sov.\ Phys.\ JETP \textbf{46}, 369 (1977)];
see also
M. Steiner, M. Fabrizio, and A. O. Gogolin,
Phys.\ Rev.\ B \textbf{57}, 8290 (1998).

\bibitem{Eggarter78}
T. P. Eggarter and R\ Riedinger,
Phys. Rev. B {\bf 18}, 569 (1978).

\bibitem{Stone81}
A.\ D.\ Stone, and J.\ D.\ Joannopoulos,
Phys.\ Rev.\ B \textbf{24}, 3592 (1981).

\bibitem{Witten81}
E. Witten,
Nucl.\ Phys.\ B \textbf{188}, 513 (1981).

\bibitem{Comtet95}
J. P. Bouchaud, A. Comtet, A. Georges, and P. Le Doussal,
Ann.\ Phys.\ (N.Y.) \textbf{201}, 285 (1990);
A. Comtet, J. Desbois, and C. Monthus,
Ann.\ Phys.\ (NY) \textbf{239}, 312 (1995).

\bibitem{Shelton98}
D. G. Shelton and A. M. Tsvelik,
Phys.\ Rev.\ B \textbf{57}, 14242 (1998).

\bibitem{Balents97}
L.\ Balents and  M.\ P.\ A.\ Fisher,
Phys. Rev. B \textbf{56}, 12970 (1997).

\bibitem{Bocquet99}
M. Bocquet,
Nucl.\ Phys.\ B \textbf{546}, 621 (1999).

\bibitem{Gruzberg01} I. A. Gruzberg, N. Read,
    and A. W. W. Ludwig, Phys.\ Rev.\ B \textbf{63}, 104422
    (2001).

\bibitem{Motrunich02}
O. Motrunich, K. Damle, and D. A. Huse,
Phys.\ Rev.\ B \textbf{65}, 064206 (2002).

\bibitem{Mudry03}
C. Mudry, S. Ryu, and A. Furusaki,
Phys.\ Rev.\ B \textbf{67}, 064202 (2003).

\bibitem{Brouwer00}
P. W. Brouwer, C. Mudry, and A. Furusaki,
Phys.\ Rev.\ Lett.\ \textbf{84}, 2913 (2000).

\bibitem{Titov01}
M. Titov, P. W. Brouwer, A. Furusaki, and C. Mudry,
Phys.\ Rev.\ B \textbf{63}, 235318 (2001).

\bibitem{DMPK}
O. N. Dorokhov,
Pis'ma Zh.\ Eksp.\ Teor.\ Fiz.\
\textbf{36}, 259 (1982) [JETP Letters {\bf 36}, 318];
P. A. Mello, P. Pereyra, and N. Kumar,
Ann.\ Phys.\ (NY) \textbf{181}, 290 (1988).

\bibitem{Mudry99}
C. Mudry, P. W. Brouwer, and A. Furusaki,
Phys.\ Rev.\ B \textbf{59},13221 (1999).

\bibitem{Huffmann90}
A. H\"uffmann:
J.\ Phys.\ A \textbf{23}, 5733 (1990).

\bibitem{Brouwer00b}
P. W. Brouwer, C. Mudry, and A. Furusaki,
Nucl.\ Phys.\ B \textbf{565}, 653 (2000);
P. W. Brouwer, A. Furusaki, C. Mudry, and S. Ryu,
BUTSURI \textbf{60}, 935 (2005)
(see cond-mat/0511622 for english version).

\bibitem{Furusaki00}
P. W. Brouwer, A. Furusaki, I. A. Gruzberg, and C. Mudry,
Phys.\ Rev.\ Lett.\ \textbf{85}, 1064 (2000).

\bibitem{Motrunich01}
O. Motrunich, K. Damle, and D. A. Huse,
Phys.\ Rev.\ B \textbf{63}, 224204 (2001).

\bibitem{Gruzberg05}
I. A. Gruzberg, N. Read, and S. Vishveshwara,
Phys.\ Rev.\ B \textbf{71}, 245124 (2005).

\bibitem{Bunder01}
J. E. Bunder and R. H. McKenzie,
Nucl.\ Phys.\ B \textbf{592}, 445 (2001).

\bibitem{Abramowitz65}
M.~Abramowitz and I.~A.~Stegun,
\emph{Handbook of Mathematical Functions},
Dover, New-York (1965).

\bibitem{Gradshteyn94}
I. S. Gradshteyn and I. M. Ryzhik,
\emph{Tables of Integrals, Series, and Products},
5th Ed., Academic Press, London (1994).

\bibitem{Frappat00}
L. Frappat, A.\ Sciarrino, and P. Sorba,
\textit{Dictionary on Lie Algebras and superalgebras},
(Academic Press, London 2000).

\bibitem{DeWitt92}
B. DeWitt,
\textit{Supermanifolds, Second edition},
(Cambridge University Press, New York 1992).

\bibitem{Freund86}
P. G. O. Freund,
\textit{Supersymmetry},
(Cambridge University Press, New York 1986).

\bibitem{Zhang05}
Y.-Z. Zhang and M. D. Gould,
J.~Math.~Phys. \textbf{46}, 013505 (2005).

\bibitem{Goetz07}
G. G\"otz, T. Quella, and V. Schomerus,
J.\ Algebra \textbf{312}, 829 (2007).

\bibitem{Frappat98}
L. Frappat, V. Hussin, and G. Rideau,
J.\ Phys.\ A: Math Gen. \textbf{31}, 4049 (1998).

\end{thebibliography}
\end{document}